\documentclass[aps,prx,twocolumn]{revtex4-2}
\usepackage[utf8]{inputenc}
\usepackage[english]{babel}
\usepackage[T1]{fontenc}
\usepackage{lmodern}
\usepackage{graphicx,color,comment}
\usepackage{amsmath}
\usepackage{amsfonts}
\usepackage{amssymb}
\usepackage{microtype}
\usepackage{braket}
\usepackage{subfigure}
\usepackage{etoolbox}
\usepackage[breaklinks=true,colorlinks,citecolor=blue]{hyperref}

\newcommand{\ct}{\cite}

\newcommand{\beq}{\begin{eqnarray}}
\newcommand{\eeq}{\end{eqnarray}}
\newcommand{\be}{\begin{equation}}
\newcommand{\ee}{\end{equation}}

\newcommand{\asp}[1]{\textcolor{blue}{#1}}

\begin{document}

\title{Late-time critical behavior of local  string-like observables under quantum quenches}
\author{Souvik Bandyopadhyay$^1$} 
\email{souvik@iitk.ac.in}
\author{Anatoli Polkovnikov $^2$}
\author{Amit Dutta$^1$}
\affiliation{$^1$Department of Physics, Indian Institute of Technology Kanpur, Kanpur 208016, India}
\affiliation{$^2$Department of Physics, Boston University, Boston, Massachusetts 02215, USA}

\begin{abstract}
	In recent times it has been observed that signatures of equilibrium quantum criticality surprisingly show up in many-body systems which are manifestly far from equilibrium.  We explore such scenarios in interacting spin systems subject to a quench and develop a robust method to systematically probe ground state critical physics through nonequilibrium post-quench dynamics. Analyzing late-time behavior of finite string-like observables, we find emerging sharp signatures of equilibrium criticality.  Specifically, these observables accurately detect equilibrium critical points and universal scaling exponents after long times following a quench. This happens despite the fact that the analyzed systems are strongly chaotic/ergodic and is interestingly due to a strong memory of the initial conditions retained by these observables after quench. We find that our results can also be used to explain critical signatures in post-quench domain formation, seen in a recent experiment with trapped ion quantum simulators. 
\end{abstract}

\maketitle

\section{Introduction}
\label{sec:intro}

Understanding phase transitions have been a central theme in the study of many body statistical mechanics. A plethora of studies into the nature and classification of phase transitions \cite{chaikin_lubensky,goldenfeld,stanley} has led to a comprehensive theoretical and experimental understanding of the phenomena. Phase transitions are known to manifest as nonanalyticities in free energy densities at finite temperatures and energy at very low temperatures. Furthermore, it has been well established that a critical system exhibits a diverging correlation length in the thermodynamic limit and obeys universal scaling laws which are not dependent on microscopic details of the system. Similar to thermal phase transitions driven by thermal fluctuations, a class of phase transitions in the zero temperature ground state of quantum many-body systems have also been discovered \cite{sachdev_qpt,dutta15_book}. These quantum phase transitions (QPTs) are solely driven by quantum fluctuations arising from competing mechanisms, resulting in different favored ground states in different phases. Interestingly,   {continuous} QPTs are also accompanied by a diverging correlation length near criticality and follow universal scaling laws (see Ref.~\cite{ROSSINI2021} for a review). Apart from equilibrium quantum phase transitions, recent studies have led to the discovery of a new type of quantum criticality exclusive to out of equilibrium systems \cite{anatoli11}. Known as dynamical quantum phase transitions \cite{heylpol13,karrasch13,kehrein14,sraddha16,heyl_review18,sourav18,jurcevic17,peotta21}, these extend the concept of criticality and universality to the early time dynamics of nonequilibrium systems, which are far from their ground states.\\

{Even so, it is usually difficult to understand how equilibrium critical phenomena can manifest in generic high energy excited states of a system.  In this light, there has been an extensive search for footprints of quantum criticality in excited and particularly out of equilibrium systems \cite{zanardi07,Zhang2017,heyl18,keesling19,bag19,titum19,paraj20,song21,halimeh21,ceren21,asmi21,villa21,chinni21,ikeda21,paul22,kheiri22}.} Particularly,  Refs.~\cite{halimeh21,ceren21} and very recently Refs.~\cite{asmi21,paul22}, reported signatures of equilibrium criticality following a quench, which manifest in late time observables and higher order correlations. It is remarkable that some of these signatures of quantum phase transitions persist even after chaotic scrambling in quenched systems. On the other hand, a recent quenching experiment in a trapped ion quantum simulator, has also detected signatures of equilibrium QPTs; these are manifested in the late time domain statistics of a quenched chaotic system of interacting spins \cite{Zhang2017}. Nonetheless, a generic theoretical origin of these surprising signatures of equilibrium criticality in systems manifestly far from equilibrium has remained elusive as yet.\\

In this paper, we set out to develop a unified theory connecting several recent experimental and theoretical observations on such nonequilibrium signatures of quantum criticality, particularly those reported in Refs.~\cite{Zhang2017,asmi21}. We note that the theory also naturally explains the recently observed critical behavior of local probes, such as magnetization density, following chaotic quenches (see Ref.~\cite{ettore20,halimeh21}). We begin by demonstrating that the time averaged Loschmidt echo (LE), a nonequilibrium analog of the partition function,  {develops sharp signatures following a sudden quench when the quenched Hamiltonian crosses an equilibrium critical point in both integrable (also seen in Refs.~\cite{schmitt15,zhou19}) and chaotic systems.} Additionally, we observe that the time averaged LE satisfies universal finite size scaling laws as the quenched Hamiltonian approaches the critical point following a generic quantum quench. To understand the definition of the LE, consider a system in a many body initial state $\ket{\psi(0)}$ allowed to evolve unitarily with a quenched Hamiltonian $H$; the Loschmidt echo following the sudden quench is then defined as, $\mathcal{L}=\left|\braket{\psi(0)|\psi(t)}\right|^2=|\braket{\psi(0)|\exp{[-iHt]}|\psi(0)}|^2$. However, as evident from its definition, due to an exponentially growing basis with system size, the LE is very difficult to probe experimentally in many body systems.\\

We therefore proceed to systematically construct experimentally accessible and time-insensitive local probes, out of string type observables, which capture critical behavior in quenched quantum systems. We find that mimicking the behavior of the LE,  {sharp signatures indicative of the equilibrium critical point can also emerge as a function of the quenching parameter in time averaged string observables}. Recently, such operators have been shown to efficiently detect dynamical critical behavior and early time singularities in both integrable and chaotic out of equilibrium systems \cite{jurcevic17,jad21,pol21}. By construction, these local string operators eventually approach the projector onto the complete initial state as their spatial support (string length) increases.  {Following a quantum quench, the string observables are expected to reach a  {steady state} after sufficiently long times.} Specifically, for integrable quenches,  {generic} local observables are known to eventually relax into a generalized Gibbs ensemble in a thermodynamically large system \cite{rigol_gge07,das_gge16}. On the other hand, if $H$ is chaotic, the system eventually thermalizes and expectation values of local observables can be described by a microcanonical ensemble. This is in accordance with the eigenstate thermalization hypothesis (ETH) \cite{srednicki99,Rigol2008_review,luca16,Russomanno21}.  Thus, expectation values of generic local observables are expected to be smooth functions of energy, containing no information of the initial state \cite{alba19} except energy density. In contrast to the long time behavior of LE, it is therefore unlikely for local observables to develop true nonanalyticities in their bare expectations following chaotic quenches. Nevertheless, we find that, even after a sufficiently long ergodic evolution, the observables remarkably hold strong and stable memory of the initial state. Besides, the string observables are also found to have much longer life-times in chaotic eigenstates in comparison to single-site observables such as magnetization density. Consequently, after  {sufficiently long} times following the quench, we see robust and sharp signatures of the equilibrium QCP in the local probes, even with bilinear observables, as found numerically in Ref.~\cite{asmi21}. We therefore show that string observables indeed allow us to systematically access zero temperature ground state properties, such as phase transitions in quenching experiments \cite{Zhang2017}, without having to actually prepare the system in its ground state. \\

 Interestingly, we further observe that the long time averaged string observables also satisfy critical scaling relations with the string length  following near critical quenches even in chaotic systems. This allows us to directly determine critical exponents associated with the equilibrium QCPs in the long time behavior of the local observables. Moreover, we find that the simple time averaged expectation of the string operators reveal a much deeper connection with the probability of domain formation in the post quench system. To elaborate, the time averaged projectors are nothing but the late-time probability distribution of domains, thus enabling us to perform a complete statistical analysis of polarized and   domain formation, in the post-quench system. This connection in turn, assigns a direct experimental meaning to the string observables and at the same time shed light on the origin of these sharp critical signatures in quenching experiments. We find that not only does the domain statistics reveal the position of equilibrium QCPs accurately, they also theoretically explain the recent experimental findings \cite{Zhang2017} on a quantum simulator \cite{monroe03}.\\
 
  To further probe the sensitivity of the results on the choice of the initial state of the system, we analyze the long time averaged infinite temperature correlators of the string observables. Particularly, we show that the time averaged infinite temperature auto correlation and out of time ordered correlators (OTOC) also detect the equilibrium QCP sharply. This establishes that the sharp signatures in string observables detecting equilibrium QCPs, are in general insensitive to any specific choice of initial states in quenching experiments. In recent times, infinite temperature OTOCs of local observables have also become instrumental in the study of information scrambling \cite{maldacena16,swingle19,swingle20} and dynamical phases \cite{pol21} in chaotic many body systems. Furthermore, OTOCs have been experimentally measured in ultracold atomic lattices through various echo protocols \cite{patrick16,silva20,fine14,tarek15,lewis19,nie20}. As we demonstrate our ideas through the experimentally accessible quantum Ising model of spins \cite{bloch17}, our claims are readily verifiable in state of the art quantum simulators. \\

The paper is organized in the following sections: In Sec.~\ref{sec:model} we introduce the  model system and elaborate on the quenching protocol employed in the rest of the paper to demonstrate the results. Section.~\ref{sec:lo_analytic} contains a study of the complete analytic structure of the time averaged Loschmidt echo when the system is quenched to a QCP in the integrable limit. In Sec.~\ref{sec: local_lo},  we define the local string operators and demonstrate critical behavior of their long time average following both integrable and chaotic quenches. In Sec.~\ref{sec: critical_scaling}, we proceed to show that the time averaged LE along with the local string observables follow appropriate scaling laws near the equilibrium QCPs and calculate universal critical exponents characterizing the QCPs. In Sec.~\ref{string_stat},  we probe the late time statistics of finite strings and show how it connects with recent experimental data on ultracold ion lattices measuring domain statistics following a quench in interacting spin systems. Sec.~\ref{sec:otoc} describes the infinite temperature two-time correlators of the string operators as they develop sharp signatures at equilibrium QCPs, following both integrable and chaotic quenches. In Sec.~\ref{Sec:memory} we elaborate on the origin of the strong initial state memory in string observables following ergodic evolution. Finally, we conclude in Sec.~\ref{sec:conclusion} with a brief summary of the important results and possible future directions of investigation. We further declare that everywhere in the paper we will adopt the natural system of units unless otherwise specified.

\section{Model system and quenching protocol}
\label{sec:model}

To demonstrate the efficacy of string operators and the LE in capturing quantum critical physics in quenched systems, we use  a ferromagnetic axial next-nearest-neighbor  {(ANNNI) model} \cite{dutta15_book}, with nearest-neighbor interactions and next-nearest-neighbor interactions, both ferromagnetic,  and a  {transverse} external field $h$. Such a system having $L$ spins can be described completely by the many-body Hamiltonian,
\begin{equation}\label{eq:ham}
H=-\sum\limits_{i=1}^{L}\sigma^x_i\sigma^x_{i+1}-J_2\sum\limits_{i=1}^{L}\sigma^x_i\sigma^x_{i+2}-h\sum\limits_{i=1}^{L}\sigma^z_i,
\end{equation}
where $\sigma_i$ represent Pauli matrices at the ${\rm} i{\rm th}$ site. For numerical advantage, we have used periodic boundary conditions, such that $\sigma_{L+i}\equiv\sigma_{i}$.The line $J_2=0$ in the parameter space is the well known integrable transverse field Ising model with quantum critical points  {(QCPs)} $h=\pm 1$, separating paramagnetic and ferromagnetic phases in a thermodynamically large ($L\rightarrow\infty$) system. Interestingly, the system still hosts an Ising critical line separating the ferromagnetic phase from a paramagnetic phase in the plane of parameters $J_2$ and $h$. For perturbative integrability breaking strengths, the Ising critical line is given approximately by a transcendental equation for $J_2<0$ and $h$ (see  {Refs}.~\cite{matteo06,karrasch13}),
\begin{equation}\label{eq:nonin_qcp}
1+2J_2=h+\frac{J_2h^2}{2\left(1+J_2\right)}.
\end{equation}
As one approaches the integrable point $J_2\rightarrow 0$, the critical point drifts towards the critical point of the transverse Ising model, i.e., $h=1$. The Ising phase transition is known to reflect a spontaneous breaking of a global ${\rm Z}_2$ symmetry  {due to spin inversion}, in the ordered (ferromagnetic) phase ground state. This allows the longitudinal magnetization density in the ground state to serve as an appropriate order parameter, which is nonzero in the ordered phase while vanishes in the paramagnetic phase.\\

 {We analyze quench dynamics starting from an initial state $\ket{\psi(0)}$ and inflict a sudden quench in the transverse field at a time $t=0$, thereby allowing the system to evolve with the final quenched Hamiltonian}.  {Following such a quench, the long time average of local observables $O$ in the thermodynamic limit approaches the average given by the diagonal ensemble (see Ref.~\cite{luca16})},
\begin{equation}\label{eq:de}
\lim\limits_{T\rightarrow\infty}\int_0^T dt\braket{\psi(t)|O|\psi(t)}=\sum\limits_{\alpha}\left|\braket{\psi(0)|\phi_{\alpha}}\right|^2\braket{\phi_{\alpha}|O|\phi_{\alpha}},
\end{equation}
where the states $\ket{\phi_{\alpha}}$ are the eigenstates of the quenched Hamiltonian $H$.  {For nonintegrable quenches, it is also overwhelmingly probable that the expectations $\braket{\psi(t)|O|\psi(t)}$ coincide with the diagonal ensemble average at almost all times. Furthermore, because the system is ergodic, the eigenstate expectation values of local observables $\braket{\phi_{\alpha}|O|\phi_{\alpha}}$ are essentially given by corresponding microcanonical averages}.  By the central limit theorem, the probabilities to occupy different energy states $|\braket{\psi(0)|\phi_{\alpha}}|^2$ are expected to be peaked around mean energy,  i.e. near $E_{\alpha}=\overline{E}\equiv \langle \psi_0 | H |\psi_0\rangle$.  From this argument, any memory of the initial state except for the total mean energy of the system and perhaps its variance, determining the width of the energy distribution, should be lost. However, we will show that for string observables these simple considerations are not completely valid. Even though in accordance with ETH, the mean state to state fluctuations of these observables in nearby eigenstates decrease with system size, we see a very rapid rise in their relative variance within a narrow energy shell with increasing string size. These increasing fluctuations prevent a purely thermal description of comparatively longer strings and encode a persistent memory of the initial state even after chaotic quenches. Because of these large fluctuations, we further show that long string observables take comparatively longer to relax and have highly oscillating late-time connected auto-correlation functions in chaotic eigenstates.

\section{Time averaged Loschmidt echo following an integrable quench}
\label{sec:lo_analytic}

In this section, we discuss the complete analytic structure of the time averaged Loschmidt echo following a quench at the integrable point. We start from the ground state of  {$H_0\equiv H(J_2=0, h_i)$} and quench the system at time $t=0$ to a final set of parameters  {$H\equiv H(J_2=0, h)$}. Allowing the system to evolve with the quenched Hamiltonian, we then probe the long time averaged LE,
\begin{equation}
	\bar{\mathcal{L}}=\lim\limits_{T\rightarrow\infty}\frac{1}{T}\int_{0}^{T}dt\left|\braket{\psi(0)|\psi(t)}\right|^2.
\end{equation} 
Expanding in terms of the eigenstates of the Hamiltonian $H\ket{\phi_{\alpha}}=E_{\alpha}\ket{\phi_{\alpha}}$ and carrying out the integration one therefore arrives at  {(as usually we assume there are no extensive degeneracies in the spectrum)}
\begin{equation}\label{eq:ol}
	\mathcal{O}_L\equiv-\frac{1}{L}\log{\bar{\mathcal{L}}}=-\frac{1}{L}\log{\sum\limits_{\alpha}\left|\braket{\psi(0)|\phi_{\alpha}}\right|^4}.
\end{equation}

\begin{figure}
	\centering
	\includegraphics[width=0.79\columnwidth,height=4.65cm]{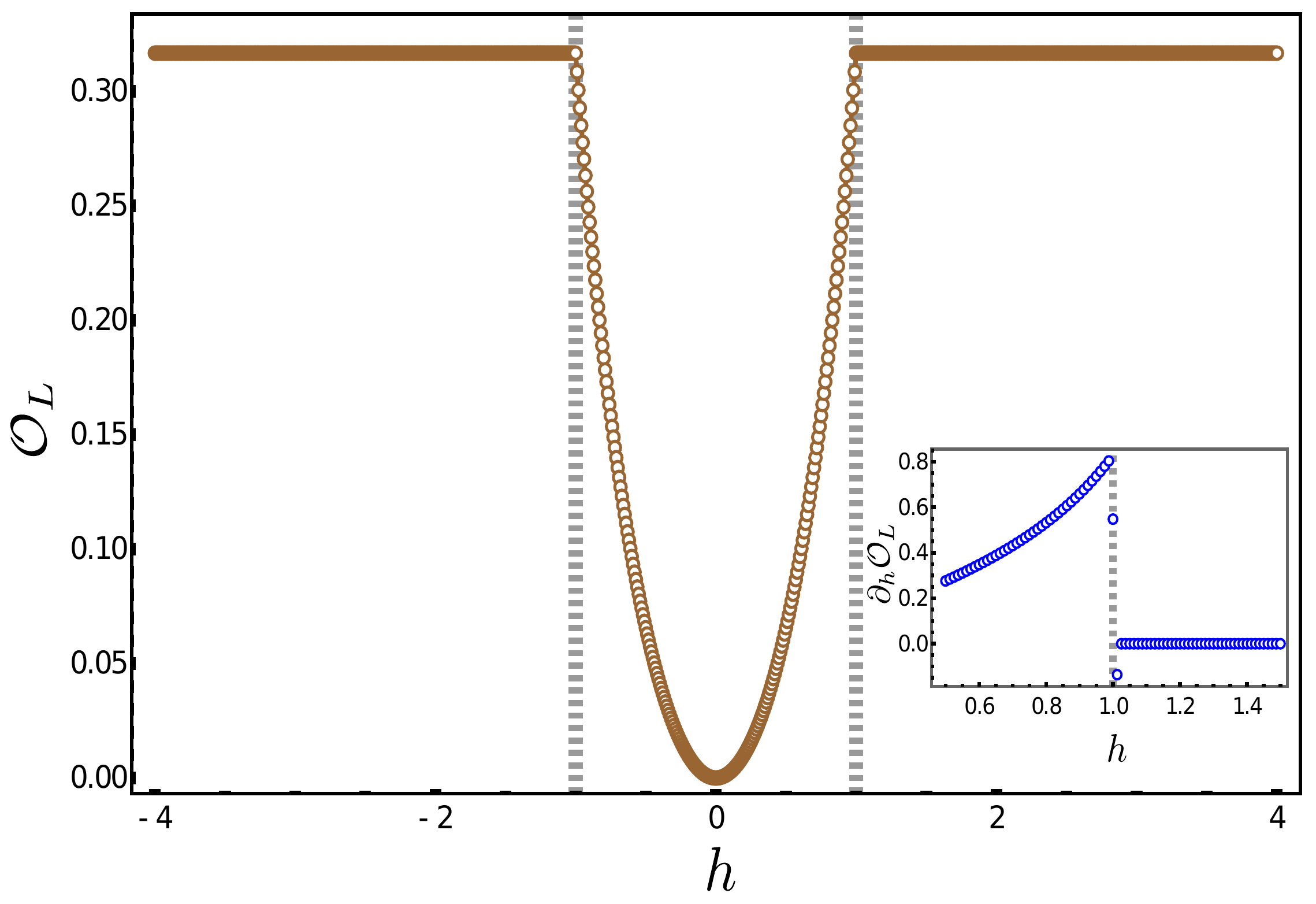}
	
	\caption{ {The rate function of the time averaged Loschmidt echo $\mathcal{O}_L$  {at the integrable point and in the limit of $L\to\infty$} develops nonanalyticities at equilibrium QCPs, in the thermodynamic limit as a function of the quenched (final) transverse field starting from the ground state of the integrable Hamiltonian $H_0$.}}
	\label{fig:1} 
\end{figure}
In Fig.~\ref{fig:1} we  {plot $\mathcal O_L$ as a function of $h$ and observe that it becomes nonanalytic at the equilibrium critical points, where its first derivative with respect to $h$ develops a discontinuous jump}.
\begin{figure*}[ht]
	\subfigure[]{
		\includegraphics[width=5.8cm,height=5.6cm]{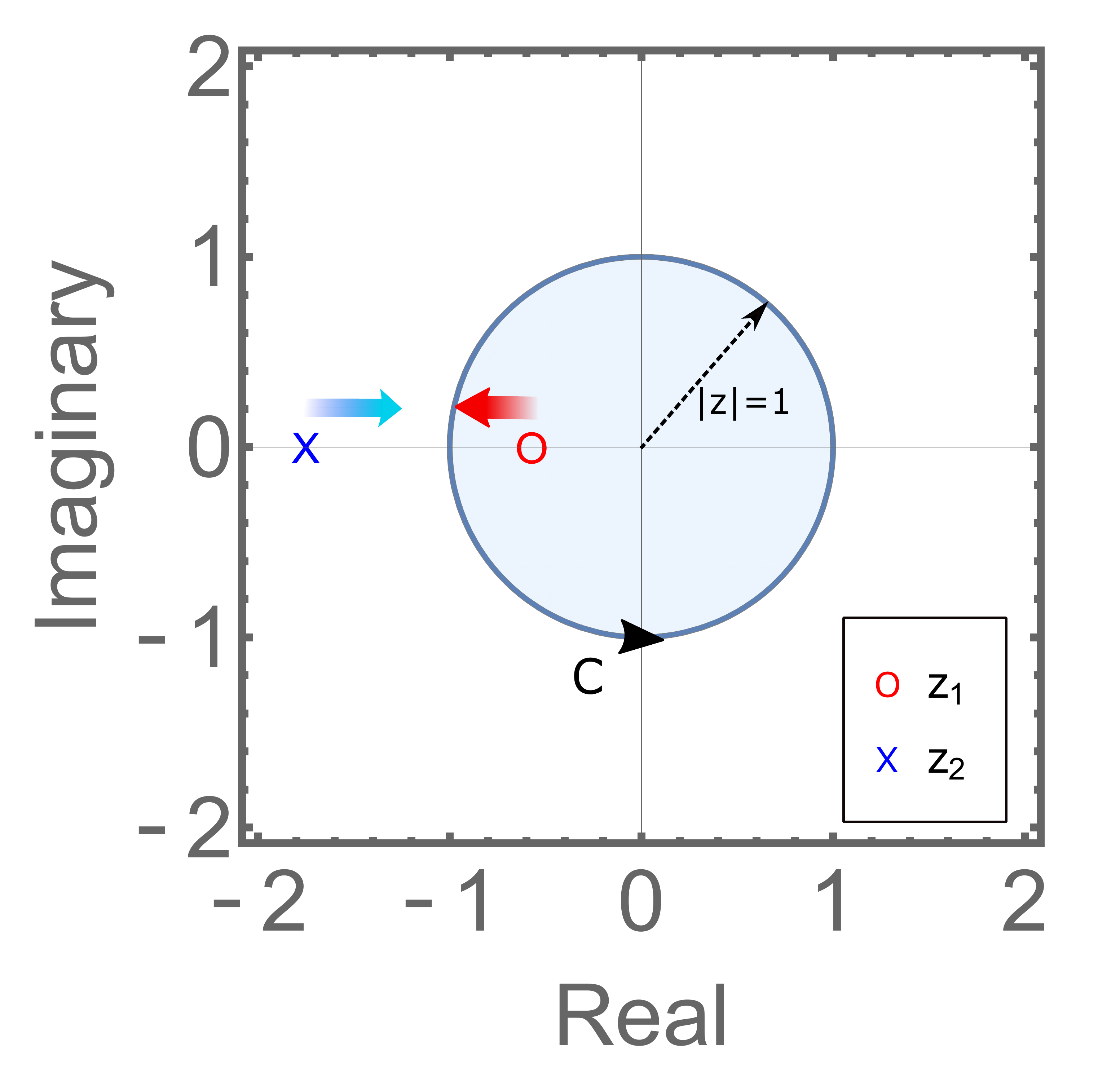}
		\label{fig:2a}}
	\subfigure[]{
		\includegraphics[width=5.4cm,height=5.5cm]{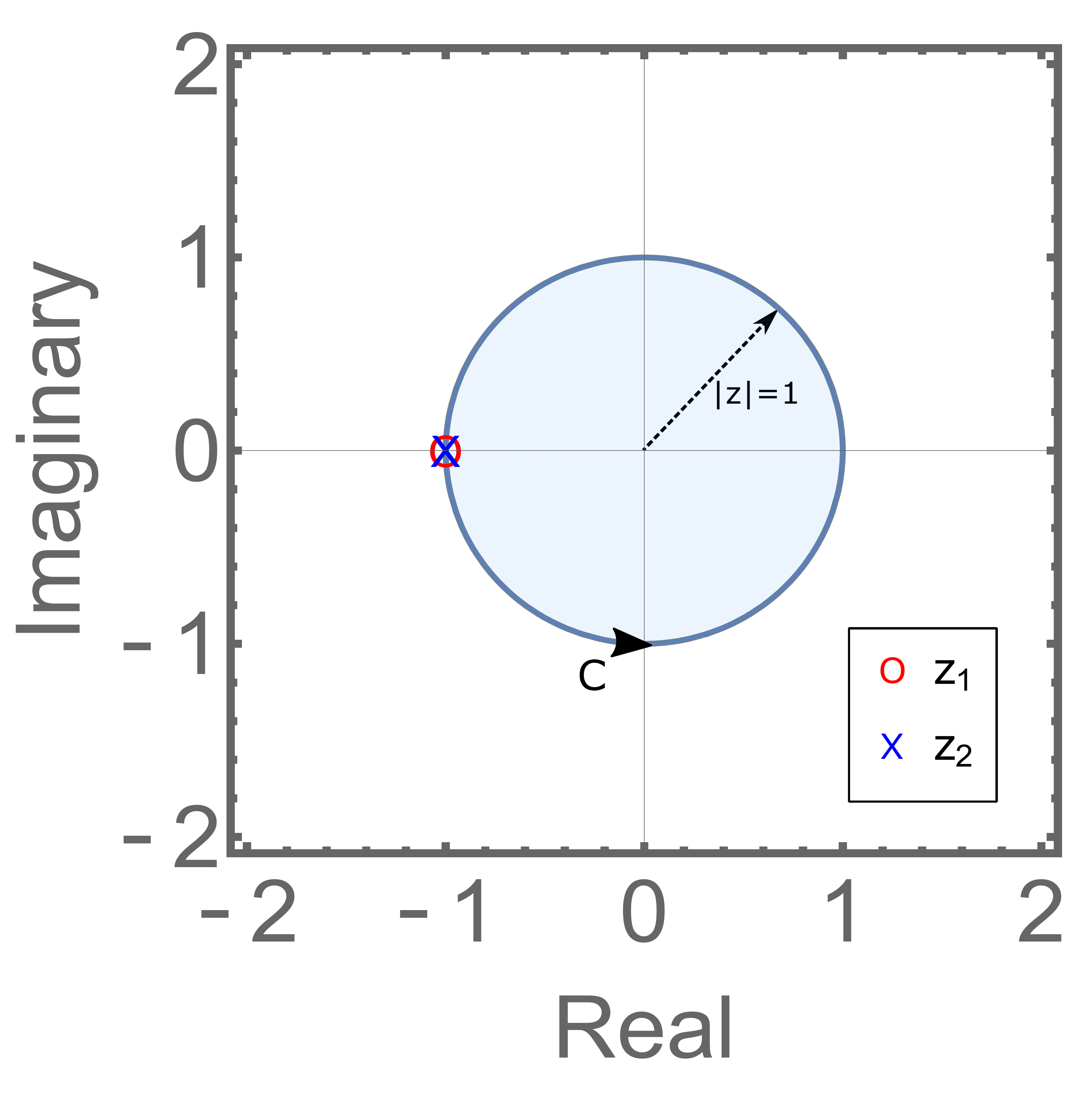}
		\label{fig:2b}}
	\subfigure[]{
		\includegraphics[width=5.8cm,height=5.75cm]{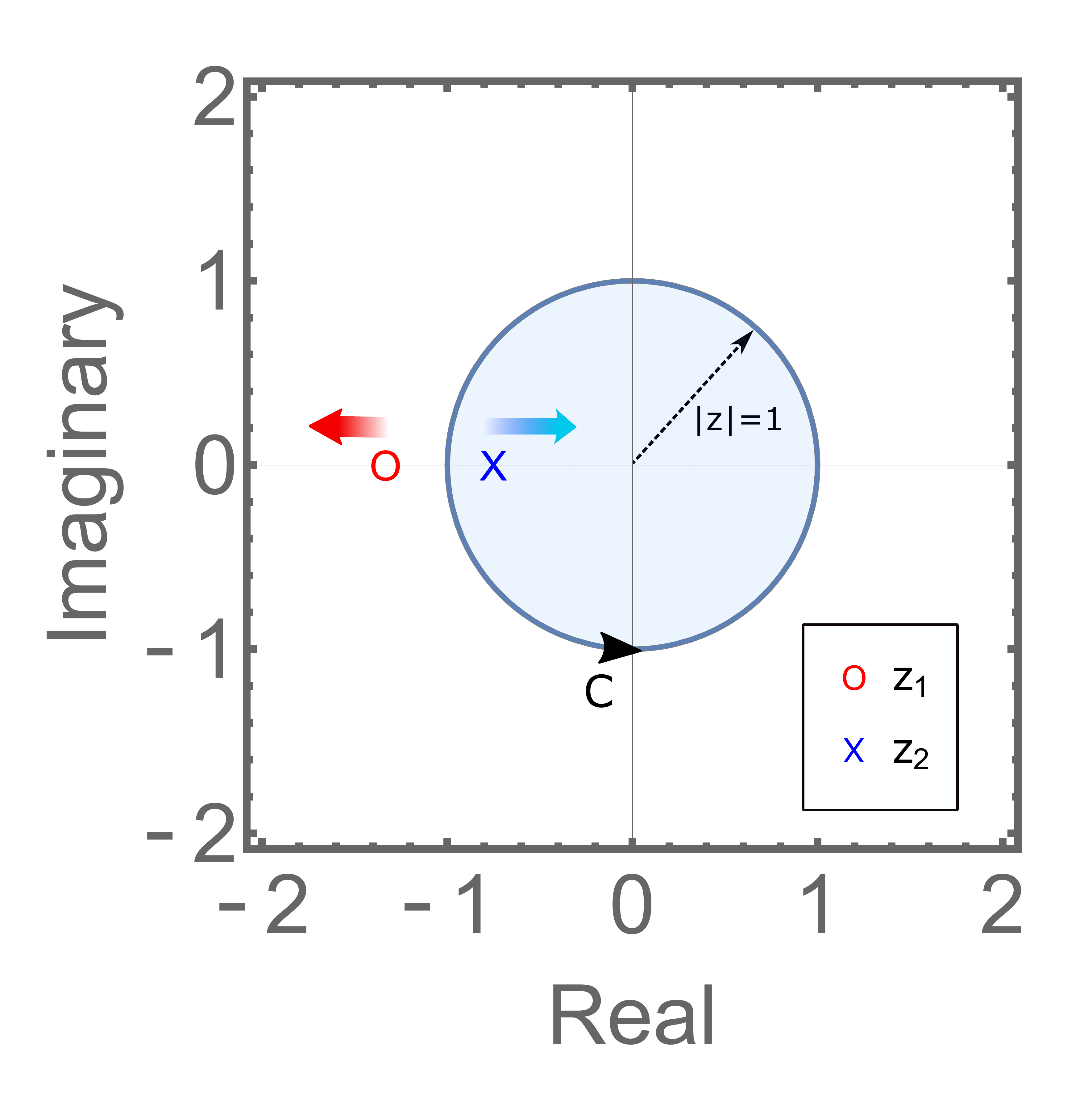}
		\label{fig:2c}}
	\caption{The positions of the poles $z_1$ and $z_2$  {of the function  {$\partial_hF(h,z)$} (see Eq.~\eqref{eq:logle}),  which elucidate the analytic structure of the Loschmidt echo $\mathcal{O}_L$}  {in the thermodynamic limit}.  The arrows indicate the drift of the poles as one approaches the ferromagnetic phase from the paramagnetic phase, i.e., as $h$ decreases. The three panels correspond to an integrable quench to (a) $h=1.5$ (paramagnetic), (b) $h=1$ (critical) and (c) $h=0.75$ (ferromagnetic), starting from the ferromagnetic ground state at $h_i=0$. The unimodular circle $C$ represents the contour of integration employed in evaluating the derivative in Eq.~\eqref{eq:exactle}.}
	\label{fig:2cover}
\end{figure*}
 {In Appendix~\ref{Sec:Appendix_1} we show that in the thermodynamic limit Eq.~\eqref{eq:ol} can be represented through the following contour integral of the complex variable $z=\mathrm e^{ik}$ over the circle $C\equiv|z|=1$}:
\begin{equation}\label{eq:exactle}
\mathcal{O}_L=\frac{1}{2\pi i}\oint_{C}F(h,z)dz,
\end{equation}
where 
\begin{equation}\label{eq:logle}
	F(h,z)=\frac{1}{z}\log\left[1-\frac{h^2 \left(z^2-1\right)^2}{8 z (h+z) (h z+1)}\right].
\end{equation}
 {To understand the origin of the nonanalyticity of $\mathcal O_L$ it is easier to analyze its derivative with respect to $h$:
\begin{equation}\label{eq:der}
\partial_h\mathcal{O}_L=\frac{1}{2\pi i}\oint_{C}\partial_hFdz.
\end{equation}
As evident from Eq.~\eqref{eq:logle}, the function $\partial_h F$ has simple poles on the real axis at $z_1=-1/h$ and $z_2=-h$, which cross at the QCP $h=h_c=1$,  as well as some additional poles,  which are not sensitive to $h_c$.  {In  Fig.~\ref{fig:2cover} we show the movement of these poles in the complex plane with the quenched field $h$, near the critical point corresponding to the singularity at $h_c=1$ and $z=-1$. The residues corresponding to these poles are ${\rm Res}(z=z_1)=-1/h$ and ${\rm Res}(z=z_2)=1/h$ (see Appendix~\ref{Sec:Appendix_1} for details). }As $h$ crosses $h_c$ the poles entering the integration contour switch causing the integral to jump leading to the cusp singularity in $\mathcal O_L(h)$ shown in Fig.~\ref{fig:1}.
}

\begin{figure*}
	\subfigure[]{
		\includegraphics[width=7.5cm,height=5.5cm]{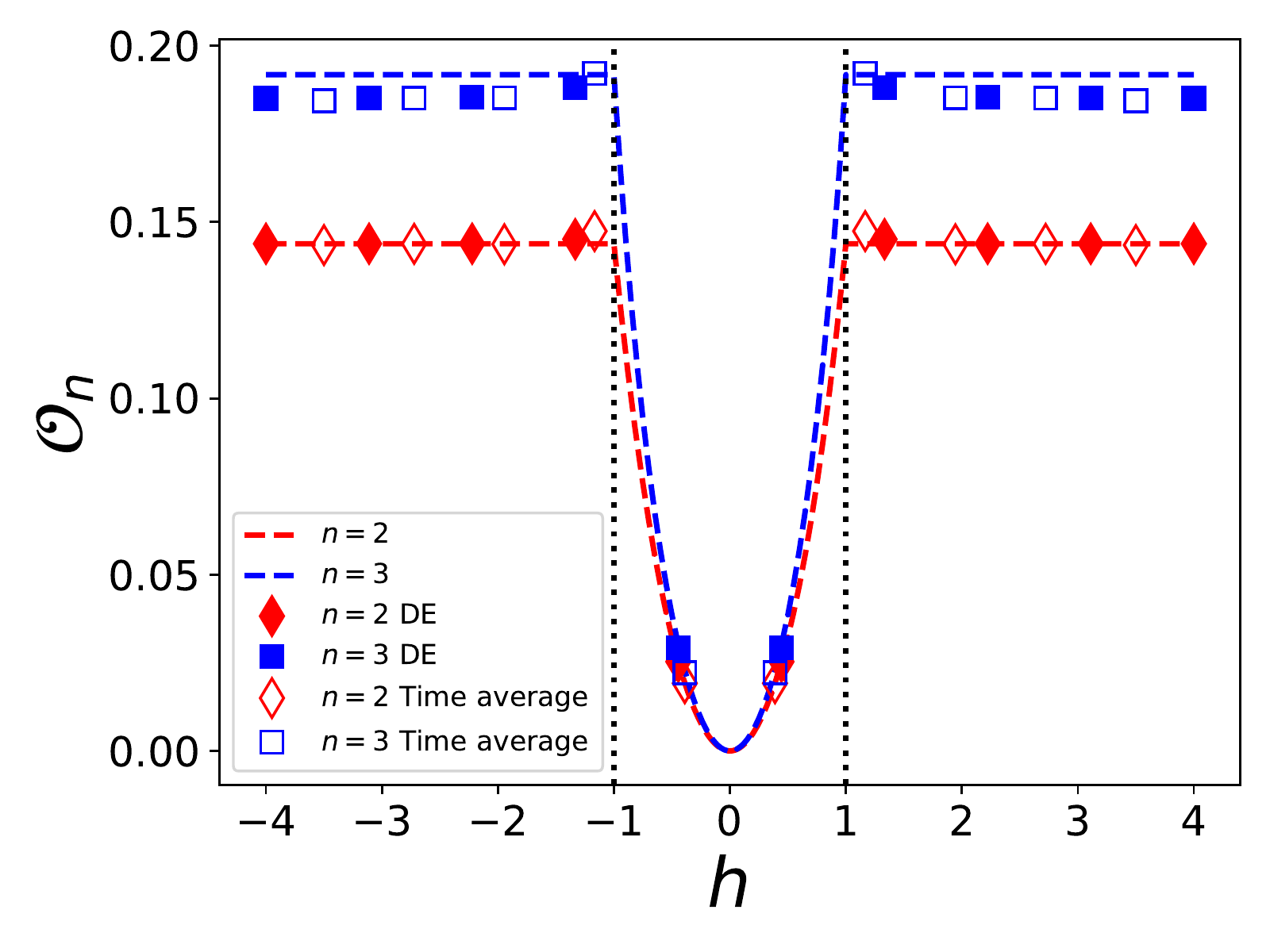}
		\label{fig:3a}}
	\hspace{1.0cm}
	\subfigure[]{
		\includegraphics[width=7.6cm,height=5.5cm]{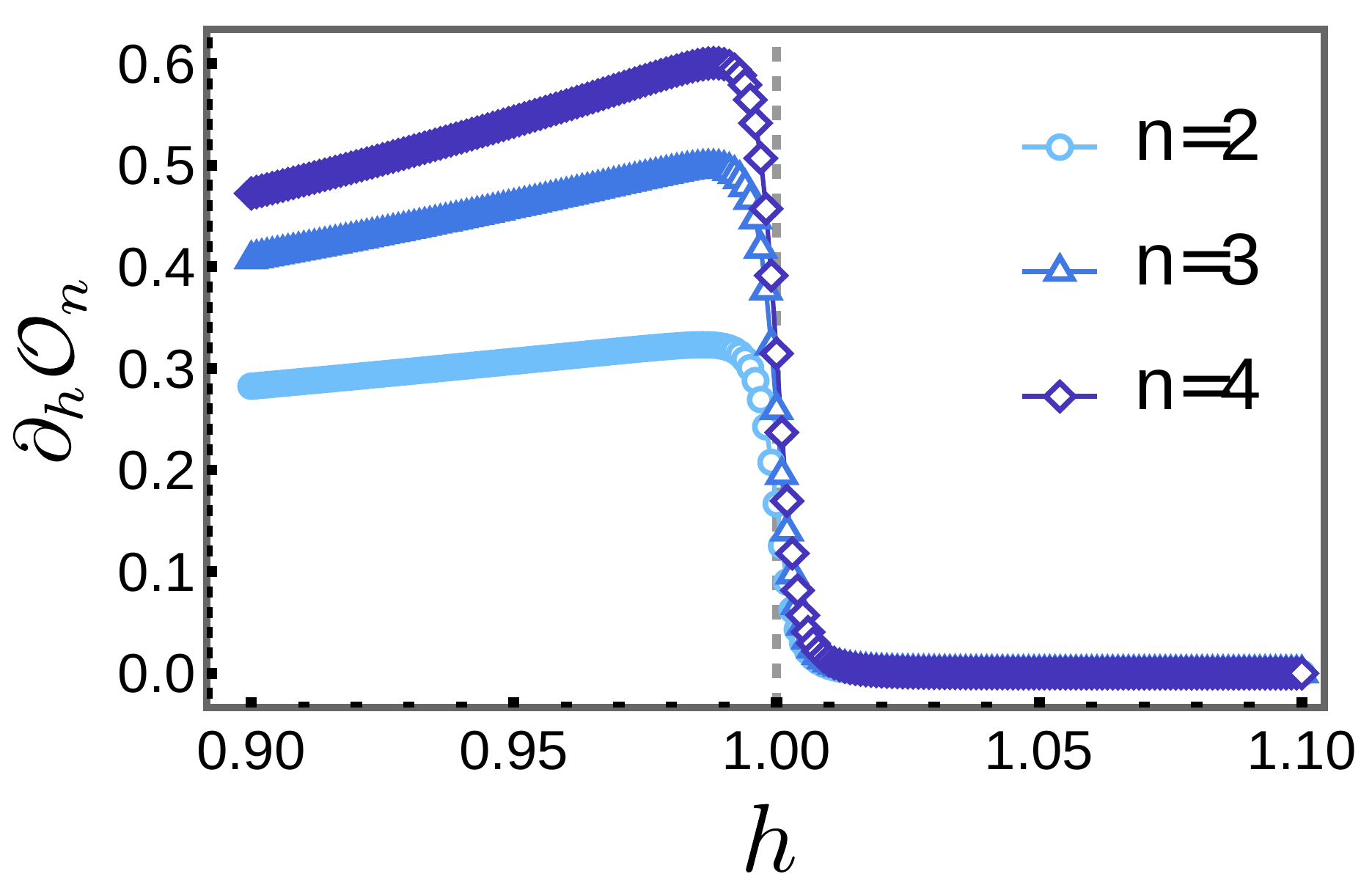}
		\label{fig:3b}}
	
	\caption{(a) The observables $\mathcal{O}_n$ develop nonanalyticities at quantum critical points, following a quench of the transverse field in an integrable model in the thermodynamic limit (shown in dashed lines). The discrete points show the corresponding quantities in a system with $L=16$ spins, calculated using exact diagonalization The solid (hollow) symbols represent the diagonal ensemble averaged (time averaged) quantities. To simplify calculations, only the even spin-parity sector has been considered and all time averaging has been performed up to $T=30$ starting from the ferromagnetic ground state at $h_i=0$. (b) Jump singularities at the critical point $h=1.0$ in the first derivative of the local observables $\mathcal{O}_n$, following an integrable quench in the thermodynamic limit. Exact analytical calculations of the discontinuous jumps  {are} presented in Appendix~\ref{Sec:A3}.}
\end{figure*}

\section{Time averaged expectation of local projectors}
\label{sec: local_lo}

As discussed in Sec.~\ref{sec:intro}, it is in general difficult to experimentally probe the LE in macroscopic systems. We therefore proceed to define local scalable string operators $P_n$ of finite length $n$  {defined such that as $n$ increases their expectation values $\mathcal L_n$ approach the Loschmidt echo.  Specifically we define} 
\begin{equation}\label{eq:ln}
\mathcal{L}_n(t)=\braket{\psi(t)|P_n|\psi(t)},
\end{equation}  
where,
 {
\begin{equation}\label{eq:string}
P_n=\frac{1}{L}\sum\limits_{i=1}^{L}\frac{1}{2^n}\prod\limits_{j=i}^{i+n-1}\left(\mathbb{I}_j+\sigma_j^x\right).
\end{equation}
}
Clearly for $n=L$ we recover the complete Loschmidt echo  {for the fully polarized initial state corresponding to $h_i=0$,}
\begin{equation}
\mathcal{L}_L(t)=\braket{\psi(t)|P_L|\psi(t)}=\left|\braket{\psi(0)|\psi(t)}\right|^2.
\end{equation}
{Note that one can add to $P_n$ a similar operator projecting to the  {left} spin state composed of products of $\left(\mathbb{I}_i-\sigma_i^x\right).$}\\

Following an integrable quench ($J_2=0$) at $t=0$ starting from the ground state at $h_i=0$ to $h$,   {we next analyze} the long time averages of the projectors $\mathcal L_n$ {:
\begin{equation}
\label{eq:On_def}
\mathcal{O}_n=-\frac{1}{n}\log \overline {\mathcal L_n(t)},\quad \overline {\mathcal L_n(t)}=\lim\limits_{T\rightarrow\infty}\frac{1}{T}\int_{0}^{T}dt \mathcal L_n(t).
\end{equation}
According to Eq.~\eqref{eq:de} the function $\overline {\mathcal L_n(t)}$ can be expressed as a weighted expectation value of the string operator $P_n$ in the eigenstates of the quenched Hamiltonian $H$}. In Fig.~\ref{fig:3a} we observe sharp nonanalyticities developing in $\mathcal{O}_2$, $\mathcal{O}_3$ and $\mathcal{O}_4$ in the thermodynamic limit,   {which we also compare} to exact diagonalization results in finite systems with numerically accessible system sizes,~ {obtained using the QuSpin package~\cite{bukov1,bukov2}.  Further, Fig.~\ref{fig:3a} also compares the infinite and finite time averages of the observables $\mathcal L_n(t)$ showing robustness of the observed nonanalyticities to the duration of the time-averaging window.} Fig.~\ref{fig:3b} shows the jump discontinuities following critical quenches in the first derivative response $\partial_h\mathcal{O}_n$ for finite strings (see Appendix~\ref{Sec:A3} for a detailed analytical derivation).

\vspace{1cm}
\begin{figure}[ht]
	
		\includegraphics[width=6.3cm,height=5.0cm]{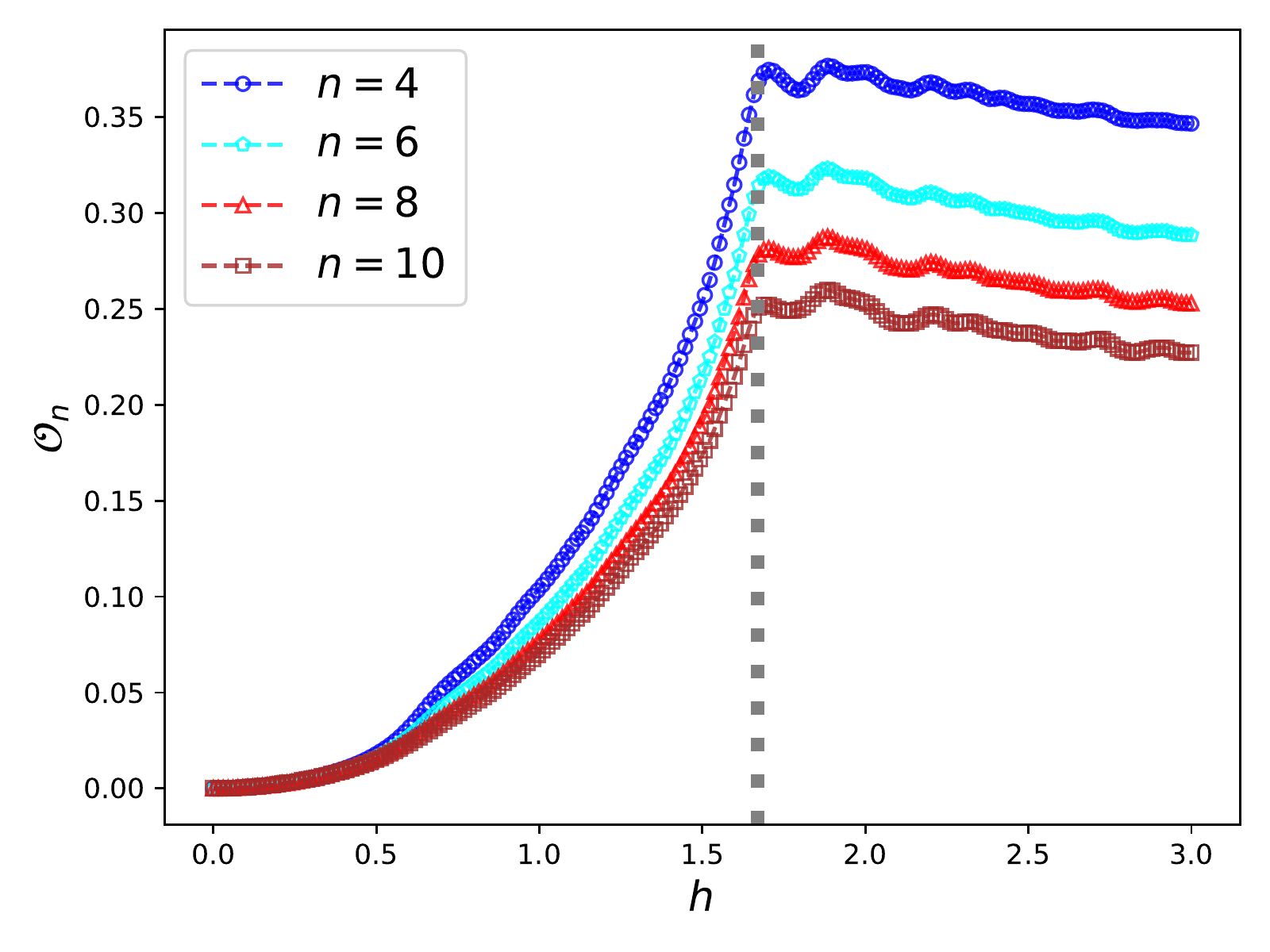}
		\label{fig:4a}
	
	\caption{ {The observables $\mathcal{O}_n$ for finite strings of length $n$ in the nonintegrable ANNNI model ($J_2=0.5$) exhibit sharp transitions at the Ising QCP ($h\approx1.6$) in a system of $L=16$ spins. All the time averages have been performed up to $T=30$ following the quench  {from} a completely polarized initial state. }}
	\label{fig:4}
\end{figure}

It is also evident that the sharp signatures detecting the QCP in the string observables are not exclusive to integrable systems.  {To demonstrate this,   {we numerically analyze quenches to a nonintegrable Hamiltonian ($J_2=0.5$) stating from the same polarized initial state}.  In Fig.~\ref{fig:4}, we show the results for $\mathcal{O}_n$  {when averaged over sufficiently long times} and the full time averaged Loschmidt echo corresponding to $n=L$.  Both the LE and the observable $\mathcal O_n$ develops a sharp singularity at the transition point $h=h_c\approx 1.6$ for the chosen parameters.}  We note that using the string observables, we are able to accurately detect the corresponding position of the QCP   {despite high energy dumped into the system during the quench.} \\

\section{Universal scaling laws following critical quenches}
\label{sec: critical_scaling}

Given the sharp behavior of string observables detecting equilibrium criticality, it is natural to ask whether the time-insensitive local probes also capture universal physics near equilibrium QCPs following a quench. Starting with the long time averaged Loschmidt echo, we address this question by studying how the string observables behave under scaling transformations near critical quenches.
\begin{figure}[ht]
	\subfigure[]{
		\includegraphics[width=6.5cm,height=4.4cm]{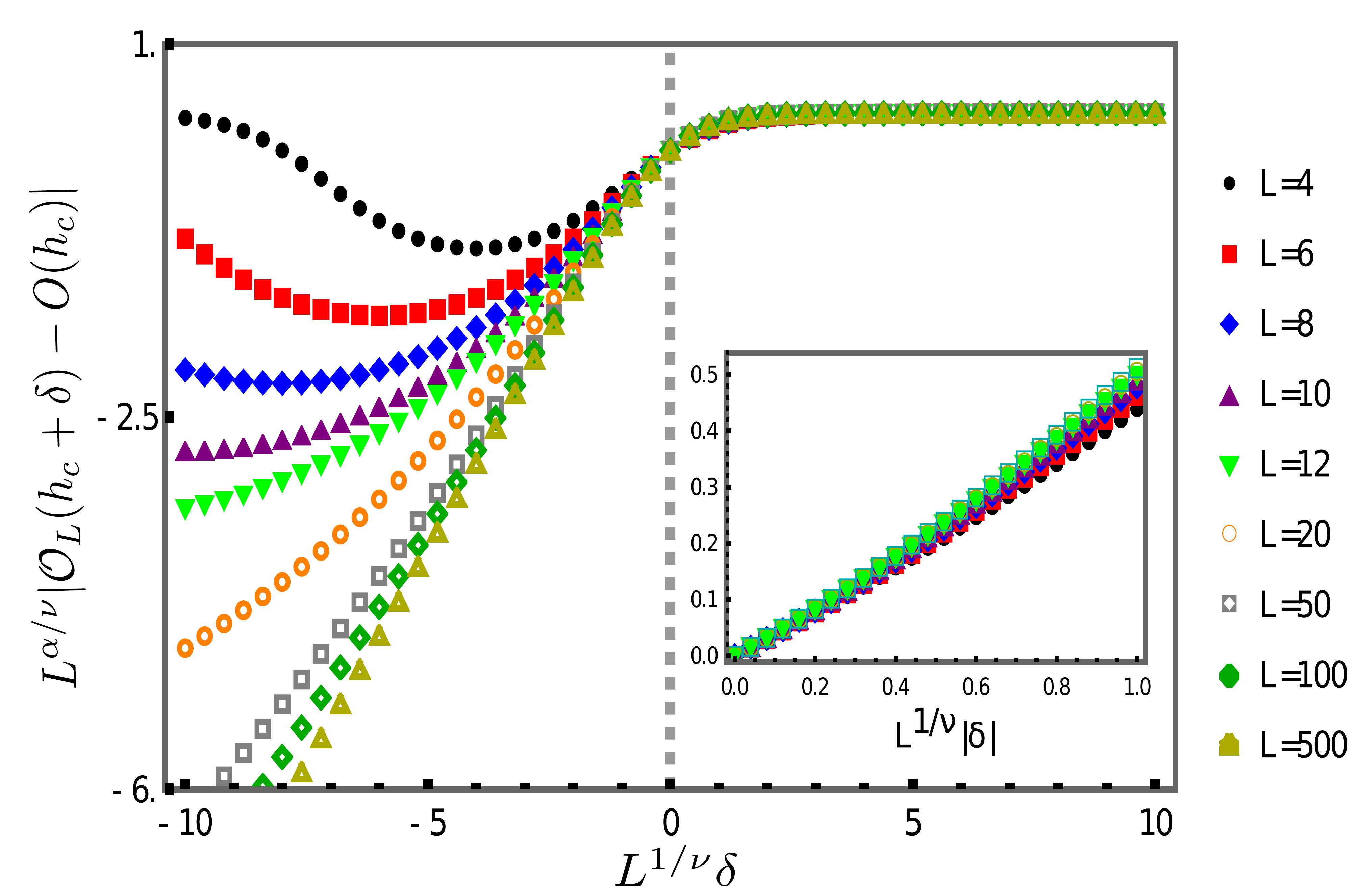}
		\label{fig:5a}}
		\hspace{0.25cm}
	\subfigure[]{
		\includegraphics[width=6.5cm,height=4.5cm]{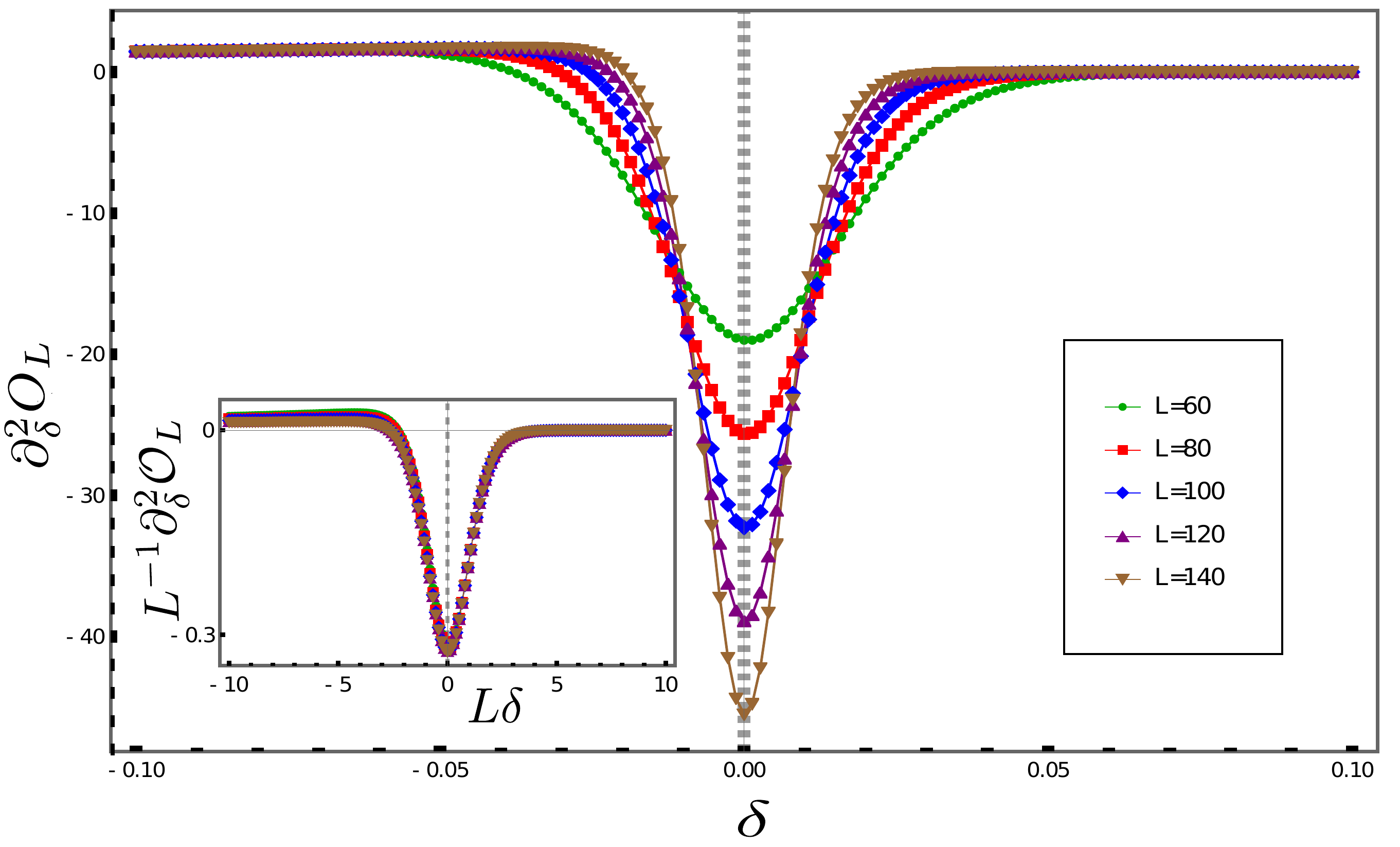}
		\label{fig:5b}}

\caption{(a) Finite size scaling of the observable $\mathcal{O}_L$ when the averaging is done over the diagonal ensemble, following an integrable quench starting from the ground state at $h_i=0$. On choosing the critical exponents to be $\nu=\alpha=1$ [see  Eq.~\eqref{eq:scaling_1}], all data points for different $L$ collapse into a single scaling function near the critical point. (Inset) Zoomed-in image of the finite-size scaling collapse near the critical point ($L^{1/\nu}\delta\ll 1$) for the same critical exponents. (b) The second derivative of $\mathcal{O}_L$ diverges with increasing system size near the QCP. (Inset) Scaling collapse of the rescaled derivative response $L^{-1}\partial_{\delta}^2\mathcal{O}_L$ near the critical point [see Eq.\eqref{eq:chi_fs}].}
\end{figure}
 Specifically, close to the equilibrium critical point $h=h_c+\delta$, we estimate the singular part of the observable $\mathcal{O}_L$ by a universal function under scaling transformations $\delta\rightarrow L^{1/\nu}\delta$. Namely, in the finite size scaling regime ($L^{1/\nu}|\delta|\ll 1$), the singular part of the time averaged LE is observed to scale as,
\begin{equation}\label{eq:scaling_1}
|\mathcal{O}_L(h_c+\delta)-\mathcal{O}_L(h_c)|\sim L^{-\alpha/\nu}\Phi(L^{1/\nu}\delta),
\end{equation}
where $\nu$ and $\alpha$ are critical exponents and $\Phi$ is an universal scaling function (see Appendix.~\ref{Sec:A2} for a detailed discussion on the scaling form). 
\begin{figure}[ht]
	\includegraphics[width=7.0cm,height=5.5cm]{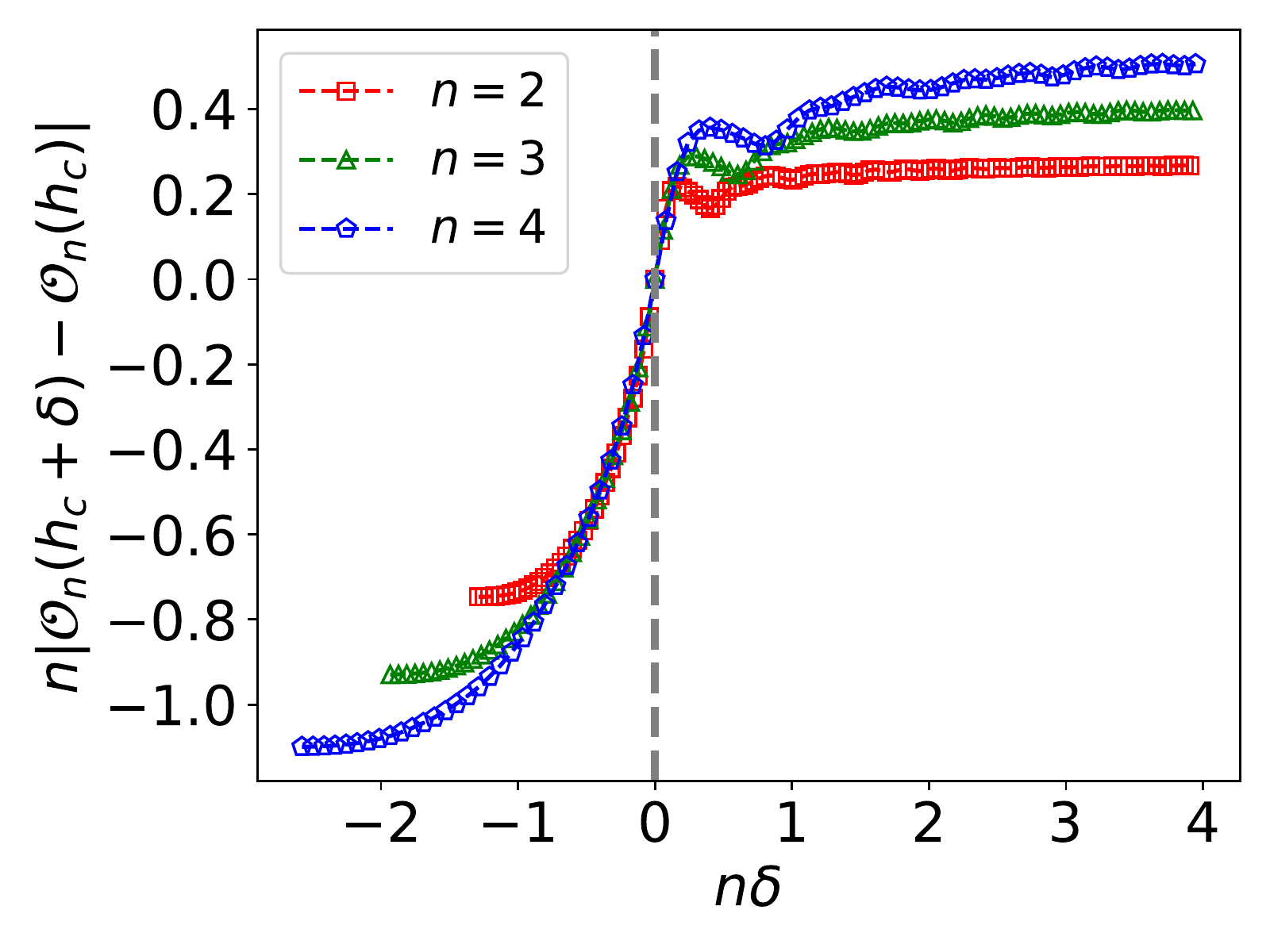}
	
	\caption{ The finite string observables $\mathcal{O}_n$ collapse near the equilibrium QCP in the nonintegrable  {regime} ($J_2=-0.2$) following a quench from a completely polarized ferromagnetic ground state at $h=0$. The time averaging has been done  {over the interval $[0,T=30]$} in a chain of $L=16$ spins. The exact critical point is at $h_c=0.6393$ as established in previous works using DMRG calculations (see Ref.~\cite{matteo06}).}
\label{fig:6}
\end{figure}
 In Fig.~\ref{fig:5a} we show a finite size scaling collapse following a nonintegrable and integrable quench, respectively. In the finite-size scaling regime, the transition in $\mathcal{O}_L$ for different system sizes is seen to collapse near the equilibrium critical point into a single function for the scaling exponents $\nu=\alpha=1$, thus verifying the scaling relation in Eq.~\eqref{eq:scaling_1}. We therefore find that the sharpness of the transition detected by $\mathcal{O}_L$ exhibit a universal approach to a true nonanalyticity as one approaches the thermodynamic limit.\\

Also, due to the emergence of a nonanalyticity at equilibrium QCPs, the second derivative $\left|\partial_\delta^2\mathcal{O}_L\right|$ diverges in the thermodynamic limit. In the finite size scaling regime, owing to the scaling form in Eq.~\eqref{eq:scaling_1}, we expect the second derivative near the QCP to diverge as,
\begin{equation}
\left|\partial_{\delta}^2\mathcal{O}_L\right|\sim L^{\frac{2-\alpha}{\nu}}\Psi(\delta^{1/\nu} L),
\end{equation}
in the regime $L^{1/\nu}\delta \ll 1$, for some dimensionless scaling function $\Psi(x)$ satisfying $\Psi(0)={\rm const\neq 0}$. Therefore, for an Ising critical point, the quantity $\left|\partial_{\delta}^2\mathcal{O}_L\right|$ diverges polynomially in the thermodynamic limit as,
\begin{equation}\label{eq:chi_fs}
\left|\partial_{\delta}^2\mathcal{O}_L\right|\sim L\Psi( L\delta),
\end{equation}
near the critical point. This scaling behavior has been verified in Fig.~\ref{fig:5b}, where the rescaled derivative response $L^{-1}\partial_{\delta}^2\mathcal{O}_L$ when plotted against $L\delta$ shows a good collapse near the critical point for different system sizes.  {We analytically derive this scaling form for integrable quenches in Appendix~\ref{Sec:A2}}.\\

Moving on to the finite projectors $P_n$ ($n<L$), due to the introduction of another length scale $n$, we assume a modified scaling ansatz near an equilibrium QCP in the limit $n\ll L$,
\begin{equation}\label{eq:scaling_n}
|\mathcal{O}_n(h_c+\delta)-\mathcal{O}_n(h_c)|=n^{-\alpha/\nu}\Pi(n^{1/\nu}\delta),
\end{equation}
where $\Pi$ is the associated  scaling function. Similar to the finite size scaling of the full projectors $\mathcal{O}_L$, in Fig.~\ref{fig:6} we show that the data for $\mathcal{O}_n$ having different string sizes $n$, collapse under scaling transformations with the exponents $\alpha=\nu=1$ for strongly chaotic quenches.  {Again, we derive this scaling of the observables with $n$ in the integrable regime in Appendix~\ref{Sec:A3}}. These scaling forms allow us to extract universal critical information following a quench using the string observables in a scalable approach. Furthermore this suggests that by measuring string expectations accessible to measurements which are local in space and time, it is possible to accurately determine the universality class of equilibrium QCPs in generic quenching experiments.  {In Appendices~\ref{Sec:sr_int} and \ref{Sec:acc} we show  {how this scaling behavior is modified} under stronger integrability breaking.   {In particular,  we show} that these critical signatures scale with both the system size and the string length and  {gradually disappear in the thermodynamic infinite time limit for finite $n$ in agreement with ETH.  Nevertheless, the scaling given in Eq.~\eqref{eq:scaling_n} works very well~ even far away from integrability if we preform very long but finite time averages and one can still extract very accurately the position of QCP even with short strings in quenches to the strongly chaotic regime (see also Fig.~\ref{fig:drift}).}}

\section{Statistics of finite strings}
\label{string_stat}

 {To further assign an experimental relevance to the string observables, we look for post quench signatures of the critical point in the statistics of local strings with a close connection to recent experimental results.   {The string observables are very natural to detect in local projective measurement schemes typically used in various quantum simulators.  For example $P_3$ in Eq.~\eqref{eq:string} measures the density of instances, where three or more spins polarized along the $x$-axis are detected next to each other. } Particularly,  using site resolved florescence imaging of trapped-ion lattices, Ref.~\cite{Zhang2017} showed that it is indeed possible to directly measure domain formation probabilities in a post-quench many-body spin system. We numerically simulate such quenching experiments and probe the connection of the proposed string operators with domain formation probabilities. }\\

\begin{figure*}

	\subfigure[]{
		\includegraphics[width=5.6cm,height=5cm]{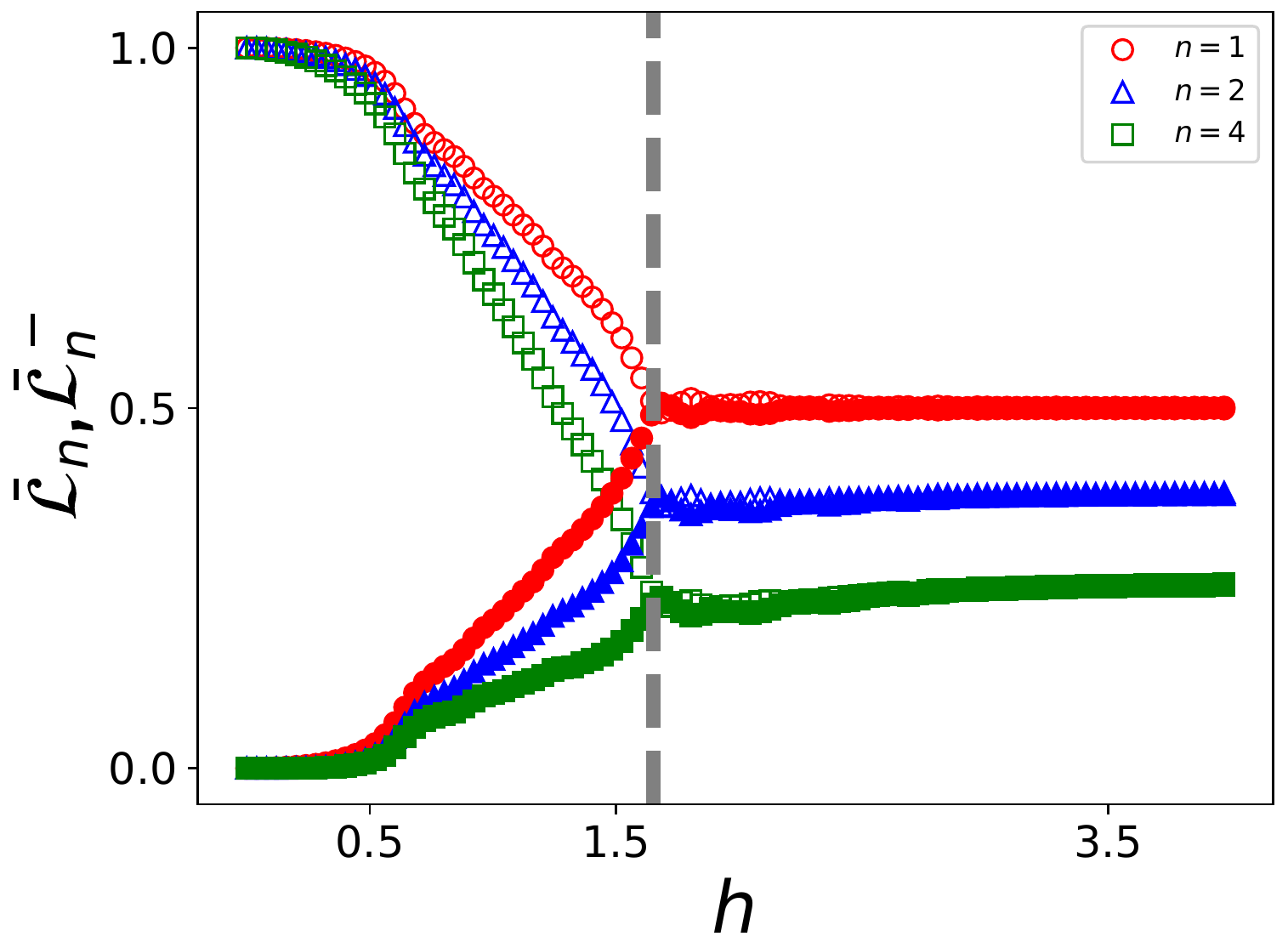}
		\label{fig:8b}}
	\subfigure[]{
    	\includegraphics[width=5.6cm,height=5cm]{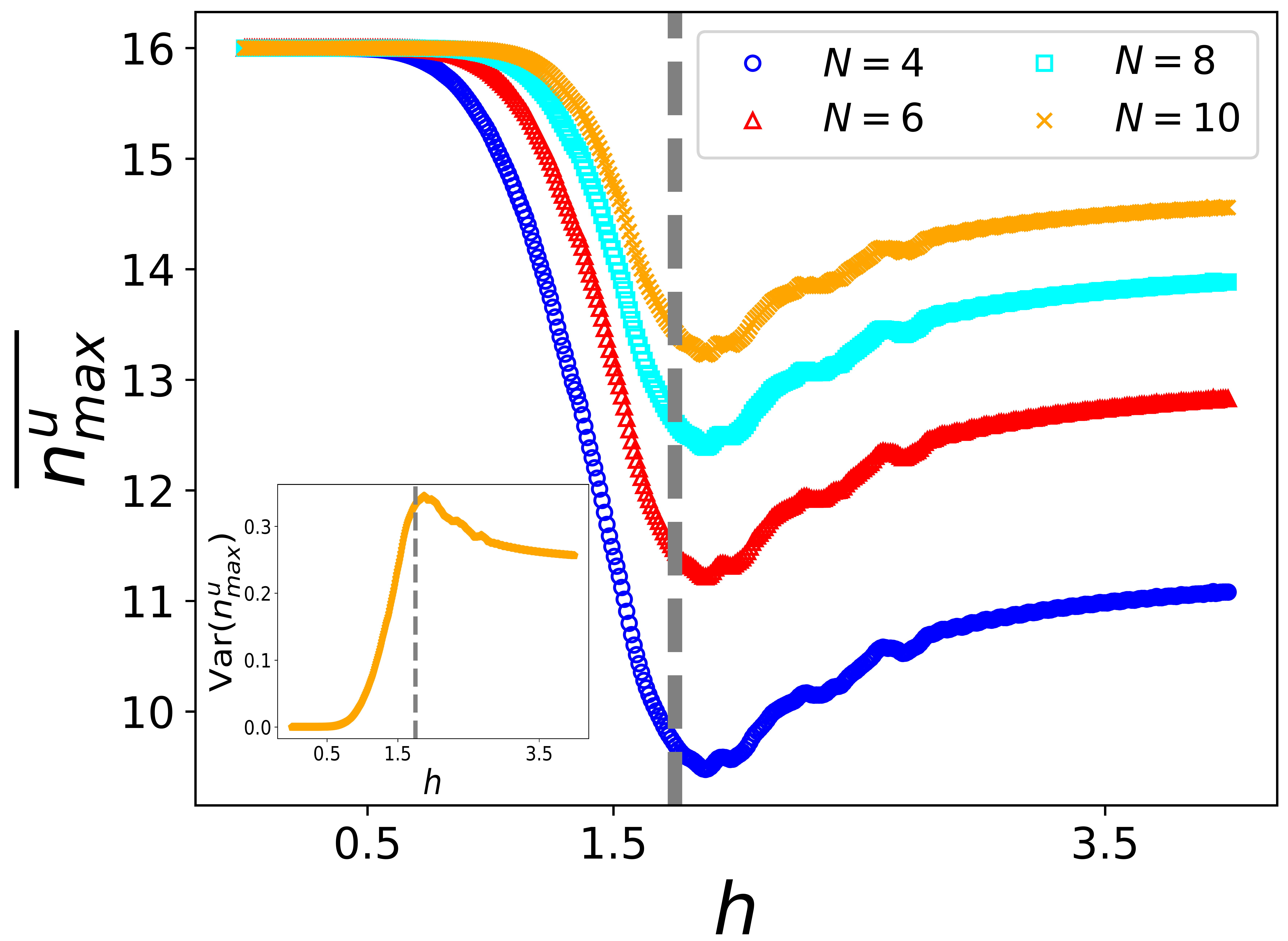}
	    \label{fig:9a}}
    \subfigure[]{
	    \includegraphics[width=5.6cm,height=5cm]{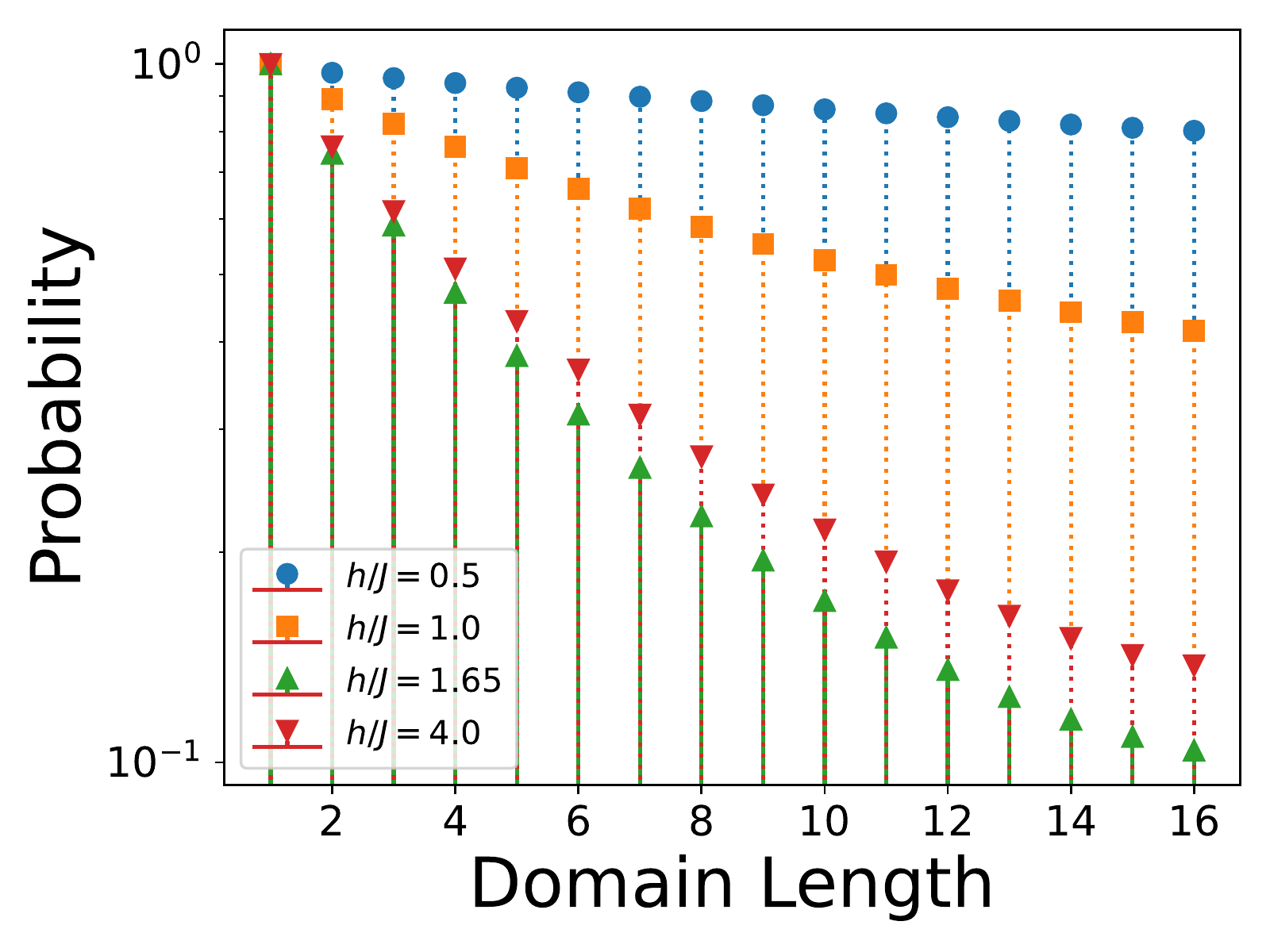}
	    \label{fig:9b}}

	\caption{ (a) For quenches into the paramagnetic phase both $\rightarrow$ and $\leftarrow$-polarized strings contribute equally after long times for $L=16$ spins. The hollow (solid) symbols representing the single measurement probabilities $\bar{\mathcal{L}}_n$ ($\bar{\mathcal{L}}_n^-$), collapse for a particular string length $n$  {when the quenched field lies in the disordered phase}. The noise in the data near the critical point is a finite-size effect (see Appendix~\ref{sec:noise}). (b) The mean longest size of   domains after $N$ measurements show a prominent dip precisely near the critical point in a system of size $L=16$. (Inset) The coefficient of variation of the maximum domain length from $N=10$ uncorrelated measurement data showing maxima for a critical quench in the nonintegrable setup.  {(c) Time averaged probability distribution of domains of either polarization. The probabilities of formation of long domains are smallest for the quench close to the critical point (green triangles) in a qualitative agreement with the experimental observations of Ref.~\cite{Zhang2017}.  In all plots time averaging was done over the window of time $T=30$ following the quench starting from the completely polarized state $\ket{\rightarrow\rightarrow...}$ in the nonintegrable regime corresponding to $J_2=0.5$. } 
	}
	\label{fig:9_cover}
\end{figure*}

 {The observable  $\mathcal L_n(t)$ as defined in Eq.~\eqref{eq:ln} is the average density of the projectors 
\[
P_n^i={1\over 2^n}\prod_{j=i}^{i+n-1} \left(\mathbb{I}_j+\sigma_j^x\right).
\]
For translationally invariant states we are considering, this average is nothing but the probability of having a string of length $n$ or more starting from any given site $i$ and continuing to the right of this site.  Correspondingly
$\bar {\mathcal L}_n$ defined in Eq.~\eqref{eq:On_def} represents a time average of the probability of observing such a string after the quench,  which is also equivalent to the probability of observing this string in the time-averaged post quench density matrix.  We now show that the operators $P_n$ provide  {us with} sufficient information to extract post-quench domain statistics. For this purpose, it is convenient to introduce the following notation:
\begin{equation}
\bar{\mathcal{L}}_n=\rho(l^+\geq n),
\end{equation}
with $\rho(l^+\geq  n)$,  as we just discussed,  representing the late-time probability of finding $\rightarrow-$strings having length $l^+=n$  {or longer to the right (or to the left) of an arbitrary site~\footnote{In what follows, while referring to probabilities of obtaining string lengths of a given site we will always imply a fixed position of the string in the chain}}.   This observation in turn implies that  {$1-\bar{\mathcal{L}}_n=1-\rho(l^+\geq  n)=\rho(l^+<n)$} is {the probability of finding polarized strings of  length $n$ or smaller}.  Therefore
\begin{equation}\label{eq:prob}
\rho(l^+\leq n)=1-\bar{\mathcal{L}}_{n+1}.
\end{equation}
Now consider $N$ independent measurement sequences on the system determining the time-averaged spin configurations following a sudden quench, as it was employed in the experiment~\cite{Zhang2017}.  In this setting, we proceed to find the probability that the longest $\rightarrow-$ polarized string starting at some fixed position, say $i=0$,  observed after all such independent measurements has a length $n$ as
\begin{multline}\label{eq:Mprob}
\mathcal{M}_n^+(N) =\rho(l^+\leq n)^N-\rho(l^+\leq n-1)^N\\=
 \left(1-\bar{\mathcal{L}}_{n+1}\right)^N- \left(1-\bar{\mathcal{L}}_n\right)^N.
\end{multline} 
It is easy to see that as expected
\begin{equation}\label{eq:norm_u}
\sum\limits_{s=0}^{L}\mathcal{M}_s^+(N)=1.
\end{equation}
}\\

%
Similar to the distributions $\bar{\mathcal{L}}_n$ of $\rightarrow$-polarized domains, we now construct the post-quench distribution of $\leftarrow$-polarized strings following a quench as $\bar{\mathcal{L}}_n^-$. The probabilities $\bar{\mathcal{L}}_n^-$ are simply given by time averaged expectation of the strings,
\begin{equation}
	P_n^-=\frac{1}{L}\sum\limits_{i=1}^{L}\frac{1}{2^n}\prod\limits_{j=i}^{i+n-1}\left(\mathbb{I}_j-\sigma_j^z\right).
\end{equation}
projecting onto $\leftarrow$-polarized clusters of length $l^{-}\ge n$. Following Eq.~\eqref{eq:Mprob}, one can now proceed to define probability distributions $\mathcal{M}_n^-$ that the longest $\leftarrow$-string detected has a length $l^-= n$ in the post quench state. In Fig.~\ref{fig:8b},   {we observe such probability distributions $\bar{\mathcal{L}}_n$ and $\bar{\mathcal{L}}_n^-$ averaged over a long finite time window} following a chaotic quench starting from the completely polarized initial state $\ket{\rightarrow\rightarrow...}$.   {Interestingly, in contrast to quenches into the ordered phase, $h<h_c$  where there is a very large anisotropy in the occurrence of left and right strings,  for quenches into the paramagnetic phase $h>h_c$ the isotropy is restored after long waiting times.  At small amplitude quenches starting from a completely polarized ground state $\ket{\rightarrow\rightarrow...}$ such anisotropy is not surprising as only large domains of the same polarization are likely to appear (see Appendix~\ref{Sec: Appendix_2}).  Interestingly, we find that the probabilities of creating left- and right- domains meet exactly at the  {zero temperature transition point despite pumping large energy into the system during the quench} and remain essentially identical after that.  The crossover between the anisotropic and isotropic regimes is very even for non-perturbative integrability breaking quenches and for short strings.  We note that this behavior of domain distributions is also consistent with recent studies in quenched systems~(see also \cite{halimeh21}), studying the signatures of QCP through finite but long-time behavior of single-spin observables (see Appendix~\ref{Sec:Appendix_3})}.\\

 {Equipped with the independent distributions of detecting polarized $\rightarrow/\leftarrow$-domains, we next proceed to construct joint distributions of detecting longest $\rightarrow$-polarized string having length $l^+=n$ and longest $\leftarrow$-polarized string having length $l^-=m$ in the late-time state,
\begin{equation}
\mathcal{M}(n,m)=\rho\left(l^+\leq n\cap l^-\leq m\right)=\mathcal{M}_n^+\mathcal{M}_m^-.
\end{equation}
This in turn allows us to explicitly write down post-quench probability distributions that the longest   string observed after $N$ independent measurements has a length $n$ as,
\begin{eqnarray}\label{eq:exclusive_u}
	\nonumber\mathcal{M}_n(N)&=&\left(\mathcal{M}_n^+\mathcal{M}_n^-\right)^N-\left(\mathcal{M}_{n-1}^+\mathcal{M}_{n-1}^-\right)^N,
\end{eqnarray}
irrespective of its polarization.
We now find the mean longest   domain length observed after $N$ independent measurements in the post quench state as,
\begin{equation}
\overline{{\rm n_{max}^u}}=\sum\limits_{s=0}^Ls\mathcal{M}_s(N).
\end{equation}
In Fig.~\ref{fig:9a}, we show that the mean longest domain size exhibits a minima close to the equilibrium QCP, which allows for an accurately experimentally accessible determination of the critical point in quenching experiments. In fact, such minima of the mean maximum domain length near a critical quench,  were recently observed in Ref.~\cite{Zhang2017} using a quantum simulator. The string observables therefore offer a comprehensive understanding of the recent experimental findings.  At the same time, the coefficient of variation/relative deviation of the maximum domain length,
\begin{equation}
{\rm Var}({\rm n_{max}^u})=\frac{\sigma({\rm n_{max}^u)}}{\overline{{\rm n_{max}^u}}},
\end{equation}
$\sigma$ denoting the standard deviation of the longest domain length after long times, also shows a sharp maxima when the system is quenched to the critical point, reflecting large fluctuations near critical quenches [see the inset of  Fig,~\ref{fig:9a}].  In Fig.~\ref{fig:9b}, we plot $\bar{\mathcal{L}}_n+\bar{\mathcal{L}}_n^-$ defining probabilities of finding strings of any polarization and of length $n$ or longer.  The probabilities of large domains clearly reach a minimum for quenches close to the ciritcal point (green triangles) again in a qualitative agreement with the experiment~\cite{Zhang2017}.  Recently, the domain statistics of polarized spins have also been seen \ct{jad21,pol21,heyl21} to detect dynamical quantum phase transitions at early times developing sharp cusp singularities. Our results suggest that similar singularities near equilibrium QCPs, now as a function of quench amplitude, develop in the long time behavior of finite string observables.}\\

\section{Infinite temperature unequal time correlations}
\label{sec:otoc}

Apart from the equal time expectations, we now proceed to examine the various two-time probes constructed out of string operators which are manifestly insensitive to the choice of the initial state in quenching setups. The infinite temperature auto-correlation and OTOC of the string operators provide a good testbed for this purpose, as they are experimentally measurable in optical lattices and at the same time contain no information of the initial state of the system.
\begin{figure}
	\centering
	\includegraphics[width=7.5cm,height=6cm]{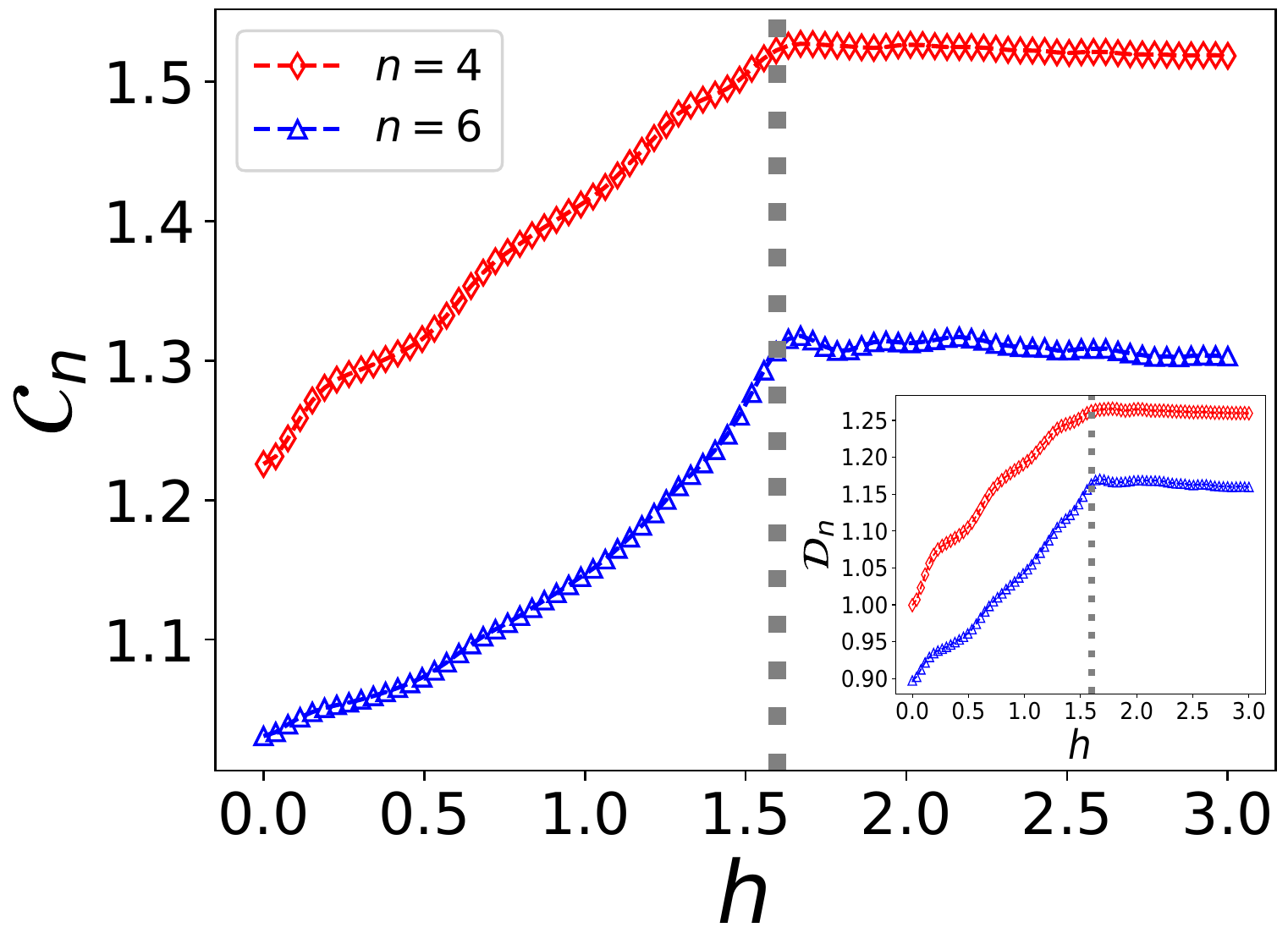}
	
	\caption{ {Logarithmic infinite temperature time averaged OTOC following a quench to a transverse field $h$ starting from a completely polarized state $\ket{\rightarrow\rightarrow...}$ in the nonintegrable ANNNI model. The OTOC is seen to develop a sharp transition at the equilibrium critical point $h_c\approx 1.6$ for $J_2=0.5$}. (Inset) Sharp signatures in the corresponding logarithmic auto correlation function. In both the simulations, the time averaging has been done up to $T=20$ in a chain of $L=10$ spins.}
	\label{fig:10} 
\end{figure}
 Particularly, we consider the long time averages of the infinite temperature auto-correlation and OTOCs,
\begin{eqnarray}\label{eq:inf_corr}\
\nonumber\mathcal{D}_n=-\frac{1}{n}\log\lim\limits_{T\rightarrow\infty}\frac{1}{T}\int_{0}^{T}dt\frac{1}{2^L}\rm{Tr}\left[P_n(t)P_n(0)\right],\\
\mathcal{C}_n=
-\frac{1}{n}\log\lim\limits_{T\rightarrow\infty}\frac{1}{T}\int_{0}^{T}dt\frac{1}{2^L}\rm{Tr}\left[P_n(t)P_n(0)P_n(t)\right].
\end{eqnarray}
Note that for extensive projectors, i.e., when the string sizes span the complete chain, we recover the full Loschmidt echo from these infinite temperature probes,
\begin{equation}
\begin{split}
\rm{Tr}\left[P_L(t)P_L(0)\right]=\braket{\psi_0|P_L(t)|\psi_0}=\mathcal{L}_L(t),\\
\rm{Tr}\left[P_L(t)P_L(0)P_L(t)\right]=\rm{Tr}\left[P_L(0)P_L(-t)P_L(0)\right]\\
=\braket{\psi_0|P_L(-t)|\psi_0}=\mathcal{L}_L(-t).
\end{split}
\end{equation}

Also, for the local strings ($n<L$), the  {infinite} time averages in Eq.~\eqref{eq:inf_corr} can be replaced by averages over all eigenstates $\ket{\phi_{\alpha}}$ of the quenched Hamiltonian, such that 
\begin{eqnarray}\label{eq:d}
e^{-n\mathcal{D}_n}=\overline{\braket{\phi_{\alpha}|P_n|\phi_{\alpha}}^2},
\end{eqnarray}
represents the second moment of the distribution of diagonal matrix elements $\braket{\phi_{\alpha}|P_n|\phi_{\alpha}}$ over the spectrum of the quenched Hamiltonian $H$. Eq.~\eqref{eq:d} further reflects the fact, that the averaged correlators are completely insensitive to both initial state information and time. Similarly, the long time averaged OTOC in Eq.~\eqref{eq:inf_corr} reduces to a time-insensitive weighted average over all eigenstates $\ket{\phi_{\alpha}}$,
\begin{equation}
e^{-n\mathcal{C}_n}=\overline{\braket{\phi_{\alpha}|P_n|\phi_{\alpha}}\braket{\phi_{\alpha}|P_n^2|\phi_{\alpha}}}
\end{equation}
\begin{figure*}
	\subfigure[]{
		\includegraphics[width=6.9cm,height=5.4cm]{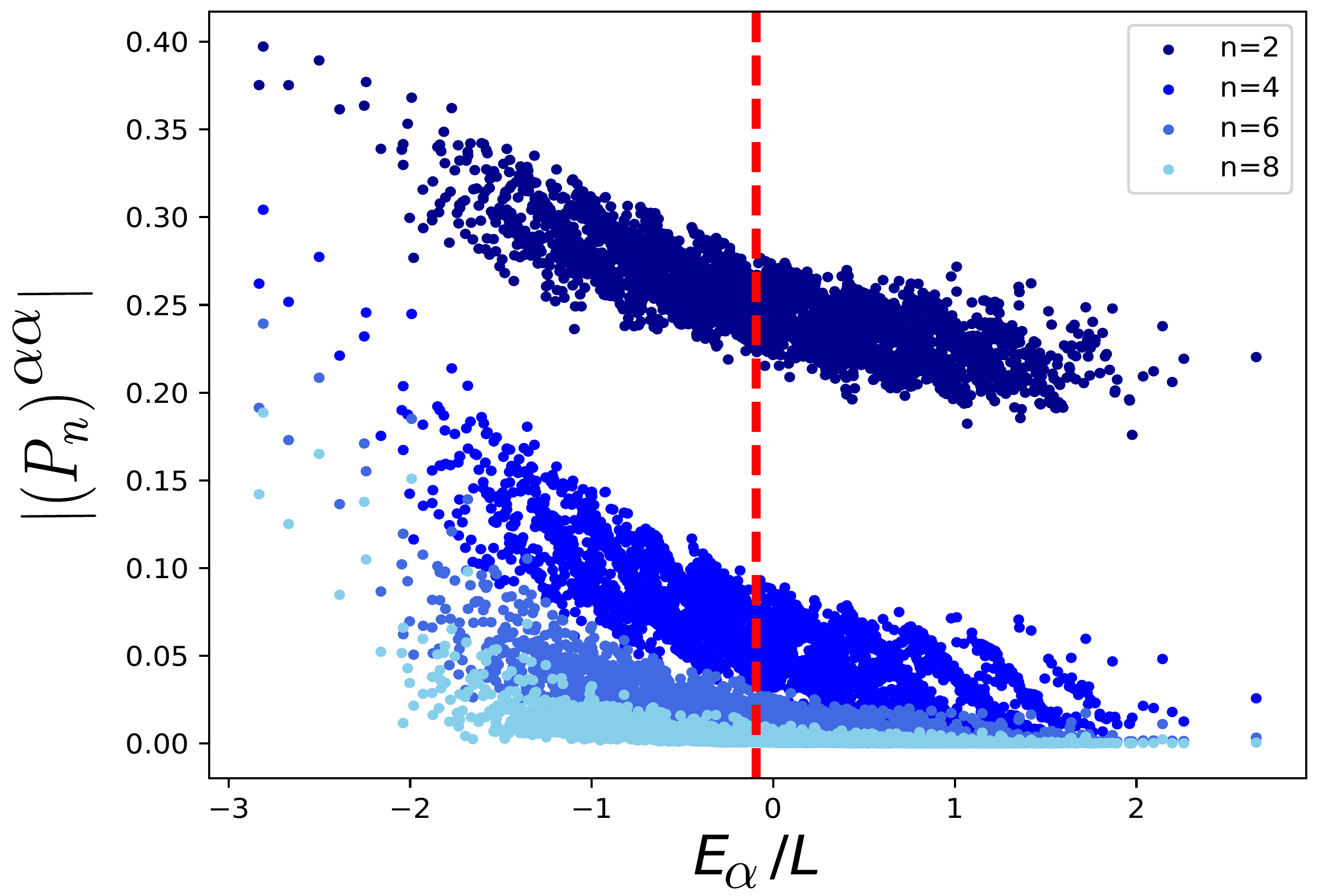}
		\label{fig:11a}}
	\hspace{1.5cm}
	\subfigure[]{
		\includegraphics[width=7.0cm,height=5.5cm]{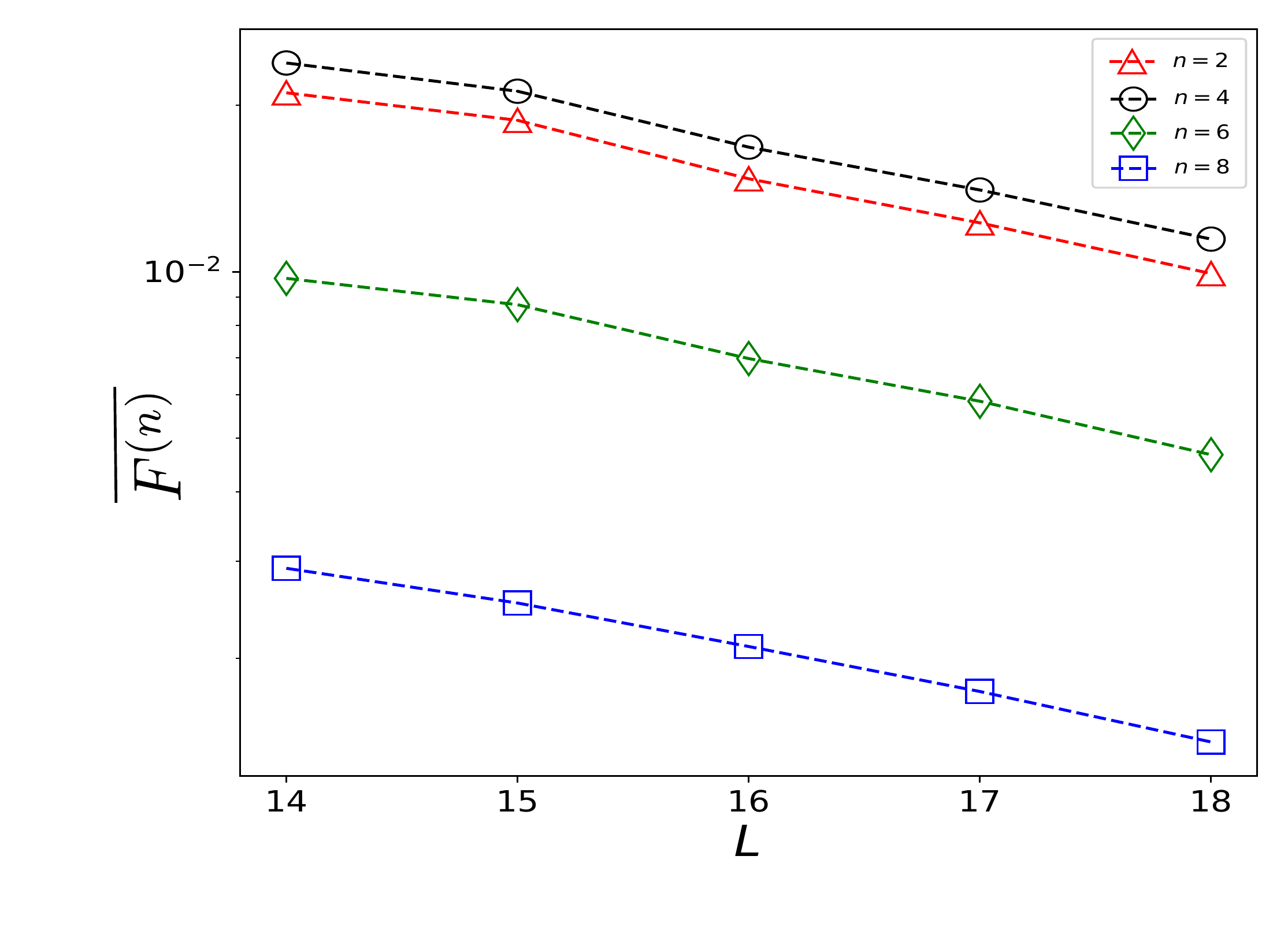}
		\label{fig:11b}}

	\caption{(a) The distribution of diagonal matrix elements of the local projectors $P_n$ when plotted against energy density in the eigenspace of a quenched chaotic ANNNI Hamiltonian, shows a diminishing contribution of excited states with increasing string lengths. Starting from an initial ground state of the Hamiltonian [see Eq.~\eqref{eq:ham}] for $J_2=1.0$, $h=2$, the system is quenched to $J_2=1.0$, $h=-2.5$ in a chain of $L=16$ spins. These quench parameters have been chosen such that the mean energy after the quench (marked by the vertical dashed line) lies near the middle of the spectrum, where we have a sufficiently high density of states. (b) Mean state-to-state fluctuation of the operators $P_n$ in nearby eigenstates, averaged over $50\%$ of the eigenstates centered about the mean energy after quench.}
	\label{dist}
\end{figure*}
In Fig.~\ref{fig:10}, we see that similar to $\mathcal{O}_n$, finitely long time averaged correlations $\mathcal{C}_n$ and $\mathcal{D}_n$ develop sharp signatures near the equilibrium critical point following a nonintegrable quench. Particularly, the observables are seen to be sensitive to even a small shift in the critical point due to an integrability breaking perturbation.   {Physically the observables $C_n(t)$ and $D_n(t)$ reveal a similar information as $\mathcal{O}_n$ but now here is no dependence on the initial state.  Indeed $P_n(0)$ does a partial projection to a polarized string and $P_n(t)$ measures the memory effect for this string observable.  As with $\mathcal{O}_n$ we see that the long time limit of this correlation function develop a very sharp singularity with increasing string length corresponding to the QCP, even though we are averaging over the infinite temperature ensemble which in itself does not distinguish ground states from excited states.}\\

\section{Thermalization and persistent memory of finite strings}
\label{Sec:memory}

\begin{figure}
	\centering
	\includegraphics[width=7.3cm,height=5.75cm]{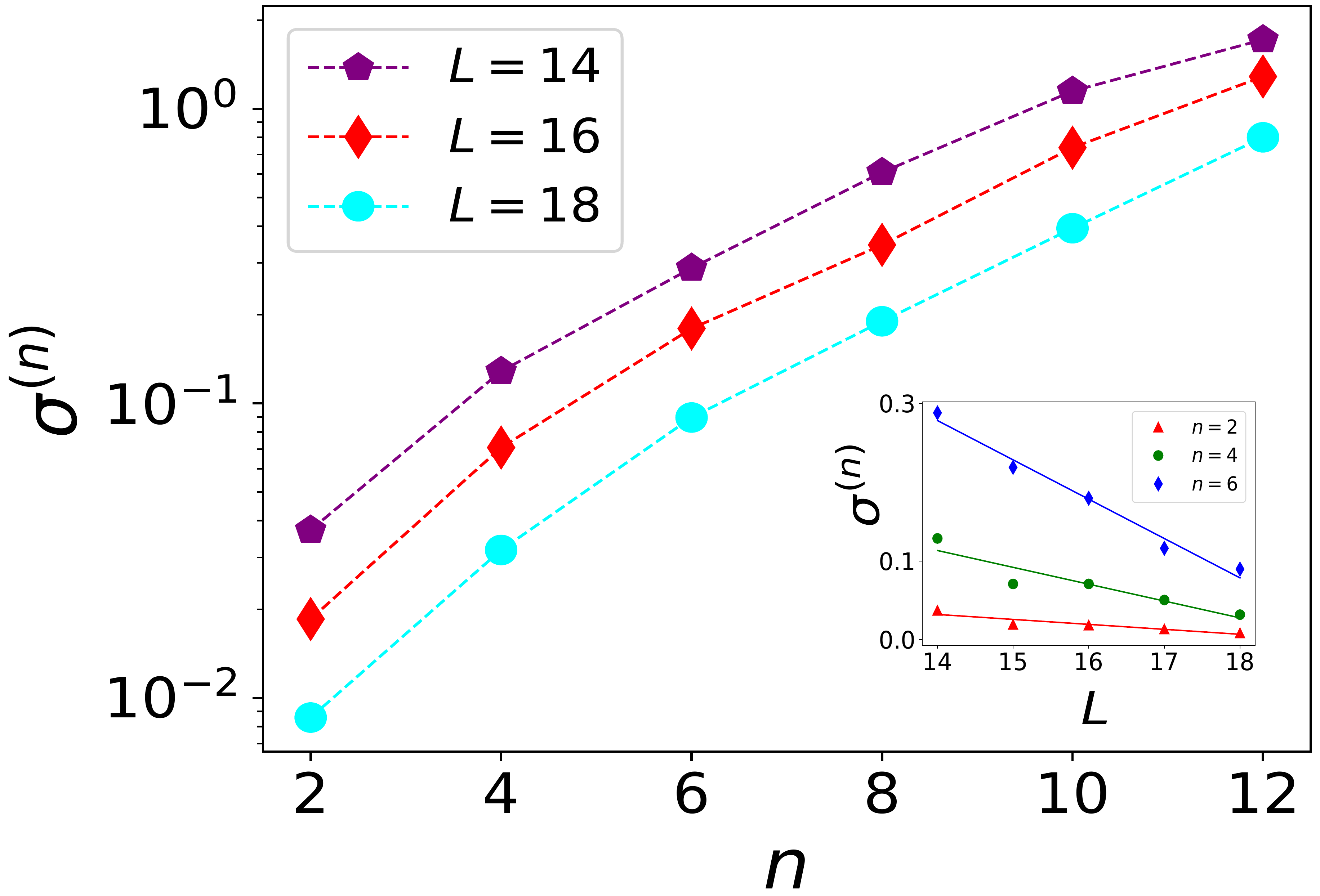}
	
	\caption{Relative fluctuation of the diagonal matrix elements of string observables against string length $n$, within a small energy shell of half-width $\delta E=0.1\sqrt{L}$ about the mean energy $\overline{E}$. (Inset) The relative fluctuations decay with system size very slowly with increasing $L$ indicating a much weaker dependence on system size than on string length. The initial and final quenching fields were chosen to be $h=-2.2$ and $h=1.8$ respectively in the ANNNI Hamiltonian with $J_2=1.0$, such that $\overline{E}$ lies near the center of the spectrum.}
	\label{fig:11c} 
\end{figure}
{From the generic behavior of the distributions $\bar{\mathcal{L}}_n/\bar{\mathcal{L}}_n^-$ and the post quench domain statistics, it is evident that string observables retain strong memory of the initial ground state after long times  {even for} a chaotic quench. In this section, we attempt to understand this seemingly conflicting behavior of such operators in light of the ETH,  which on the contrary,   {suggests that systems should reach thermodynamic equilibrium only retaining their memory of the conserved energy (in the absence of other conservation laws). We note that while ETH is generally expected to apply to all local operators,  its validity to string observables was not analyzed in prior literature.  It is clear that in order to have long memory it is necessary that these observables should have large fluctuations between different eigenstates with similar energies.  In what follows we are going to address this question.}\\

The approach of the local string projectors to the completely polarized state projector with increasing string length can be seen to manifest in the eigenstate distribution of the diagonal matrix elements $\left(P_n\right)^{\alpha\alpha}=\braket{\phi_{\alpha}|P_n|\phi_{\alpha}}$. In Fig.~\ref{fig:11a}, we show the distribution of the diagonal matrix elements in the eigenstates of a chaotic ANNNI Hamiltonian.  As the string length $n$ increases, the operators quickly become localized near the  {polarized} state,  reflecting how the probe $\mathcal{L}_n(t)$ approach the full Loschmidt echo in quenching setups. Further according to ETH,  the expectation of local observables  {should be} smooth functions of energy when it is the only conserved quantity. This implies that the state to state fluctuations between nearby eigenstates,
\begin{equation}
	F^{(n)}_{\alpha}=|\left(P_n\right)^{(\alpha+1)(\alpha+1)}-\left(P_n\right)^{\alpha\alpha}|
\end{equation}
should  {decay exponentially with the systems size: $F_\alpha^{(n)}\sim \exp[-S(L)/2]$,  where $S$ is the  entropy of the system.}\\

 {In Fig.~\ref{fig:11b}, we show the dependence of the mean fluctuations $\overline{F^{(n)}}$ on the system size.. This mean is obtained by averaging the differences $F^{(n)}_{\alpha}$ over high energy eigenstates near the average energy  $\overline{E}=\braket{\psi_0|H|\psi_0}$, following the quench. We see a clear exponential decay of $\overline{F^{(n)}}$ with system size irrespective of the string length.  Moreover while the overall magnitude of fluctuations is strongly suppressed with increasing $n$,  the slope does not depend on $n$.  Both observations are consistent with ETH~\cite{huse14, luca16,grover18} confirming that the string observables indeed satisfy ETH}.\\

\begin{figure*}
	\subfigure[]{
		\includegraphics[width=5.5cm,height=4.7cm]{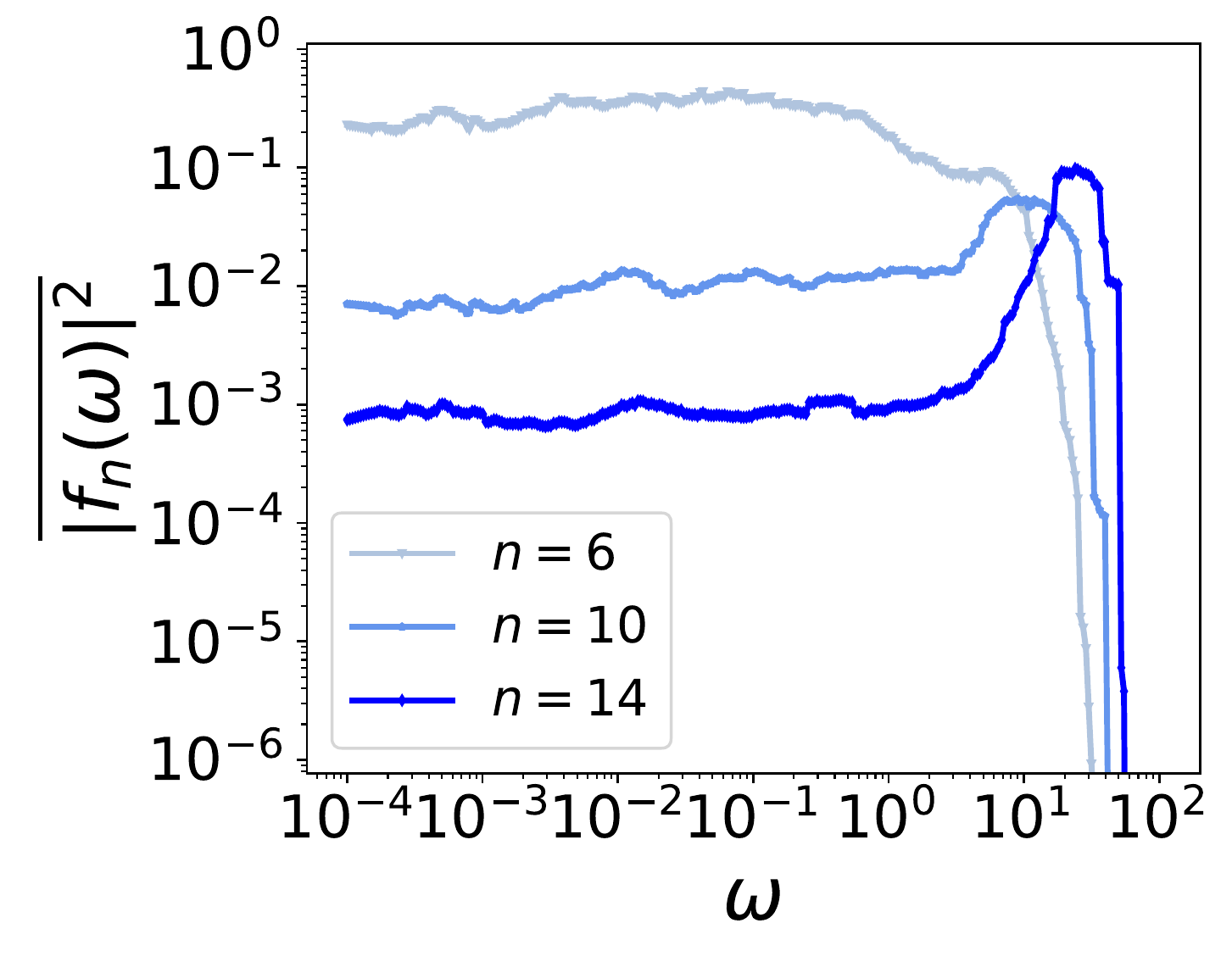}
		\label{fig:12a}}
	\subfigure[]{
		\includegraphics[width=6cm,height=4.7cm]{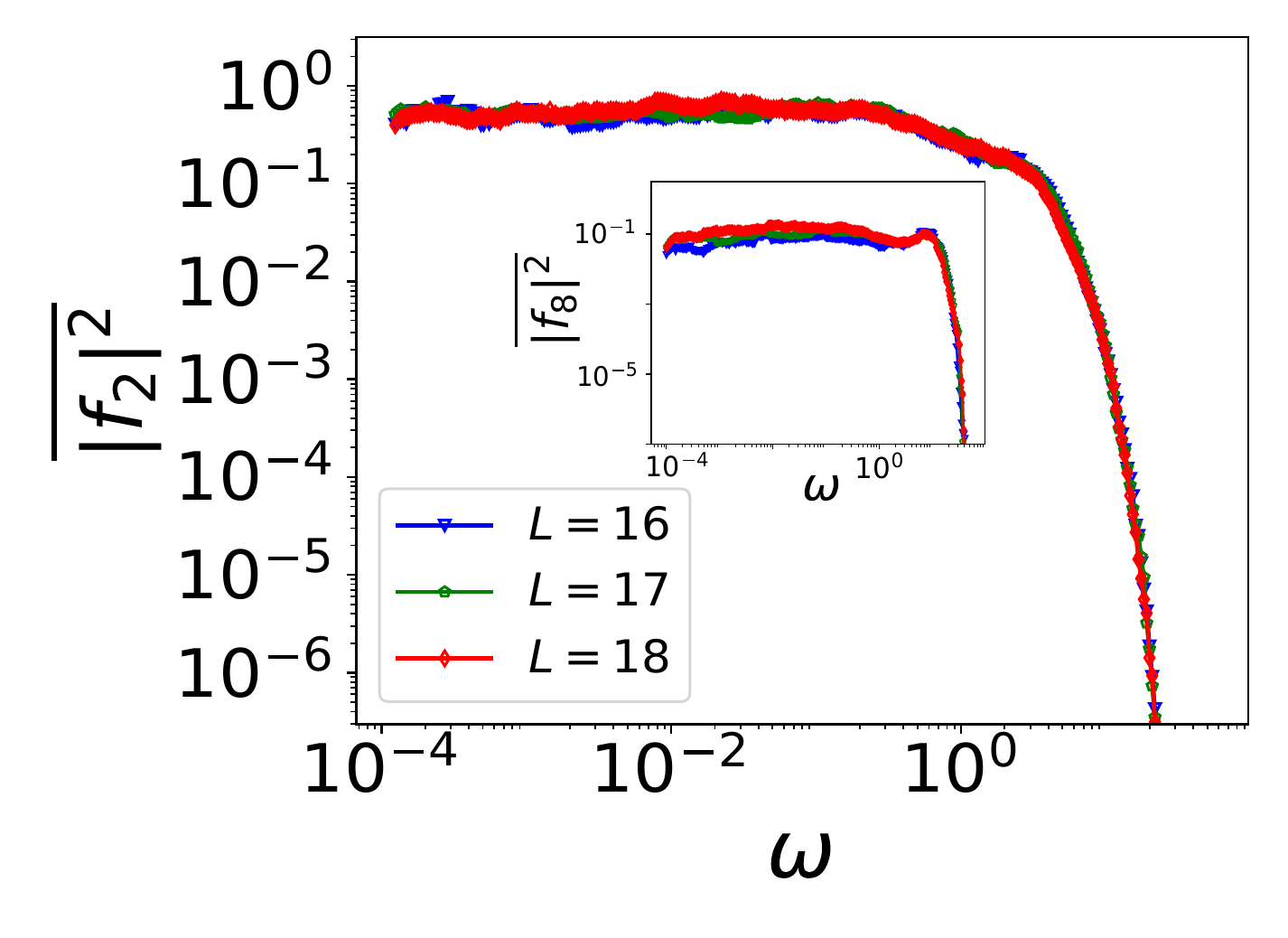}
		\label{fig:12b}}
	\subfigure[]{
		\includegraphics[width=5.5cm,height=4.5cm]{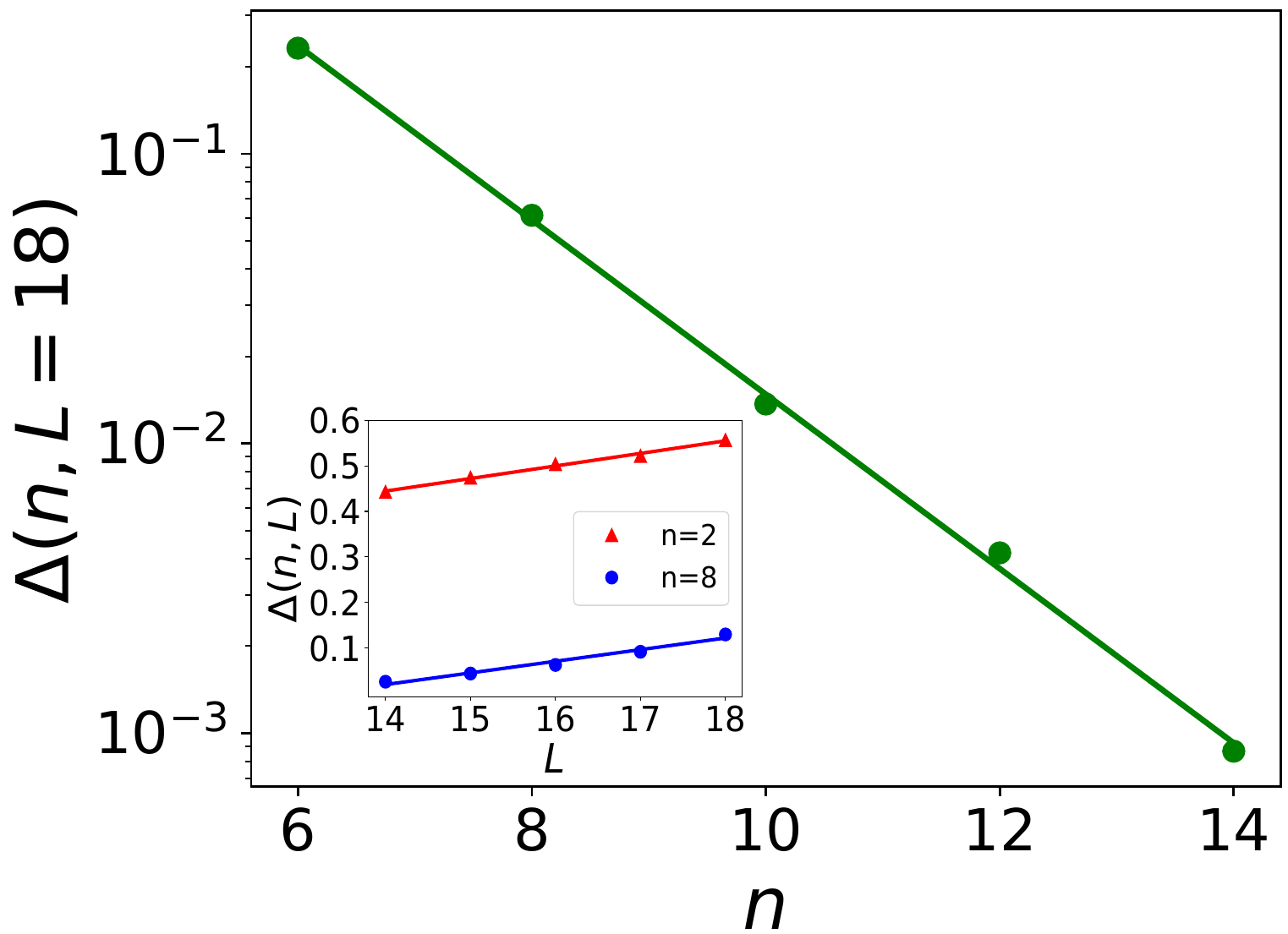}
		\label{fig:12c}}

	\caption{ {(a) The average spectral function of the string observables $P_n$, over central $50\%$ of the eigenstates in the chaotic ANNNI Hamiltonian with $J_2=1$ for $L=16$ spins. The data shown has been averaged over $50$ Hamiltonians with equispaced magnetic fields $h\in[0.9,1.1]$. With increasing string length, the low frequency spectral function is seen to approach a highly localized distribution indicating late time oscillatory behavior of connected auto-correlation functions [see Eq.~\eqref{eq:realtime}].}  {At the same time the low frequency plateau $\Delta(n,L)=\overline{|f_n(\omega\rightarrow 0)|^2}$ characteristic of chaotic random-matrix behavior of the spectral function at late times rapidly goes down with the string size indicating a rapid formation of the spectral gap. (b) The low frequency plateau of the spectral function for a particular string length rises slowly with increasing system size in accordance with ETH. However, the high frequency peak is clearly visible for sufficiently long strings even with increasing system size (inset). In panel (c), we see that while the low frequency plateau height is linearly dependent on system size consistent with late-time diffusion (inset), the spectral gap forms exponentially fast with increasing string length.}}
	\label{spectralf}
\end{figure*}

We see that the strong memory effects of the initial state in string observables are not coming from any ETH violations.  What happens instead is that for string observables their fluctuations  rapidly increase with $n$.  To quantify this effect we consider relative fluctuations of the diagonal matrix elements of $P_n$ averaged over a narrow energy window:
\begin{equation}
	\sigma^{(n)}=\frac{\sqrt{{\rm Var}[\left(P_n\right)^{\alpha\alpha}]}}{\left|\overline{\left(P_n\right)^{\alpha\alpha}}\right|}.
\end{equation}
The averaging is done in a finite energy window $\left[\overline{E}-\delta E,\overline{E}+\delta E\right]$ around the mean energy $\overline{E}$. In quenching experiments with a local Hamiltonian, the variance in energy after quench usually scales as $\sim\sqrt{L}$ with system size. We therefore choose the width of the energy shell to scale as $\delta E\sim\sqrt{L}$ while calculating fluctuations of the diagonal matrix elements.\\

We show $\sigma^{(n)}$ vs. the string size $n$ for three different system sizes in Fig.~\ref{fig:11c}.  We see that these fluctuations rapidly increase with $n$.  This increase is consistent with the exponential scaling. Moreover these fluctuations only weakly depend on the system size as expected for local observables.  Because in a system with local interactions after a global quench energy variance is extensive,  large fluctuations of the string operators allow for preserving sharp memory of the initial state  { for large system sizes scaling exponentially with $n$}.\\

 {In thermodynamic limit large fluctuations of observables within a narrow energy window allow for very long relaxation times,  i.e.  for very long lived non-equilibrium states~\cite{Dymarsky_2019}.}  To explore this physics for string operators we analyze numerically their spectral function over high-energy eigenstates
\begin{equation}\label{eq:spect}
	\left|f_n(E_{\alpha},\omega)\right|^2=\frac{\sum\limits_{\beta\neq \alpha}\left|\braket{\phi_{\alpha}|P_n|\phi_{\beta}}\right|^2\delta\left(\omega_{\alpha\beta}-\omega\right)}{\braket{\phi_{\alpha}|P_n^2|\phi_{\alpha}}_c},
\end{equation}
where $\omega_{\alpha\beta}=E_{\alpha}-E_{\beta}$ and we have used a Lorentzian filter for the delta function  $\delta(\omega)\rightarrow\mu/[2\pi(\omega^2+\mu^2)]$, such that $\mu=0.9\omega_{min}$ where $\omega_{min}$ is the minimum level spacing. To ensure proper normalization of the spectral function, we divide by the eigenstate fluctuations $\braket{P_n^2}_c=\braket{P_n^2}-\braket{P_n}^2$ which is equivalent to the spectral function summed over all frequencies \cite{dries21}. \\
\begin{figure*}
	\subfigure[]{
		\includegraphics[width=7cm,height=5.5cm]{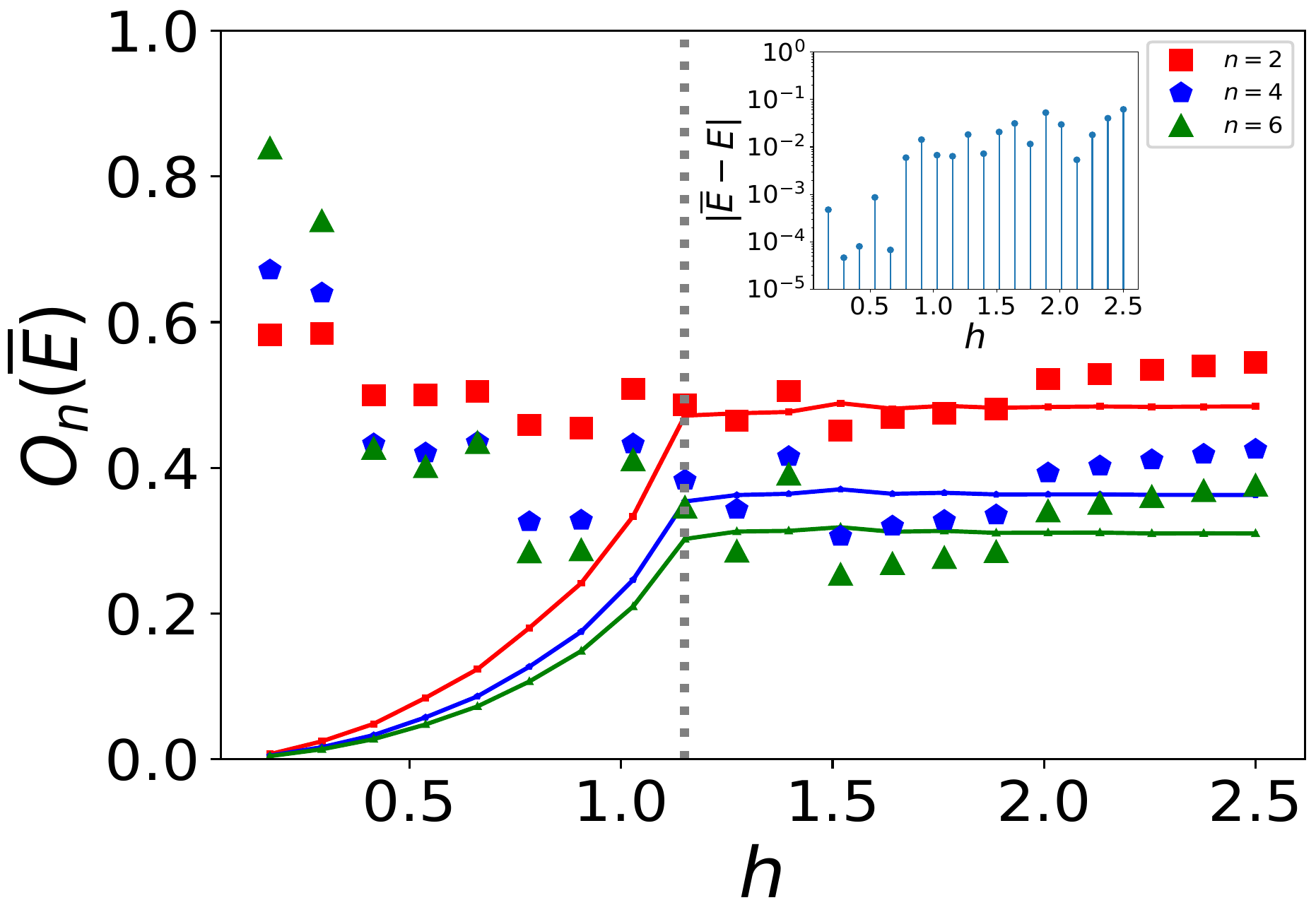}
		\label{fig:testa}}
	\hspace{1cm}
	\subfigure[]{
		\includegraphics[width=7.0cm,height=5.4cm]{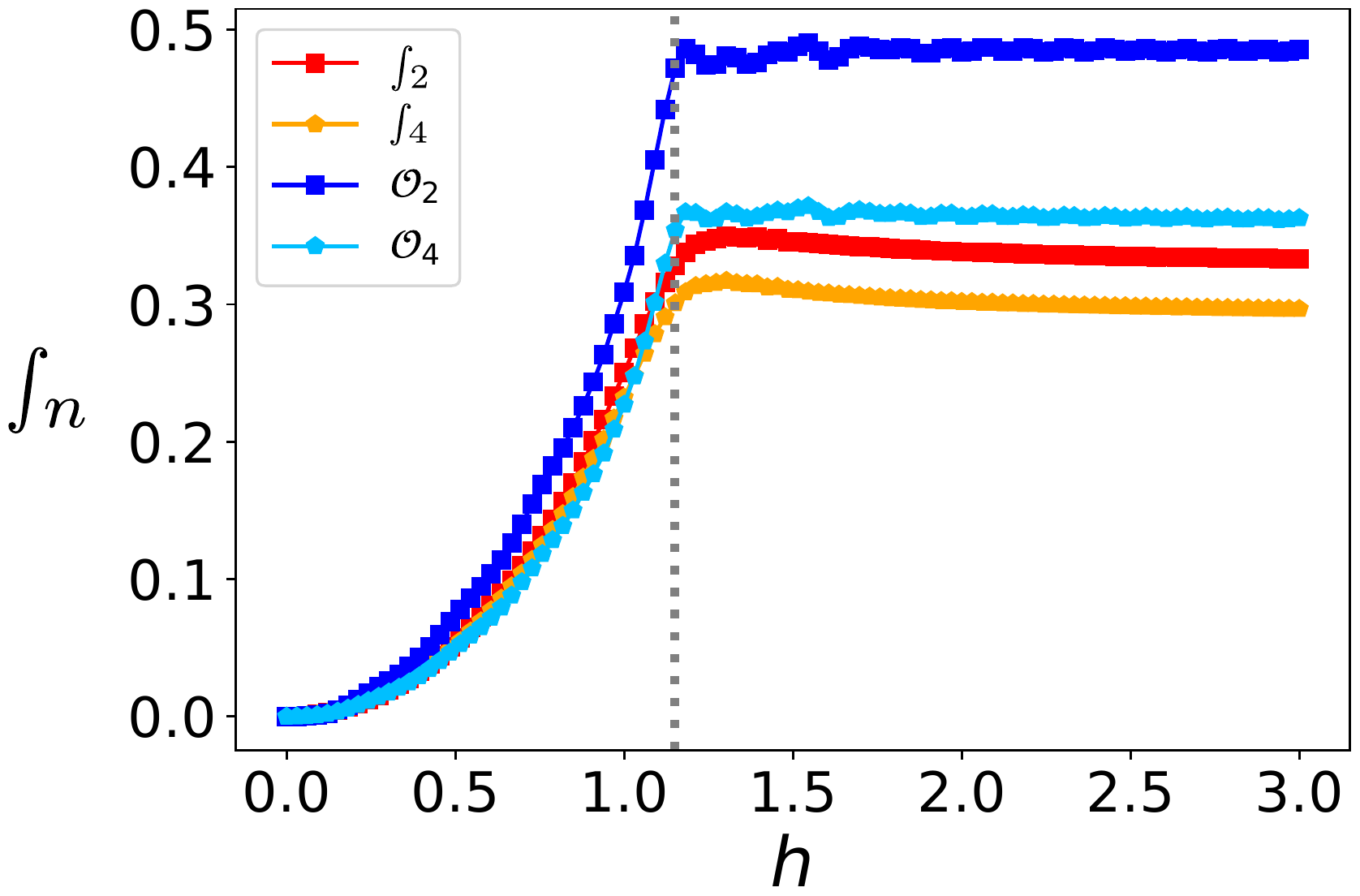}
		\label{fig:testb}}

	\caption{(a) The quantity $O(\overline{E})$ (discrete symbols) against the quenched field $h$ for different string lengths $n$. In contrast to the time averaged projectors $\mathcal{O}_n$ (solid lines), the single-eigenstate expectations $O(\overline{E})$ fail to detect the equilibrium QCP accurately. (Inset) The difference between the mean energy after quench and energy $E$ of the actual eigenstate chosen to evaluate $O(\overline{E})$. This difference is expected to decrease fast with increasing system size. (b) The string probes $\mathcal{S}_n$ contrasted with the local projectors $\mathcal{O}_n$ for different string lengths $n$ against the quenched field. Although the probes $\mathcal{S}_n$ develop a maxima near the equilibrium QCP, unlike $\mathcal{O}_n$ they fail to approach any sharp singularity with increasing string length. In both the panels the quenching simulation has been started from the polarized initial state $\ket{\rightarrow\rightarrow..}$ in a nonintegrable ANNNI model ($J_2=0.1$) of $L=16$ spins and all time averages have been performed up to $T=30$. The vertical dashed lines indicate the position of the actual QCP in equilibrium.}
	\label{test}
\end{figure*}

In Fig.~\ref{spectralf} we plot the average spectral function $\overline{|f_n(\omega)|^2}$ over central $50\%$ of the eigenstates in the nonintegrable ANNNI model. It is seen that for finite strings the average spectral function at low frequencies develop a uniform plateau as predicted by the random matrix limit of ETH.  However, with increasing string length, the  {magnitude of the plateau $\Delta(n,L)=\overline{|f_n(\omega\rightarrow 0)|^2}$ rapidly decreases,  which is indicative of forming a spectral gap characteristic of integrable systems~\cite{Pandey_2020}.  Further, the spectral weight of the string operators quickly approaches a sharply localized monochromatic distribution at frequencies increasing with the string size.  The average} spectral function is nothing but the Fourier transform of unequal time connected auto-correlation functions of the string operators \cite{srednicki99,lev21}:
\begin{equation}\label{eq:realtime}
	\left|f_n(E_{\alpha},\omega)\right|^2 = \frac{1}{4\pi}\int_{-\infty}^{\infty} e^{-i\omega t}\frac{\braket{\phi_{\alpha}|\{P_n(t),P_n(0)\}|\phi_{\alpha}}_c}{\braket{\phi_{\alpha}|P_n^2|\phi_{\alpha}}_c} dt
\end{equation}
where $\{.\}$ is the anticommutator and
\begin{eqnarray}
\nonumber\braket{\{P_n(t),P_n(0)\}}_c = \braket{\{P_n(t),P_n(0)\}}-2\braket{P_n(t)}\braket{P_n(0)}.
\end{eqnarray}
Thus, the spectral function encodes the real-time dynamics of local observables and sets the relaxation time scale of $P_n$ in eigenstates of the chaotic Hamiltonian.  {The narrowing of the spectral bandwidth with increasing string length seen in Fig.~\ref{spectralf} therefore suggests that the connected correlations of long string operators show prominent and rapid late-time oscillations rather than vanishing steadily due to loss of memory (see Appendix~\ref{Sec:Appendix_matele})}. This in turn indicates that the string observables indeed have a very long lifetime in chaotic states as compared to single-site observables.\\

{It is now evident that the projectors $P_n$ have two distinct properties, i.e., (i) with increasing string length they project states onto the completely polarized state $\ket{\rightarrow\rightarrow...}$ and (ii) in quenching experiments, they retain a strong memory of the initial state even after sufficiently long chaotic dynamics. Given this, it is natural to ask which of these features plays a dominant role in capturing the equilibrium QCPs following a quench. To address this question, we study the quantity,}
{\begin{equation}
O_n(\overline{E})=-\frac{1}{n}\log\braket{\phi_{\gamma}|P_n|\phi_{\gamma}},
\end{equation}
where $\ket{\phi_{\gamma}}$ is an eigenstate of the chaotic ANNNI Hamiltonian $H$ having energy $E$ which is closest to the mean energy $\overline{E}$ after a quench. This quantity $O_n(\overline{E})$ being closest to the typical value of $\mathcal{O}_n$ after quench, is not expected to contain any information of the initial state whatsoever. In Fig.~\ref{fig:testa}, we contrast the behavior of $O_n(\overline{E})$ with the time averaged probes $\mathcal{O}_n$ following different quenches across the equilibrium QCP. However, unlike the time averaged local probes, we do not observe any sharp signature in the quantity $O_n(\overline{E})$ detecting the equilibrium QCPs. This suggests that the memory of the initial state in a quench indeed plays an instrumental role in capturing signatures of equilibrium criticality in the string observables.}\\

 {Let us note that while we focused on particular string observables,  which are partial projectors into polarized strings, some signatures of the zero temperature QCP can be seen in other strings as well.  For example, (see also Ref.~\cite{asmi21}) we can define a product of spin operators
\begin{equation}
S_n=\frac{1}{L}\sum\limits_{i=1}^{L}\prod\limits_{i}^{i+n-1}\sigma_i^x
\end{equation}
and observe the time insensitive probe 
\begin{equation}
\mathcal{s}_n=-\frac{1}{n}\log\lim\limits_{T\rightarrow\infty}\frac{1}{T}\int_{0}^{T}dt\braket{\psi(t)|S_n|\psi(t)},
\end{equation}
following nonintegrable quenches.  In Fig.~\ref{fig:testb}, we contrast the late time behavior of such observables with that of the local projectors $\mathcal{O}_n$ after finite but sufficiently long time averaging. The observables $\mathcal{s}_n$ do indeed detect the equilibrium QCPs having a similar dependence on the quenched field as the probes $\mathcal{O}_n$.  However, unlike $\mathcal{O}_n$, the observables $\mathcal{s}_n$ do not develop nonanalyticities remaining smooth even with increasing string length.  }

\section{Summary and outlook.}
\label{sec:conclusion}

In summary,  we  studied local string observables as probes of quantum phase transitions after quenching a system into highly excited states. We particularly showed that one can accurately determine both the position of a quantum critical point and the associated universal critical exponents even long after the system equilibrates into effectively a high-temperature state. This analysis applies both to integrable and ergodic systems satisfying the ETH.  As we explain in this paper,  state to state fluctuations of local observables, which vanish
exponentially with the system size, do not prevent the system from retaining memory of the initial state.\\

 {We first analyzed} the long time averaged Loschmidt echo following a generic quench; its rate function develops sharp nonanalyticities,  which precisely detect QCPs of the equilibrium system. Along with discontinuous jump singularities in its derivative responses, the nonanalyticities of the time averaged LE can be explained by its topologically different pole structure for quenches across the equilibrium critical point in the integrable situation.  Extending the connection of the nonanalyticities with equilibrium criticality, we further proceed with a finite size scaling analysis of the time averaged Loschmidt echo and find universal critical exponents associated with the QCPs in integrable as well as non-integrable situations.  \\
 
  {Furthermore, the Loschmidt echo can also be interpreted  as an expectation value of a particular string operator given by a product of all the spins in the system.  Such an operator is obviously completely  nonlocal and is therefore hard to measure.  In order to  connect with experimentally relevant observables,  we then analyzed finite strings, i.e.,  operators involving finite products of spins.}  {Interestingly,  such string observables develop sharp signatures precisely at equilibrium QCPs of the quenched system even when the string size is very small.  Furthermore, using these local observables, we are able to extract critical scaling information and universal exponents associated with equilibrium QCPs even after a long ergodic time evolution.\\

The local string operators were also found to serve as probability distributions of domain formation in the late time state of a quenched system.  We analyzed such distributions revealing an underlying memory of equilibrium critical information encoded in the post quench domain statistics of generic quantum systems. Specifically, the mean longest domain length observed in the post-quench state, develops a sharp dip precisely near critical points of the equilibrium system.  Our results explain recent experimental findings in a trapped ion quantum simulator~\cite{Zhang2017}. They also agree with recent numerical studies~\cite{ettore20,halimeh21} observing sharp features in local single-site observables. \\

We also probed time-averaged versions of unequal time infinite temperature correlations of the local string observables following a quench. The time averaged auto-correlation and infinite temperature OTOC of the string operators were constructed such that they approach the long time averaged LE as the string size increases. Interestingly, similar sharp features near equilibrium QCPs emerge in the infinite temperature correlators in both integrable and chaotic spin chains. Furthermore, these infinite temperature correlators do not depend on any specific choice of initial states.\\

Finally, given the emergence of critical signatures following a chaotic quench, it seems evident that despite an ergodic evolution, the string observables retain strong memory of the initial ground state after long times. In fact, we showed that this persistent memory of the initial state plays a significant role in sharply detecting the equilibrium critical points in quenching experiments. We also showed that this memory are encoded in large state-to-state fluctuations near the mean energy after quench, which increase exponentially with the string length. These large fluctuations in the string expectations,  {together with a very small value of the low frequency plateau of the spectral function ensure very slow thermalization of sufficiently long string observables even when the thermodynamic limit is approached. In turn these long relaxation times allow the strings to both retain the long-time memory of the initial state and be very sensitive to the ground state properties of the quenched Hamiltonian despite the system being in a highly excited state after the quench.} These observations collectively lead to the possibility of a unified approach in which the string observables serve as a scalable probe to systematically access ground state information in generic quenching experiments.\\

A recent theoretical study \cite{diptarka21} has also reported the observation of Kibble-Zurek physics in the early time dynamics of OTOCs (see also Ref.~\cite{heyl18}), in quenched one dimensional conformal field theories (CFT). The precise connection between the late-time behavior of string observables with their early-time growth and OTOCs is yet to be understood. Another intriguing area of study might be to probe the efficacy of the string observables to retain critical information in systems undergoing topological phase transitions. For example, in short-ranged integrable systems, it has been reported \cite{Bhattacharyya15,sroy17} that the residual energy following a quench might become nonanalytic at critical points separating topologically inequivalent phases in equilibrium. The exact connection of the residual energy in such systems to the local string observables is yet to be studied.

\section*{Acknowledgments}
We acknowledge the use of QuSpin 0.3.6	for numerical computations. We thank Sourav Bhattacharjee, Michael Flynn, Stefan Kehrein, Anushya Chandran, Christopher Laumann, Anders Sandvik, Dries Sels and Diptarka Das for fruitful discussions and valuable comments. S.B. acknowledges financial support from a PMRF fellowship, MHRD, India, SPARC, MHRD India and Boston University for support and hospitality. A.P. acknowledges support from NSF under Grant DMR- 2103658, the AFOSR under Grants No. FA9550- 16-1-0334 and FA9550-21-1-0342  and SPARC program, MHRD, India.  A.D. acknowledges support from SPARC program, MHRD, India and SERB, DST, New Delhi, India.

\appendix

\section{Complete pole structure of the time averaged LE}
\label{Sec:Appendix_1}

To exactly derive the sharp nonanalyticities of the time averaged quantity $\mathcal{O}_L$ following critical quenches, we impose periodic boundary conditions on the Hamiltonian Eq.~\eqref{eq:ham} in the integrable limit $(J_2=0)$. In the thermodynamic limit, one can then rewrite the Hamiltonian in decoupled single-particle momentum sectors as,
\begin{equation}
	H=\bigotimes_{k}H_k,
\end{equation}
where,
\begin{eqnarray}\label{eq_a:hamil}
	\nonumber H_k&=&\vec{h}(k).\vec{\sigma},\\
	\nonumber h_x(k)&=&-2\sin{k},\\
	\nonumber h_y(k)&=&0,\\
	h_z(k)&=&-2h-2\cos{k}.
\end{eqnarray}

The eigenstates of the above Hamiltonian at any parameter value can be expressed as Bloch states,
\begin{eqnarray}\label{eq:hk_a}
	\nonumber\ket{+}_k=\begin{pmatrix}
		-\cos{\theta/2} \\
		\sin{\theta/2}
	\end{pmatrix},~\ket{-}_k=\begin{pmatrix}
		\sin{\theta/2} \\
		\cos{\theta/2}
	\end{pmatrix},\\
	\text{such that, } \cos{\theta}=\frac{h_z(k)}{\sqrt{h_x(k)^2+h_z(k)^2}},
\end{eqnarray}
with eigenvalues $\pm\left|\vec{h(k)}\right|$ respectively. We start from the ground state of $H_i(h=h_i)$ and quench the system at time $t=0$ to a final set of parameters $H(h)$ and probe the logarithmic time averaged LE, $\mathcal{O}_L$ following the quench.
Using the translational invariance of the problem, one arrives at the momentum resolved equivalent of Eq.~\eqref{eq:ol},
\begin{equation}
	\mathcal{O}_L=-\frac{1}{\pi}\int_0^\pi dk\log{\left[f_{+}(k)^2+f_{-}(k)^2\right]},
\end{equation}
where,
\begin{eqnarray}
	\nonumber f_{+}(k)&=&\left|\braket{\psi_k(0)|+_k}\right|^2, \\
	f_{-}(k)&=&\left|\braket{\psi_k(0)|-_k}\right|^2,
\end{eqnarray}
which can be further simplified to,
\begin{equation}\label{eq:ol_2_a}
	\mathcal{O}_L=-\frac{1}{\pi}\int_0^\pi dk\log{\left[1-\frac{1}{2}\sin^2{(\theta-\theta_i)}\right]}.
\end{equation}
We solve the integral in Eq.~\eqref{eq:ol_2_a} by an analytic continuation into the complex plane, i.e., we substitute $z=e^{ik}$, thereby changing the measure of integration to $dk=-iz^{-1}dz$. Using the definition of the Bloch angles in Eq.~\eqref{eq:hk_a} we obtain an integration over a closed contour $C$ which is a unimodular circle about the origin $z=0$,
\begin{equation}\label{eq:exactle_a}
	\mathcal{O}_L=\frac{1}{2\pi i}\oint_{C}F(z)dz,
\end{equation}
where,
\begin{equation}
	F=\frac{1}{z}\log \left[\frac{h^2 \left(z^2-1\right)^2}{8 z (h+z) (h z+1)}+1\right].
\end{equation}
 We define the measure of a nonanalyticity of $\mathcal{O}_L$ at the critical point $h_c=1$ as,
\begin{equation}\label{eq_a:delta}
	\delta=\left|\partial_{h_c^+}\mathcal{O}_L-\partial_{h_c^-}\mathcal{O}_L\right|,
\end{equation}
such that any nonzero $\delta$ signals a breakdown of continuity of $\partial_h\mathcal{O}_L$ and hence, analyticity of $\mathcal{O}_L$. To analytically probe  the jump singularities in the derivative of $\mathcal{O}_L$ at the equilibrium QCP, we must therefore focus on the first derivative of the function $F$ with respect to the quenching field $h$. The derivative $\partial_hF$ is found to be a meromorphic function,
\begin{widetext}
\begin{equation}
	\partial_hF=\frac{h \left(z^2-1\right)^2 \left(h z^2+h+2 z\right)}{z (h+z) (h z+1) \left(h^2 z^4+6 h^2 z^2+h^2+8 h z^3+8 h z+8 z^2\right)},
\end{equation}
\end{widetext}
with simple poles on the real axis at,
\pagebreak
\begin{widetext}
\begin{eqnarray}
\nonumber z_1=-\frac{1}{h},~~z_2=-h,~~z_3=0,\\
\nonumber z_4=\frac{\sqrt{2-h^2}-\sqrt{2} h \sqrt{\frac{3-2 \sqrt{2-h^2}}{h^2}-1}-2}{h},\\
\nonumber z_5=\frac{\sqrt{2-h^2}+\sqrt{2} h \sqrt{\frac{3-2 \sqrt{2-1 h^2}}{h^2}-1}-2}{h},\\
\nonumber z_6=\frac{-\sqrt{2-h^2}-\sqrt{2} h \sqrt{\frac{2 \sqrt{2-h^2}+3}{h^2}-1}-2}{h},\\
z_7=\frac{-\sqrt{2-h^2}+\sqrt{2} h \sqrt{\frac{2 \sqrt{2-h^2}+3}{h^2}-1}-2}{h}.
\end{eqnarray}
\end{widetext}
Fig.~\ref{figA:contourcover} shows all the poles of $\partial_hF$ when the transverse field $h$ lies in both the paramagnetic and ferromagnetic phase. We note that the pole structure changes abruptly across the quantum critical point $h=h_c=1$ resulting in a finite nonzero value of $\delta$ and consequently a breakdown of analyticity in $\mathcal{O}_L(h_c)$. In particular, as seen from Fig.~\ref{fig:2cover} and Fig.~\ref{figA:contourcover}, the poles represented by $z_1$ and $z_2$ exchange their positions with respect to the contour of integration $C$, as one moves across the critical point. Similarly, calculating the residues of the relevant poles contributing to the discontinues jump in the derivative $\partial_h\mathcal{O}_L$ around the critical point $h=1+\epsilon$ ($\epsilon\ll1$), we obtain,
\begin{widetext}
	\begin{eqnarray}\label{eq_a:resn}
		\nonumber		{\rm Residue}(z_1)= -1+\epsilon;~~~~
		\nonumber		{\rm Residue}(z_2)= 1-\epsilon;\\
		\nonumber		{\rm Residue}(z_3)= 1-\epsilon;\\
				{\rm Residue}(z_5)=
		\begin{cases}
			\sqrt{2}, ~\epsilon>0,\\
			-\sqrt{2}, ~\epsilon<0;\\
		\end{cases}
	{\rm Residue}(z_7)= -\sqrt{2}.
	\end{eqnarray}
\end{widetext}
To exactly calculate the discontinuous jumps in derivatives, we apply Cauchy's residue theorem,
\begin{equation}\label{eq_a:leresidue1}
	\delta=\left|\sum {\rm Residue}(\partial_{h_c^+}F,z_{in}^{para})-{\rm Residue}(\partial_{h_c^-}F,z_{in}^{ferro})\right|,
\end{equation}
where $z_{in}^{para}$ and $z_{in}^{ferro}$ are the poles lying inside the contour $C$ in the paramagnetic and the ferromagnetic phase respectively and the summation is over all the poles. Taking into consideration the residue of the poles in Eq.~\eqref{eq_a:resn}, we obtain in the limit $\epsilon\rightarrow 0$,
\begin{equation}
	\delta=2\left(\sqrt{2}-1\right)\approx 0.82,
\end{equation}
which agrees with the discontinuous jump found numerically in Fig.~\ref{fig:1}. This explains the development of a jump discontinuity (see Fig.~\ref{fig:1}) in the derivative of $\mathcal{O}_L$ at the critical points. When quenched on the critical point, the poles $z_1$ and $z_2$ lie exactly on the contour, thus making the observable $\mathcal{O}_L$ nonanalytic.\\
\begin{figure*}
	\subfigure[]{
		\includegraphics[width=8.0cm,height=7.2cm]{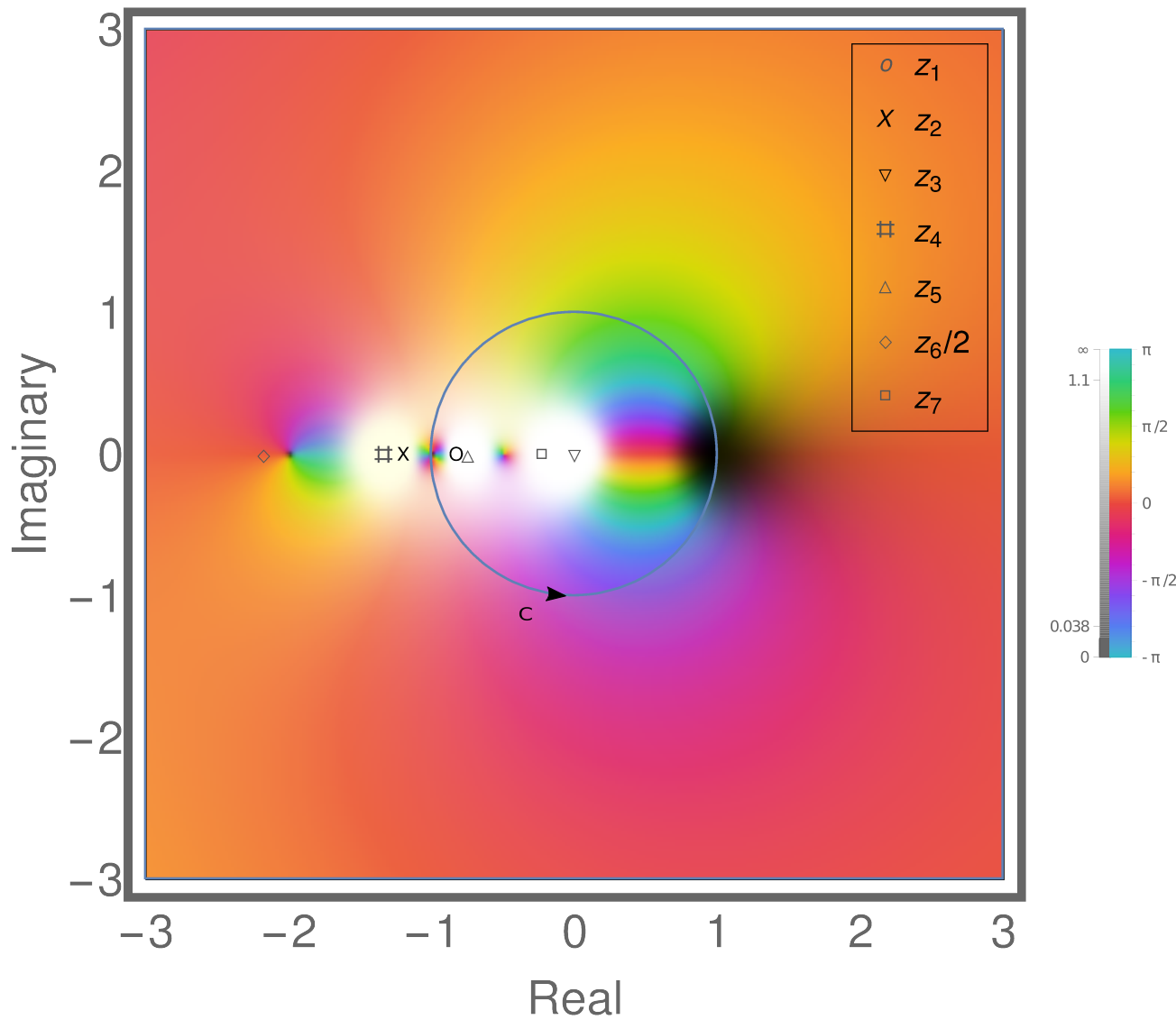}
		\label{figA:contour1}}
	\subfigure[]{
		\includegraphics[width=8cm,height=7.2cm]{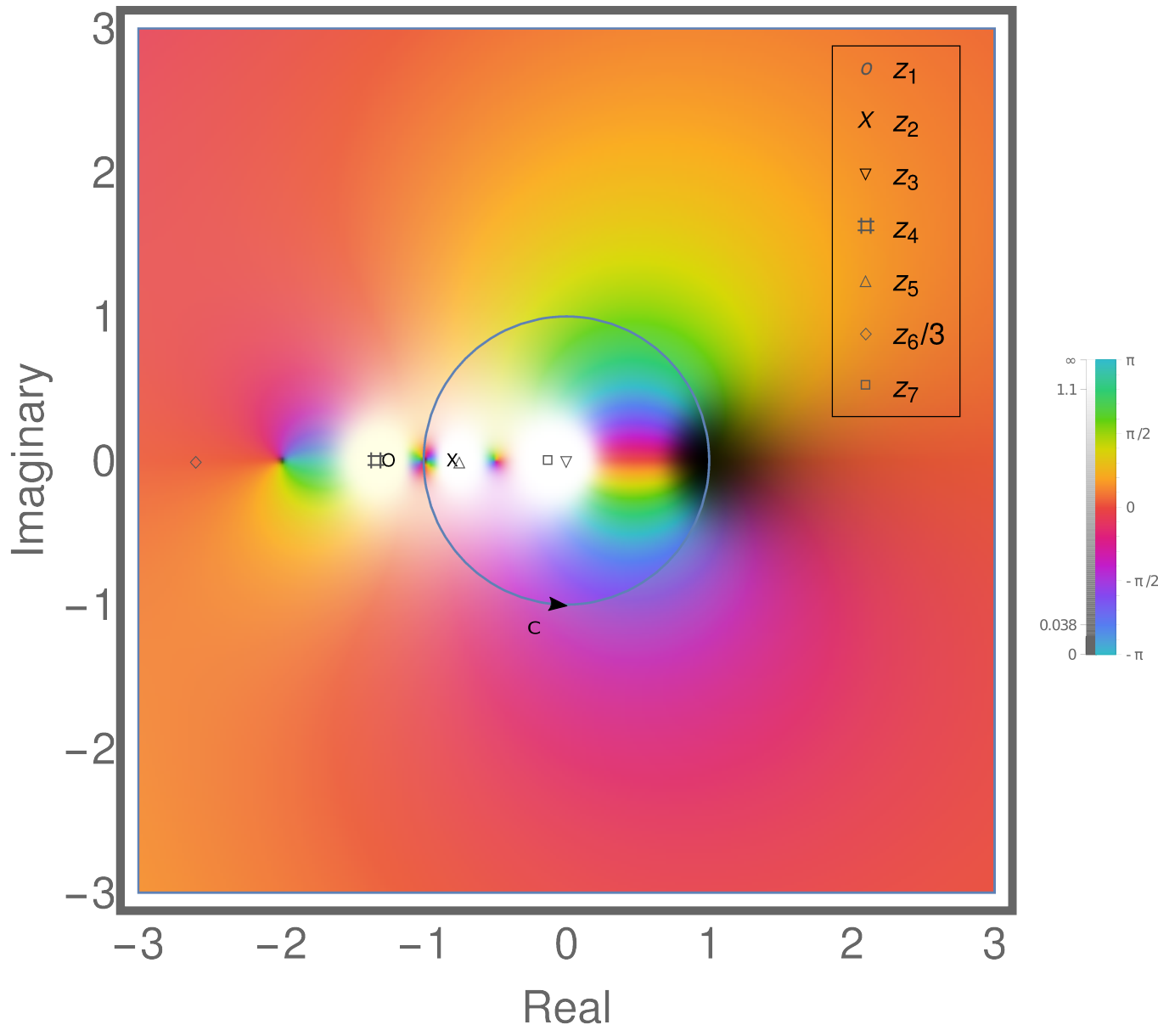}
		\label{figA:contour2}}
	\caption{The complete pole structure of the derivative $\partial_hF$ when analytically continued into the complex plane following a quench to the (a) paramagnetic phase ($h=1.2$) and the (b) ferromagnetic phase ($h=0.8$), for $J_2=0$. The color scheme represents the complex phase whereas white and black colors represent poles and zeros of the function respectively.  As is evident from both (a) and (b), the poles $z_1$ and $z_2$ exchange their positions with respect to the contour $C$ as one moves across the critical point. At the critical point the poles are seen to lie on the contour C, which in turn results in the nonanalyticity of $\mathcal{O}_L$ at the critical point.}
	\label{figA:contourcover}
\end{figure*}

\section{Critical behavior of local projectors in the thermodynamic limit}
\label{Sec:A3}

For shorter strings of length $n<L$, the long time averages $\overline{\mathcal{L}}_n$ can be evaluated exactly by solving the corresponding fermionic problem for integrable quenches in the thermodynamic limit. Through a Jordan-Wigner transformation, the Ising Hamiltonian in Eq.~\eqref{eq:ham} can be rewritten in terms of noninteracting spinless fermions,
\begin{eqnarray}
	\sigma^z_j&=&2c^{\dagger}_{j}c_j-1,\nonumber\\
	\sigma^x_i&=&\left( c_i+c_i^{\dagger}\right)\prod\limits_{j<i}\left(1-2c_j^{\dagger} c_j\right),
\end{eqnarray}
where the operators $c_i$ satisfy the standard fermionic algebra. Depending of the fermion parity specified by the operator,
\begin{eqnarray}
	P_{\pm}&=&\frac{1}{2}\left(\mathbb{I}\pm\Sigma\right), ~\text{where,}\nonumber\\
	\Sigma&=&\prod\limits_{j}\left(2c_j^{\dagger}c_j-1\right)=\prod\limits_{j}\sigma^z_j,
\end{eqnarray}
it can be shown that the Hamiltonian can be decoupled into two independent parity sectors,
\begin{equation}
	H=P_{+}HP_{+}+P_{-}HP_{-}.
\end{equation}
To simplify calculations, we choose the initial state $\ket{\psi(0)}$ to be the ground state at $h=0$ in the even parity sector. Since, the Hamiltonian $H$ respects parity, the time-evolved state of the system also resides in the same parity sector. Therefore, the expectation value of all strings in $P_n$ that do not conserve parity vanish identically. It is for this reason that $P_1$ has a trivial time independent expectation throughout the evolution and we do not consider it in this section. We now express the local projectors in terms of transformed fermionic operators,
\begin{equation}
	A_i=c_i^{\dagger}+c_i~\text{and}~B_i=c_i^{\dagger}-c_i.
\end{equation}  
In terms of the $A$ and $B$ fermions, the short projectors assume the form,
\begin{equation}\label{eq:Amajorana}
	\begin{split}
		P_2=\frac{1}{2^2L}\sum\limits_{i}1+B_iA_{i+1},\\
		P_3=\frac{1}{2^3L}\sum\limits_{i}1+2B_iA_{i+1}+B_iA_{i+1}B_{i+1}A_{i+2}.
	\end{split}
\end{equation}
Due to the noninteracting nature of the Hamiltonian for $J_2=0$ [see Eq.\eqref{eq:ham}], while evaluating the late time averages in $\overline{\mathcal{L}}_n$, we use Wick's contraction to decompose longer correlations to algebraic functions of two point correlations. Using this prescription, it can be shown that in noninteracting systems, for expectation values of strings containing generic fermionic operators $f_{i}$ one obtains,
\begin{equation}\label{eq_a:det}
	\left<\prod_{i=1}^lf_i\right>=\left|\rm{Pf}(M)\right|=\sqrt{\det{M}},
\end{equation}
where $M$ is an anti-symmetric $l\times l$ matrix of two-point correlations such that,
\begin{equation}\label{eq:wick_a}
	M_{\mu\nu}=\braket{f_{\mu}f_{\nu}};~1<\mu<\nu<l.
\end{equation}
We therefore evaluate the two point correlations comprising of the $A$ and $B$ fermions in the asymptotic steady state, which will in turn allow us to evaluate the finite projectors. Utilizing the translational invariance of the system and the consequent decoupling into conserved quasi-momenta modes $k$, it can be shown that the two point correlators in the steady state assume the form,
\begin{eqnarray}\label{eq:corr_a}
	\nonumber\braket{A_mA_{m+j}}=\braket{B_{m+j}B_m}&=&\delta_{j0},\\
	\nonumber \braket{A_mB_{m+l}}=\\
	\frac{1}{2\pi}\int_{-\pi}^{\pi}dke^{-ilk}e^{-i\theta_k}\frac{1-h\cos{k}}{\sqrt{1+h^2-2h\cos{k}}},
\end{eqnarray}
where $\theta$ is the Bloch angle of the quenched Hamiltonian as also mentioned in Eq.~\eqref{eq:hk_a}. Also, to simplify calculations, we have repositioned the momentum modes like $k\rightarrow\pi-k$ with respect to that in Eq.~\eqref{eq_a:hamil}, such that the exact gapless critical mode for the QCP $h=1$ now lies at $k=0$. To analytically calculate the expectation of $P_n$, we recast the integrals in Eq.~\eqref{eq:corr_a} for quenches near the critical point $h=1+\delta$,
\begin{equation}\label{eq:integral_a}
	\braket{A_mB_{m-1}}=\frac{1}{\pi}\int_{0}^{\pi}C_1(k,\delta)dk,
\end{equation}
where,
\begin{equation}
	C_1(k,\delta)=\frac{((\delta +1) \cos (k)-1)^2}{\delta ^2+2 \delta -2 (\delta +1) \cos (k)+2}.
\end{equation}
Since, the integrand $C_1(k,0)$ has a removable singularity at the point $k=0$, a power series expansion of $C_1$ about $\delta=0$ does not converge everywhere within the interval of integration. In the first approach, to understand the nature of the function $\mathcal{O}_2$, we evaluate the integral in Eq.~\eqref{eq:integral_a} exactly by an analytic continuation into the complex plane. Specifically, we substitute $z=e^{ik}$ to obtain an integration over a closed contour $C$ which is a unimodular circle about the origin $z=0$,
\begin{equation}\label{eq_a:contour}
	\braket{A_mB_{m-1}}=\frac{1}{2\pi}\oint_C\frac{i (\delta +z (\delta  z+z-2)+1)^2}{4 z^2 (-\delta +z-1) (\delta  z+z-1)}dz.
\end{equation}
Note that the integrand in Eq.~\eqref{eq_a:contour} is meromorphic with poles at $z=0$, $z=1+\delta$ and at $z=(1+\delta)^{-1}$. Except the pole at $z=0$, only one of the other two lie inside the contour of integration depending on the sign of $\delta$. Therefore, by summing over nonzero residues from the poles, we evaluate the correlations and hence the quantity $\mathcal{O}_2$, in the different phases separately as,\\

\begin{widetext}
	\begin{eqnarray}
		\mathcal{O}_2=
		\begin{cases}
			-\frac{1}{2} \log \left[\frac{1}{4} \{1+\frac{1}{2} \left(-\delta ^2-2 \delta +1\right)\}\right]~~\text{for}~\delta<0,\\
			-\frac{1}{2} \log\left(\frac{3}{8}\right)~~\text{for}~\delta>0.
		\end{cases}
	\end{eqnarray}
\end{widetext}
\begin{figure*}
	\subfigure[]{
		\includegraphics[width=7.5cm,height=5.5cm]{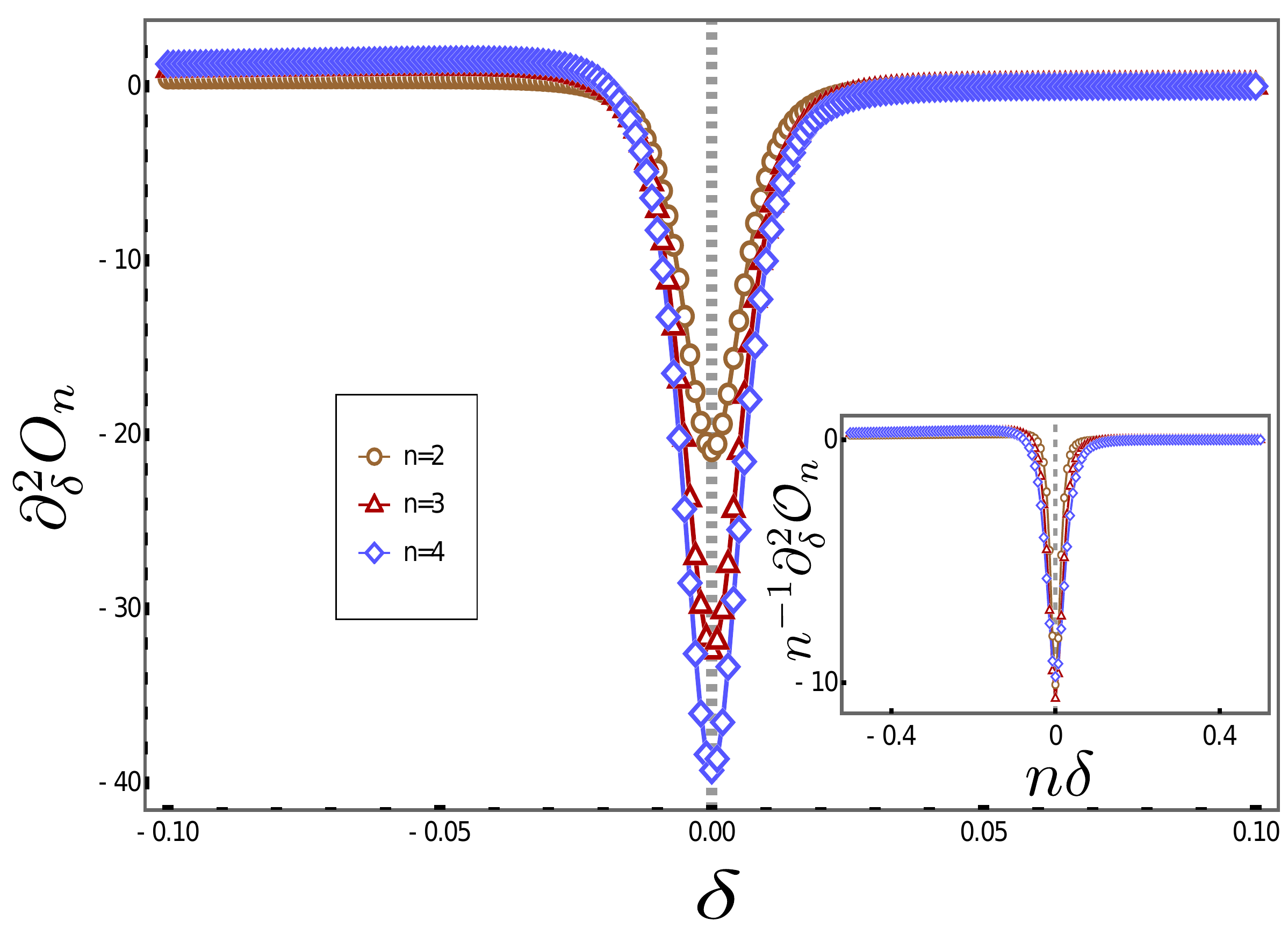}
		\label{fig:der2a}}
	\hspace{1cm}
	\subfigure[]{
		\includegraphics[width=7.65cm,height=5.5cm]{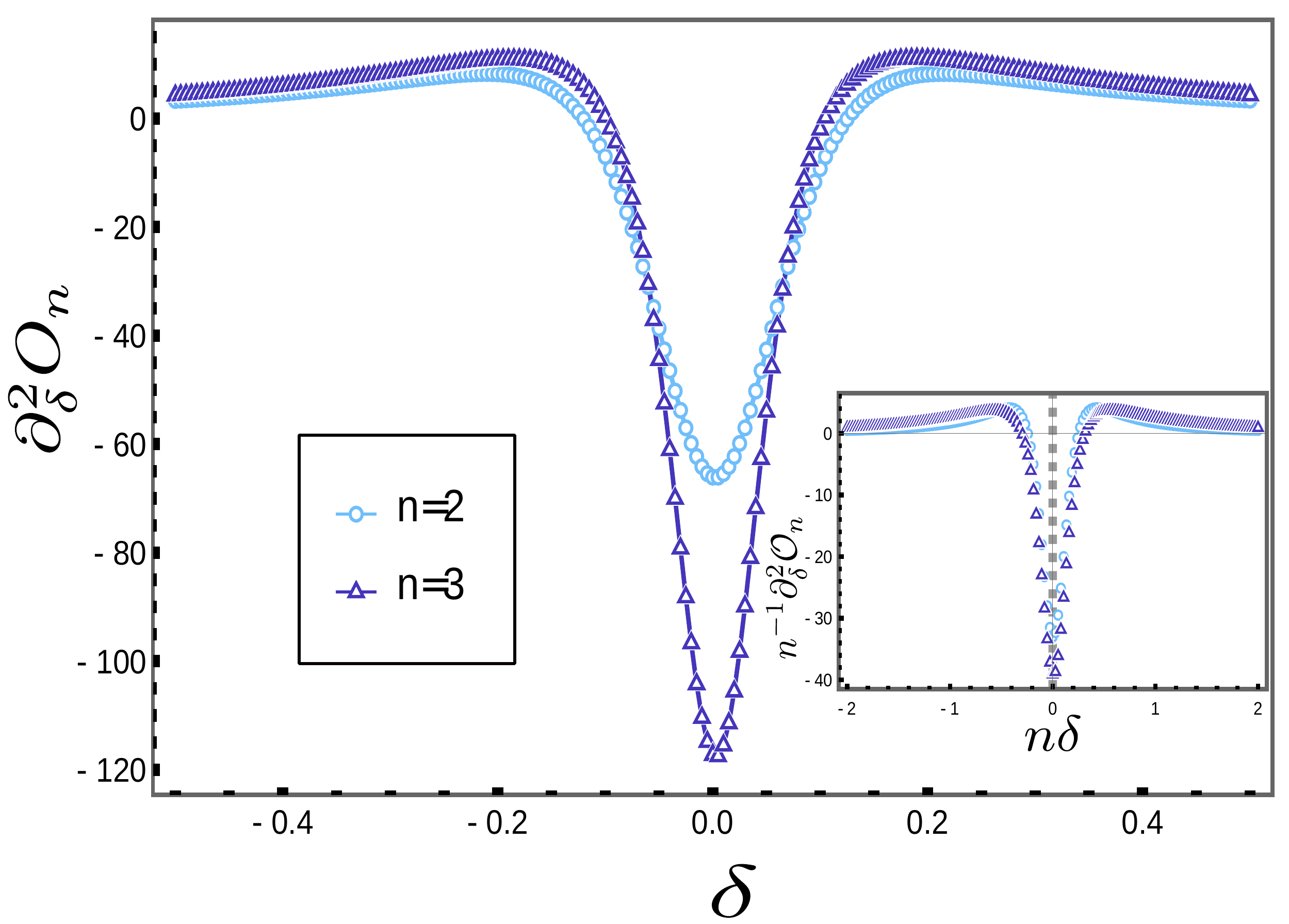}
		\label{fig:der2b}}
	
	\caption{(a) The divergence of the second derivative of the finite observables $\mathcal{O}_2,\mathcal{O}_3$ and $\mathcal{O}_4$ calculated numerically, in a thermodynamically large integrable Ising chain near the critical point $\delta=0$, following a quench starting from the initial ground state at $h_i=0,J_2=0$ to $h_f=1+\delta,J_2=0$. The cutoff scale $\epsilon$ has been chosen to be $0.01$ [see Eq.~\eqref{eq_b:corr} and the discussion following it]. (Inset) The rescaled derivatives $n^{-1}\partial_{\delta}^2\mathcal{O}_n$ collapse into a single curve for $\epsilon\ll|\delta|$ near the critical point. (b) The bare and (Inset) rescaled derivatives $n^{-1}\partial_{\delta}^2\mathcal{O}_n$ of the probes  $\mathcal{O}_2$ and $\mathcal{O}_3$ constructed from the series expansions truncated up to quadratic order, such as in Eq.~\eqref{eq_a:o2regular} with $\epsilon=0.01$.}
\end{figure*}
It is then straight forward to estimate the jump discontinuity in the first derivative of $\mathcal{O}_2$ exactly at $\delta=0$,
\begin{equation}\label{eq:bdelta2}
	\Delta_2=|\partial_{\delta^+}\mathcal{O}_2-\partial_{\delta^-}\mathcal{O}_2|= 0.33.
\end{equation}\\

Similar to the correlation function $\braket{A_mB_{m-1}}$, we evaluate the correlations contained in $P_3$ to obtain,
\begin{eqnarray}
	\mathcal{O}_3=
	\begin{cases}
		-\frac{1}{3}\log\left[\frac{1}{4} (9-7 \delta  (\delta +2))\right]~~\text{for}~\delta<0,\\
		-\frac{1}{3}\log\left[9/4\right]~~\text{for}~\delta>0.
	\end{cases}
\end{eqnarray}
This gives the discontinuous jump in the first derivative of $\mathcal{O}_3$ at the critical point $\delta=0$ as,
\begin{equation}\label{eq:bdelta3}
	\Delta_3=|\partial_{\delta^+}\mathcal{O}_3-\partial_{\delta^-}\mathcal{O}_3|= 0.51.
\end{equation}
 By performing higher order Wick's contractions [see Eq.~\eqref{eq_a:det}], it is then  possible to exactly calculate the late time expectation of $\mathcal{P}_n$ for arbitrary string lengths.\\
 
 Another approach to probe the critical nature of the observables $\mathcal{O}_n$ is to regularize the integrals like in Eq.~\eqref{eq:integral_a} near the critical point. To numerically evaluate the integrals, we introduce an infrared cutoff $\epsilon$,
\begin{equation}\label{eq_b:corr}
	\braket{A_mB_{m-1}}\rightarrow\frac{1}{\pi}\int_{\epsilon}^{\pi}C_1(k,\delta)dk,
\end{equation}
such that $\epsilon\ll|\delta|$ and $\epsilon\ll n^{-1}$. The condition $\epsilon\ll|\delta|$ ensures that the length $\epsilon^{-1}$ is much larger than the correlation length $\delta^{-1}$ near the critical point. In passing we note that a cutoff scale $\epsilon$ near the exact critical point naturally arises when dealing with finite size systems as $\epsilon\sim L^{-1}$. The condition $\epsilon\ll n^{-1}$ in such a situation ensures that we are in a regime where the observables scale with the shorter length $n$ rather than with  $\epsilon^{-1}\sim L$. Following this regularization, the function $C_1(k,\delta)$ becomes well behaved everywhere within the interval of integration $\left[\epsilon,\pi\right]$. To understand how the derivatives diverge in lowest order, we then expand the function $C_1(k,\delta)$ as a power series near $\delta=0$ up to quadratic order and perform the integral in Eq.~\eqref{eq_b:corr} to finally obtain,
\begin{widetext}
	\begin{eqnarray}\label{eq_a:o2regular}
		\mathcal{O}_2\approx-\frac{1}{2} \log \left[\frac{1}{8} \left\{\delta ^2 \left(\frac{\epsilon -\pi }{2}+\cot \left(\frac{\epsilon }{2}\right)\right)+\frac{\delta  (\epsilon +\sin (\epsilon )-\pi )}{\pi }+3\right\}\right].
	\end{eqnarray}
\end{widetext}
In Fig.~\ref{fig:3b} and Fig.~\ref{fig:der2a}, we numerically show the discontinuous jumps of the first derivative and divergences in the second derivative of $\mathcal{O}_2$, $\mathcal{O}_3$ and $\mathcal{O}_4$ near the critical point with a lower cutoff $\epsilon=0.01$. The rescaled derivatives $n^{-1}\partial_{\delta}^2\mathcal{O}_n$ when plotted against $n\delta$, collapse near the critical point for different $n$, where $0<\epsilon\ll|\delta|$. The exact jump discontinuities in the first derivatives $\Delta_2$ and $\Delta_3$ found in this appendix (Eq.~\eqref{eq:bdelta2} and Eq.~\eqref{eq:bdelta3} respectively) agree very well with the jump discontinuities shown in Fig.~\ref{fig:3b}. Also, exactly at the critical point $\delta=0$, the second derivatives diverge as,
\begin{eqnarray}
	\partial_{\delta}^2\mathcal{O}_2\left.\right|_{\delta=0}\sim-\frac{2}{3 \epsilon },\\
	\partial_{\delta}^2\mathcal{O}_3\left.\right|_{\delta=0}\sim-\frac{32}{27 \epsilon },
\end{eqnarray}
when the correlators are evaluated up to a quadratic order in $\delta$ near $\delta=0$, thus reflecting the discontinuity of the first derivatives at the critical point in the thermodynamic limit.

\section{Emergence of finite-size scaling}
\label{Sec:A2}
In Appendix.~\ref{Sec:Appendix_1}, we discussed how in the thermodynamic limit, the time averaged Loschmidt echo $\mathcal{O}_L$ following a sudden integrable quench reduces to the expression,
\begin{equation}\label{eq:scaling_a}
		\mathcal{O}_L=-\frac{1}{\pi}\int_0^\pi dk\log{\left[1-\frac{1}{2}\sin^2{(\theta-\theta_i)}\right]},
\end{equation}
where,
\begin{equation}
	\cos{\theta}=\frac{h_z(k)}{\sqrt{h_x(k)^2+h_z(k)^2}},
\end{equation}
$\theta_i$ and $\theta$ being the Bloch angles representing the single particle Hamiltonian $H_k=\vec{h}(k).\vec{\sigma}$ (see Eq.~\eqref{eq_a:hamil}), before and after the quench, respectively. At the critical point ($h=h_c$), the Bloch angle $\theta$ is not well-defined. Rather, the angles $\theta_{k=\pi,0}(h=\pm 1)$ assume indeterminate $0/0$ forms in the thermodynamic limit. Therefore, the integrand Eq.~\eqref{eq:scaling_a} is ill defined at the isolated critical points.\\
\begin{figure*}
	\subfigure[]{
		\includegraphics[width=6.3cm,height=4.8cm]{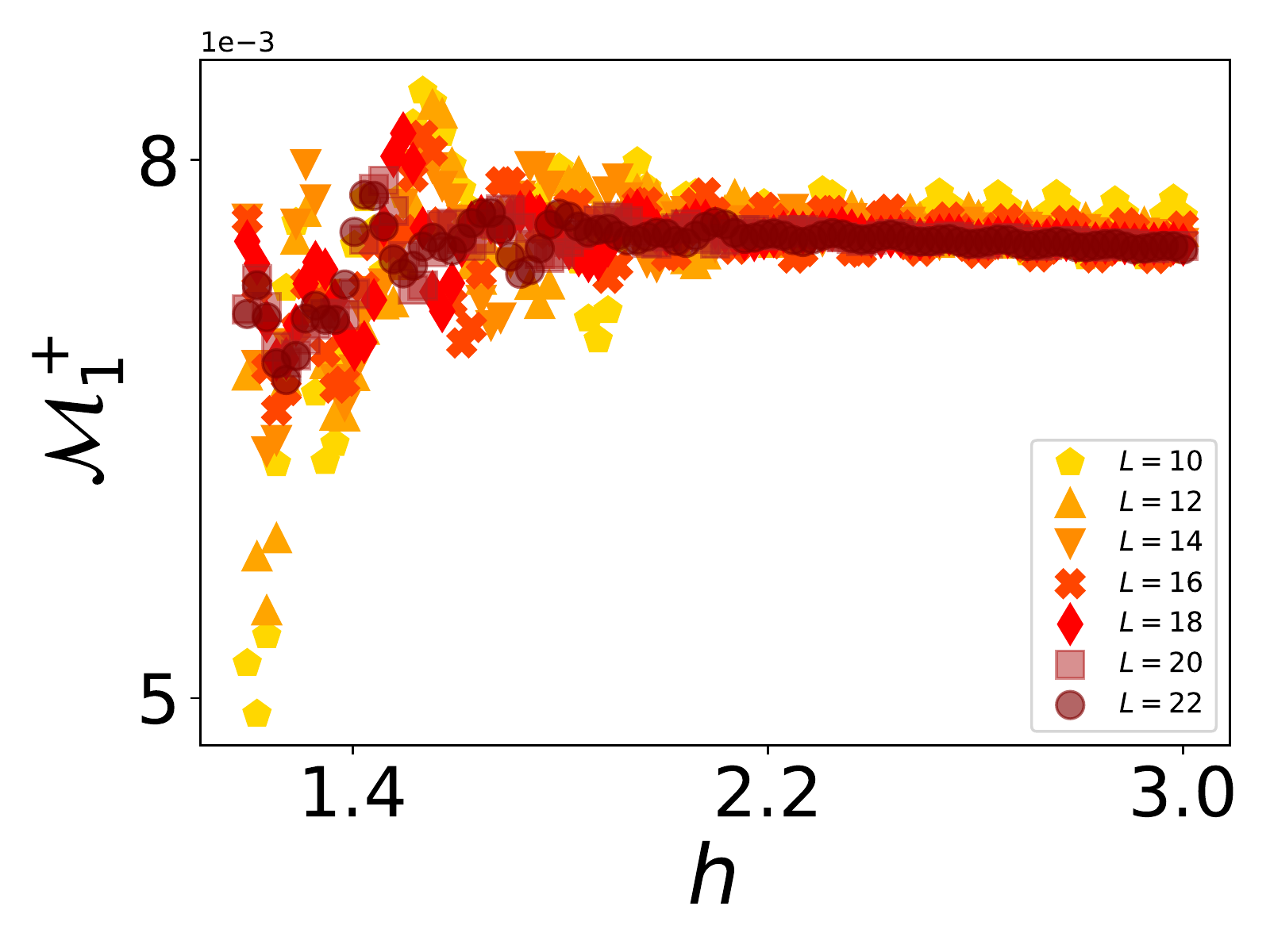}
		\label{fig:d_a}}
	\hspace{1cm}
	\subfigure[]{
		\includegraphics[width=6.5cm,height=4.5cm]{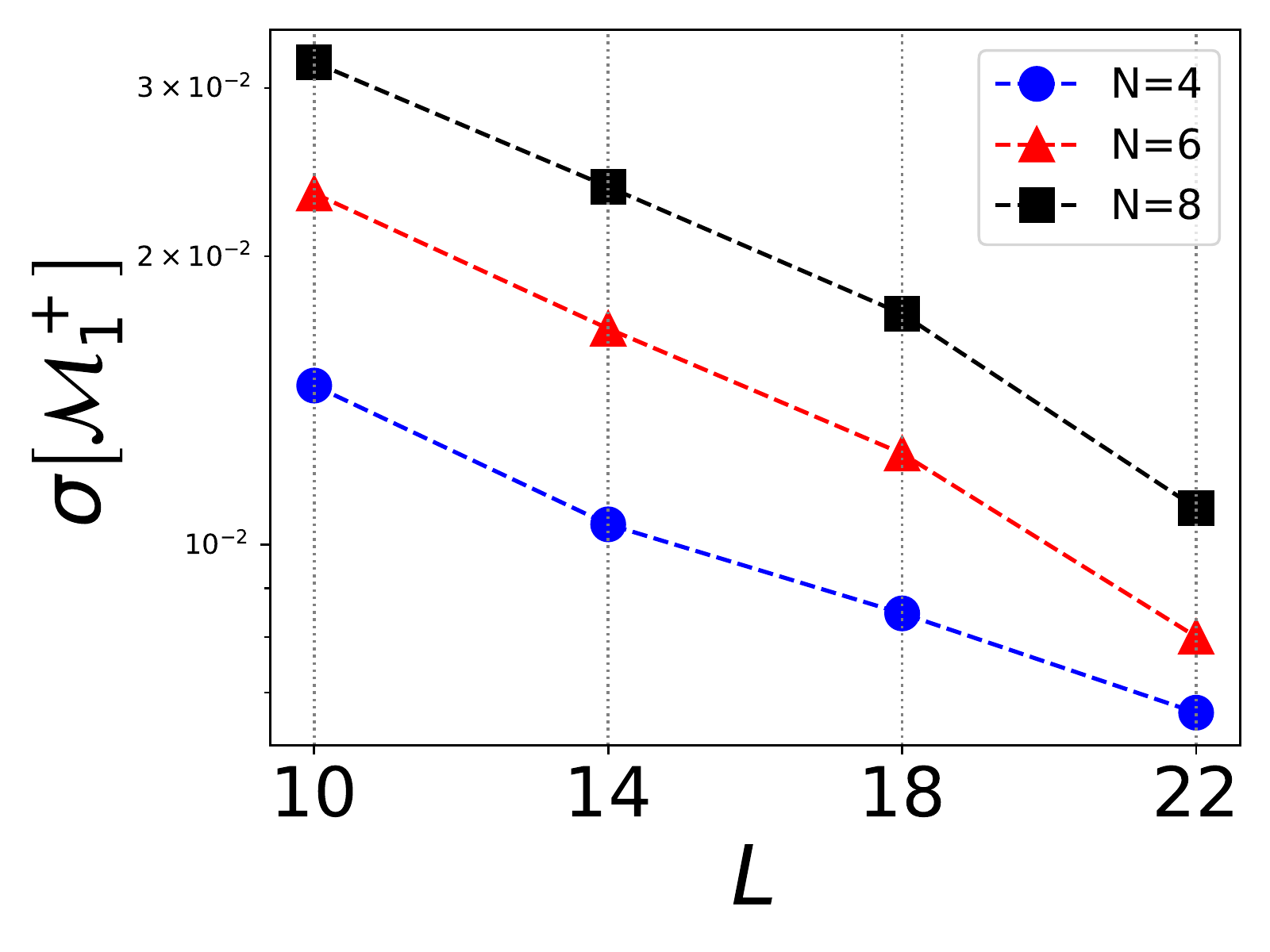}
		\label{fig:d_b}}
	
	\caption{(a) The distribution $\mathcal{M}_1^+$ for $N=10$ uncorrelated measurement sequences against the quenched field $h$, for quenches into the paramagnetic phase in a chaotic ANNNI model. The distribution is seen to approach a slowly varying function of the quenched field $h$ with increasing system size $L$ (a darker color represents a higher system size). (b) The relative deviation of $\mathcal{M}_1^+$ from its mean decreases rapidly with increasing system size irrespective of the number of measurements $N$. In both the panels the quench was started from a completely polarized initial state $\ket{\rightarrow\rightarrow...}$.}
\end{figure*}
However, as argued in Sec.~\ref{sec: critical_scaling}, the observable $\mathcal{O}_L$ for finite systems, exhibit universal critical scaling near the equilibrium critical point. To see the emergence of the scaling theory, we proceed to analytically evaluate the integral in Eq.~\eqref{eq:scaling_a} following quenches in finite size systems. As also seen in Sec.~\ref{sec:lo_analytic}, the critical singularity drifts towards the momentum $k_c=\pi$ as one approaches the critical point $h_c=1$. Expanding the integral in Eq.~\eqref{eq:scaling_a} near the critical point we obtain,
\begin{widetext}
\begin{equation}\label{eqa:expansion}
	\mathcal{O}_L\approx-\frac{1}{\pi}\int_{0}^{\pi}dk\log\left[1-\frac{1}{2}\sin ^2\left(\frac{k}{2}\right)-\frac{\delta}{2}\sin ^2\left(\frac{k}{2}\right)+\frac{1}{8} \delta ^2 \tan ^2\left(\frac{k}{2}\right)\right],
\end{equation}
\end{widetext}
where $h=1+\delta$.
From Eq.~\eqref{eqa:expansion} it is evident that exactly at the mode $k=\pi$, the term quadratic in $\delta$ in $\mathcal{O}_L$ diverges. This divergence in the second derivative reflects the emerging nonanalyticity of $\mathcal{O}_L$ near the equilibrium QCP. We therefore estimate the singular part of $\mathcal{O}_L$ by focusing on the contribution to the integral from the critical mode $k=\pi$.\\

However, since we are dealing with systems having a finite size $L$, the integral in Eq.~\eqref{eqa:expansion} has to be approximated by a finite Riemann sum in discrete momenta separated by $dk\sim\pi/L$. With increasing system size, $k$ assumes continuous values and modes near the critical mode $k=\pi-\epsilon$, approach $k=\pi$ as $\epsilon\sim L^{-1}$. 
Therefore, to study the singular part of the integral in Eq.~\eqref{eqa:expansion}, we expand it near $k_c$ by substituting $k=k_c-\epsilon$. Further using $\epsilon\sim L^{-1}$, we obtain in leading order for the singular part,
\begin{equation}\label{eqa:critical}
	\mathcal{O}_L^c\approx-\frac{1}{L}\log\left[\frac{1}{2}+\frac{x^2}{2}+{\rm O}\left(\frac{x}{L}\right)\right],
\end{equation}
where $x=L\delta$ is dimensionless. For $x=L \delta\ll 1$, $L$ being the smallest length, we expect the emergence of finite size critical scaling in the singular part  $\mathcal{O}_L^c$ with system size $L$. In this scaling regime we can hence approximate,
\begin{equation}
	\mathcal{O}_L^c\approx-\frac{1}{L}\left(x^2+\frac{x^4}{2}+{\rm O}\left(x^6\right)\right),
\end{equation}
such that,
\begin{equation}
	\mathcal{O}_L^c\approx L^{-1}\Phi(L\delta),
\end{equation}
where $\Phi$ is a scale invariant universal scaling function. This validates our scaling ansatz in Eq.~\eqref{eq:scaling_1} for $\nu=\alpha=1$. Also, from Eq.~\eqref{eqa:critical} it is clear that for sufficiently large but finite system sizes, the singular part approaches zero as the QCP is approached with $\delta$. Thus, exactly at the critical point in finite systems,  $\mathcal{O}_L$ can be estimated by the analytic nonsingular part of the integral in Eq.~\eqref{eq:scaling_a}.
Hence, to extract the critical scaling in the singular part of the time averaged Loschmidt echo in Sec.~\ref{sec: local_lo}, we show a scaling collapse of $L|\mathcal{O}_L(h_c+\delta)-\mathcal{O}_L(h_c)|$ against $L\delta$ near the QCP.

\section{Accuracy of QCP detection}
\label{Sec:sr_int}

 {In this appendix we show that it is indeed possible to extract equilibrium QCPs very accurately, using the dynamics of even comparatively short string observables in a scalable approach. Specifically, we compare our results with DMRG (density matrix renormalization group) calculations that are available in literature about the very precise location of QCPs in the ANNNI model. For this purpose, we choose the integrability breaking interaction to be antiferromagnetic ($J_2<0$) for which the location of the QCP is already well known \cite{matteo06} through various numerical and experimental studies in the past.}\\

 {As we previously demonstrated, time averaged local strings develop sharp signatures near the equilibrium QCPs. We therefore resort to their prominent second derivative response near the QCP, which is expected to capture any emerging non-analyticity in $\mathcal{O}_n$ with a divergence,
\begin{equation}\label{eq:der_min_a}
		D_n = -\partial_h^2\log\bar{\mathcal{L}}_n\left.\right|_{h_c}
\end{equation}}
\begin{figure}
	\centering
	\includegraphics[width=7.5cm,height=6cm]{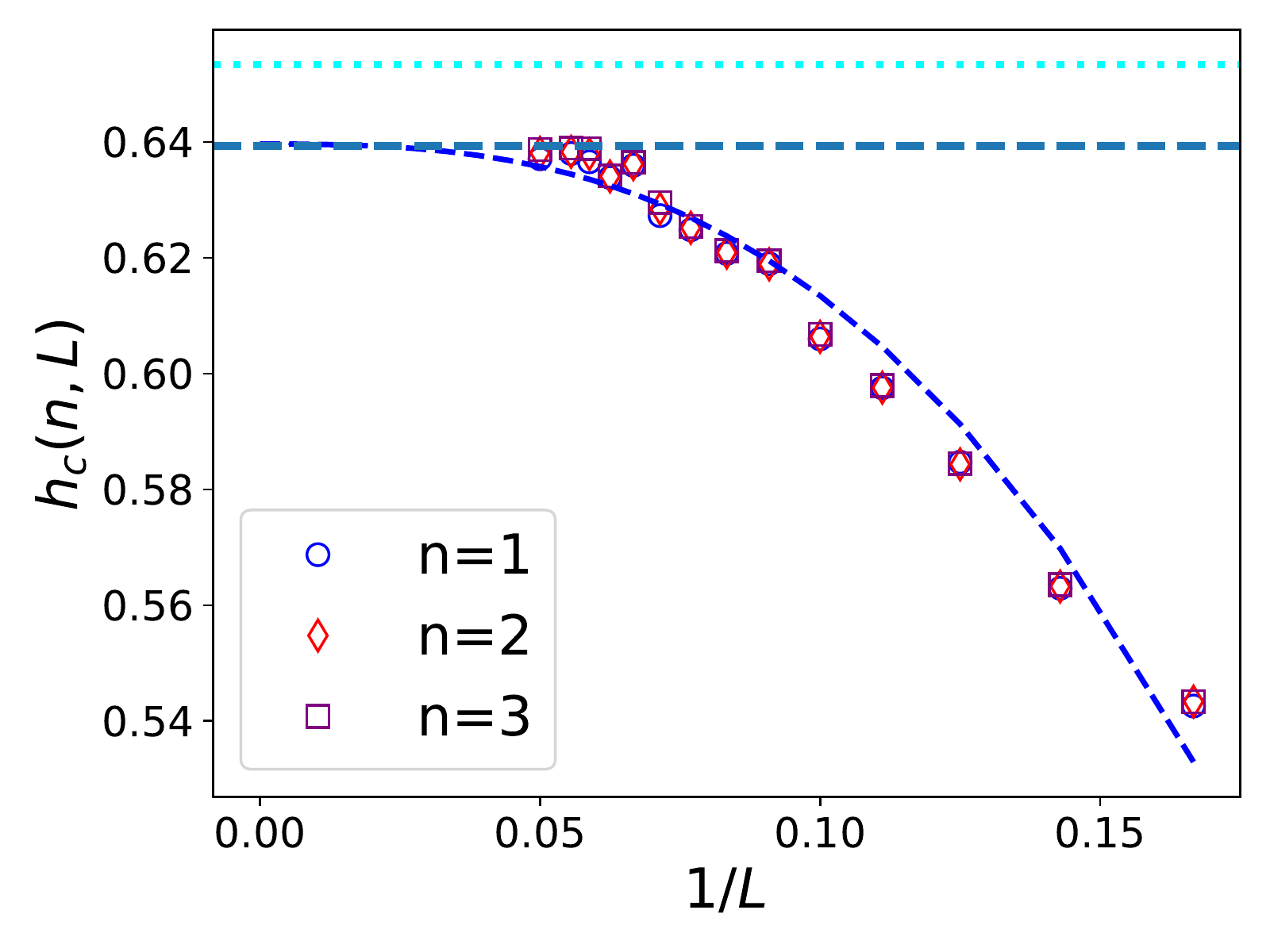}
	
	\caption{ {The flow of the QCP $h_c(n,L)$ extracted from the second derivatives of time- averaged (see Eq.~\eqref{eq:der_min_a}) finite strings having length $n$ in finite systems of $L$ spins. With increasing system size the $h_c(n,L)$ drifts fast towards the QCP $h_c=0.6393$ obtained from DMRG calculations (represented by the green horizontal line). The horizontal dotted line represents the position of the QCP obtained using a second order perturbation theory of the spectral gap vanishing (see Ref.~\cite{matteo06}). The dashed blue line corresponds to a highly non-linear numerical fit ($\sim L^{-2.75}$) of the convergence to the DMRG result. The time averaging window has been chosen to be $t\in[0,100]$ starting the quench from a completely polarized initial state $\ket{\rightarrow\rightarrow...}$, with antiferromagnetic interaction strength $J_2=-0.2$.}}
	\label{fig:drift} 
\end{figure}
 {to accurately extract QCPs following a quench. Similar to the integrable situation, the quantity $D_n$ develop very clear dips near equilibrium QCPs even for strongly chaotic quenches. We extract the critical magnetic field $h_c(n,L)$ from the minima of the second derivatives with finite system size and string length, while keeping the time averaging window sufficiently large. Interestingly, we do not detect any significant flow of the extracted QCP with string size. However, the extracted QCP shows a rapid non-linear flow (which is possibly not universal) towards the accurate DMRG value with increasing system size. In fact, for high enough system sizes, $h_c(n,L)$ quickly converges very close to the DMRG data for all finite string sizes (see Fig.~\ref{fig:drift}).}\\ 
\begin{figure*}[ht]
	\subfigure[]{
		\includegraphics[width=5.5cm,height=4.4cm]{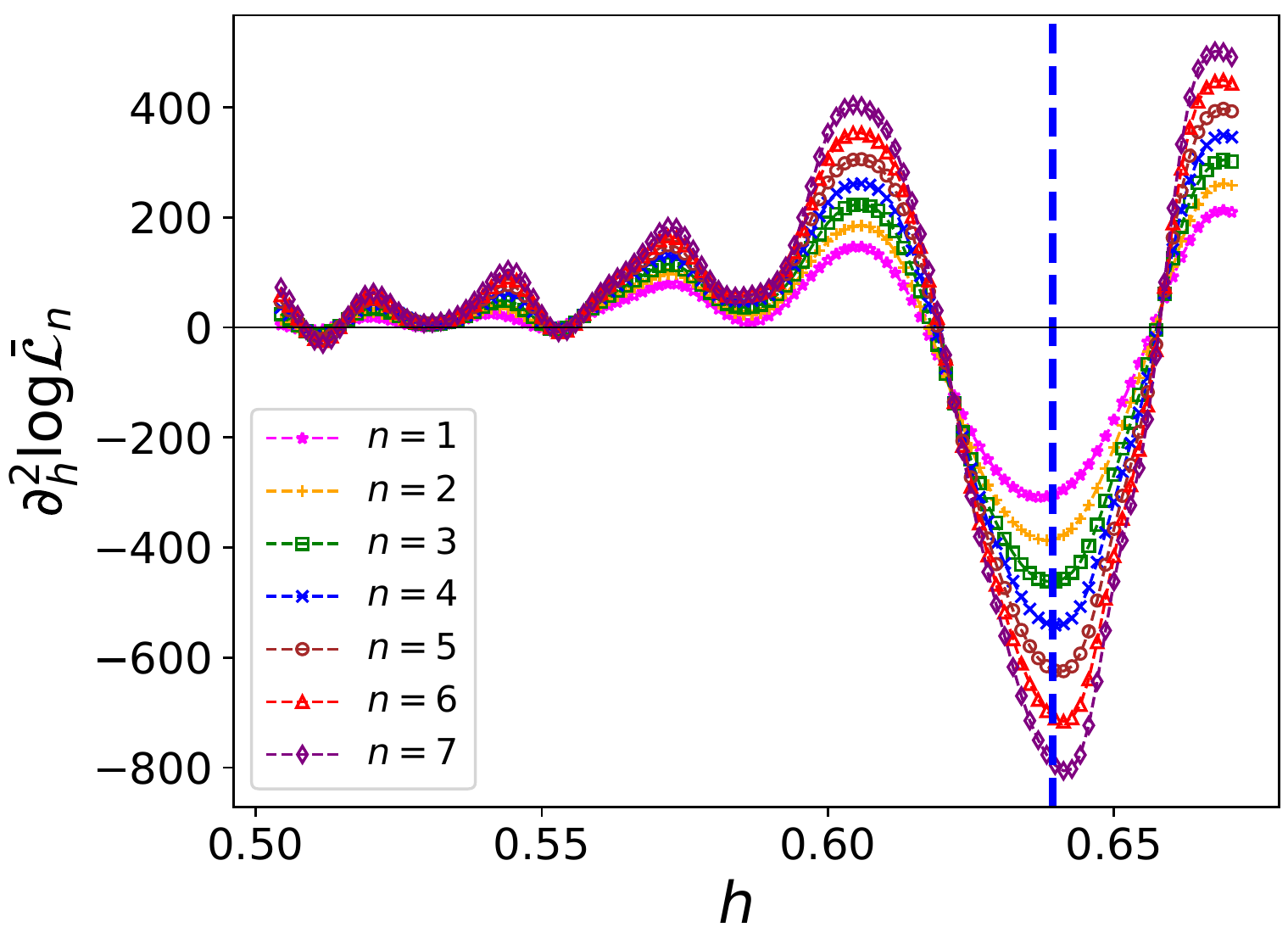}
		\label{fig:acc_a}}
	\subfigure[]{
		\includegraphics[width=5.3cm,height=4.4cm]{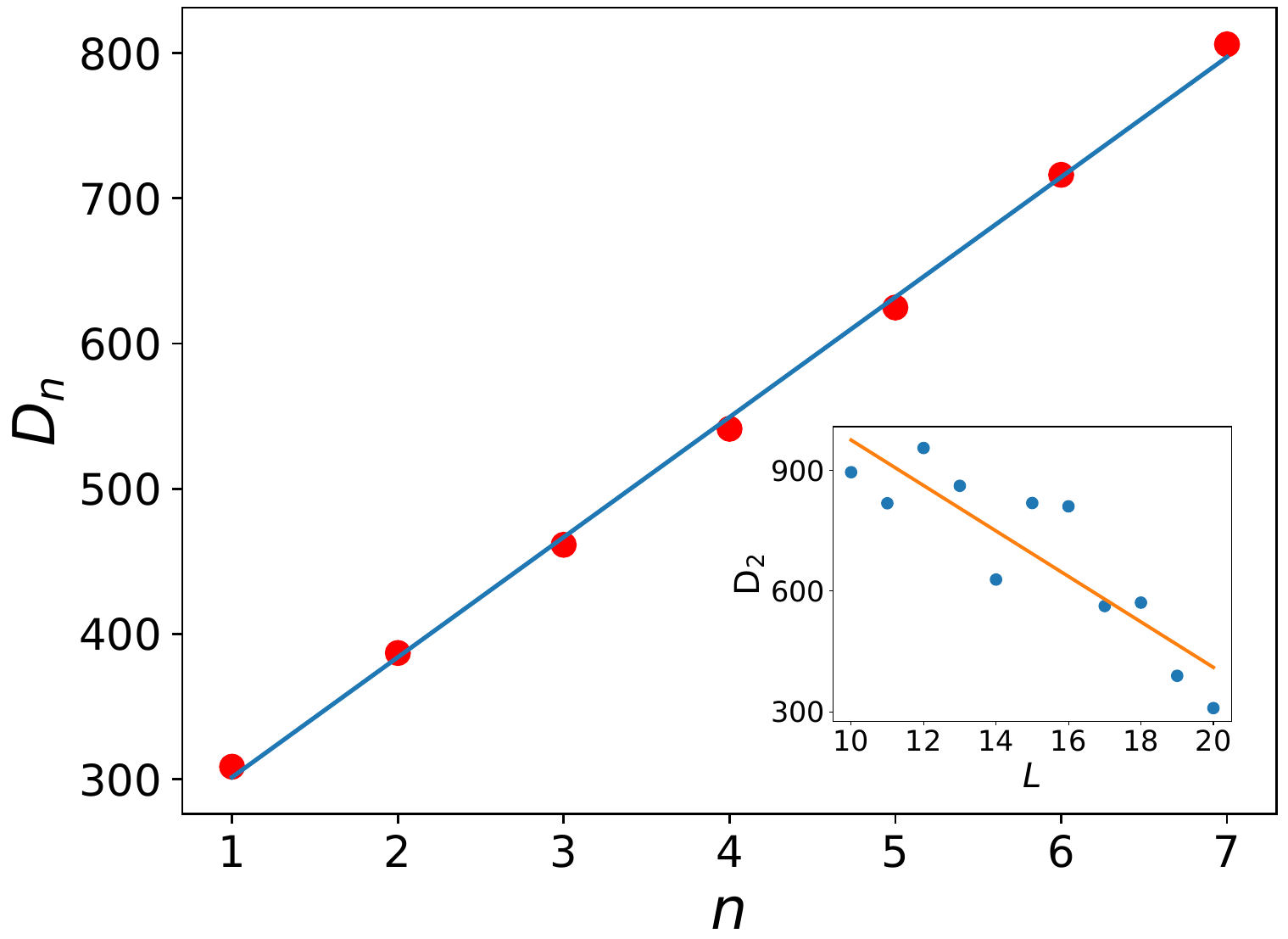}
		\label{fig:acc_b}}
		\subfigure[]{
		\includegraphics[width=5.8cm,height=4.5cm]{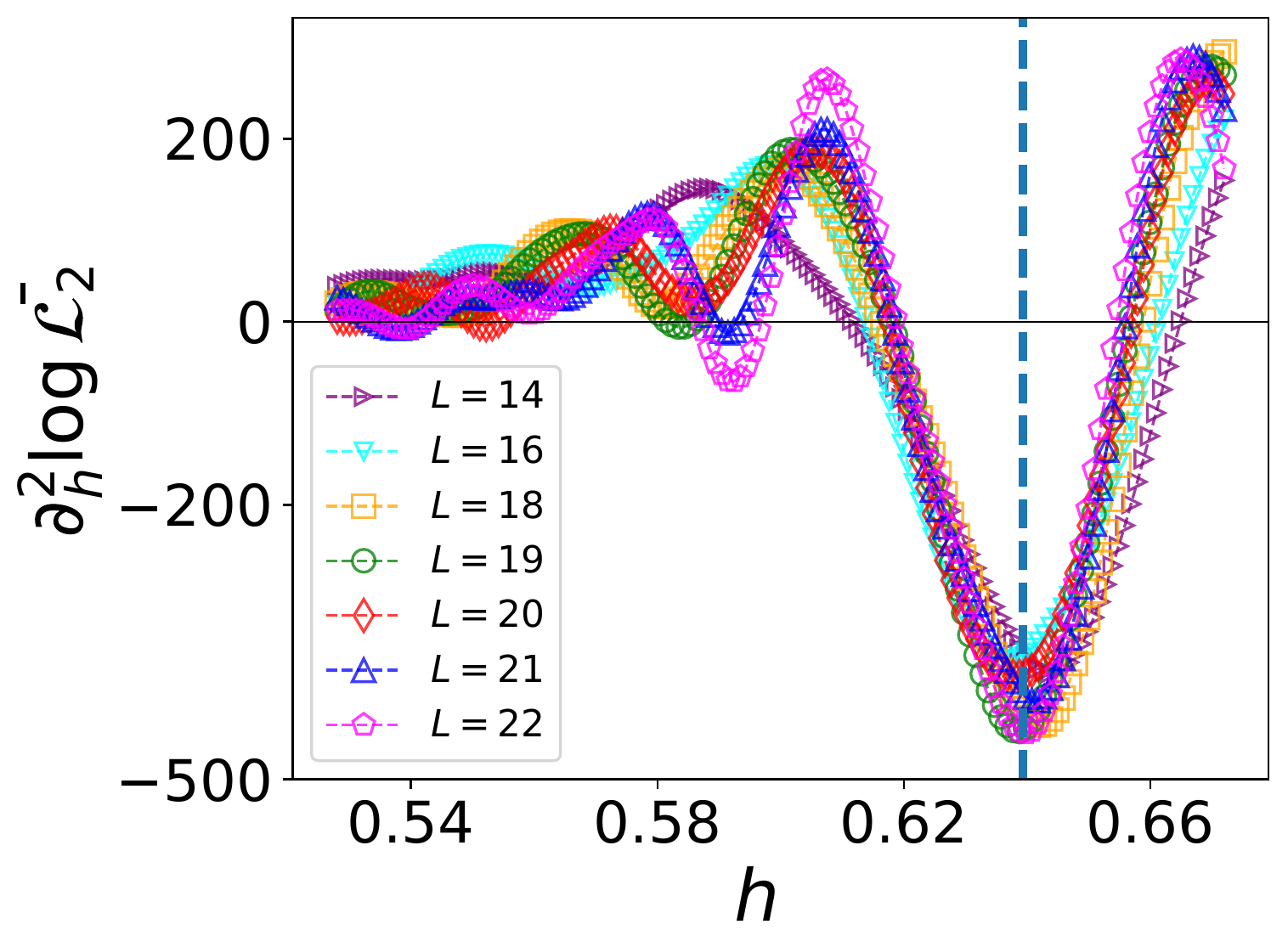}
		\label{fig:acc_c}}
	\caption{ {(a) The second derivative response of the time averaged distributions $\log\bar{\mathcal{L}_n}$ develops a sharp dip near the critical point which diverges with increasing string length $n$. (b) The derivative minima at the QCP (see Eq.~\eqref{eq:der_min_a}) scales linearly with string length and (Inset) decreases with increasing system size (shown for string length $n=2$). In both panels (a) and (b), the simulation has been done for $L=20$ spins with a time averaging period of $t\in[0,100]$. (c) The variation of the derivative dips is significantly slower with different system sizes when the time averaging period is cut-off such that $T\propto L$ (here taken to be $T=5L$) (demonstrated for $n=2$). In all the panels the quench was started from a completely polarized initial state $\ket{\rightarrow\rightarrow...}$ in the chaotic ANNNI model with $J_2=-0.2$. The vertical dashed lines indicate the position of the critical field $h_c=0.6393$ according to DMRG studies.}}
	\label{fig:scaling}
\end{figure*}

 {We further probe how the amplitude of the second derivative signal scales with string length and system size for sufficiently long time averaging. As seen in Fig.~\ref{fig:scaling}, for a fixed time averaging window, the derivative peaks indeed decrease with increasing system size. However, at the same time, they increase systematically and get sharper with the string size. This indicates that longer strings can still approach the non-analyticity at the QCP precisely even though single-site observables like magnetization might loose this information eventually due to thermalization (see for example Ref.~\cite{suchsland22}). Interestingly, in Fig.~\ref{fig:acc_c} we observe that that if the time averaging period $[0,T]$ is cut-off to finite but long times, such that $T\propto L$, the variation in the derivative signals with increasing $L$ slows down significantly.}

\section{Robustness against strong integrability breaking}
\label{Sec:acc}

 {We further probe the fate of the non-analyticities at the equilibrium QCP for strongly chaotic quenches. Particularly, in Fig.~\ref{fig:str_a} we observe that the second derivative dips detect the QCP accurately even when the integrability breaking interaction is of the same order as other energy scales in the system. Using a similar method described in Appendix.~\ref{Sec:sr_int}, we extract the critical field for $J_2=0.5$ to be $h_c^e=1.5987$, which is close to the approximately known value $1.6$, as has been verified through other methods in previous works (see for example Ref.~\cite{heyl18}). To further verify that we indeed have the correct critical field, we perform a self-consistency check by simulating an interaction quench starting from the extracted critical field $h_c^e$. By doing so, we expect a clear nonanalyticity in the observables $\mathcal{O}_n$ as a function of the quenched interaction $J_2$ near the accurately known critical value $J_2^c=0.5$, corresponding to the critical field $h_c$. In Fig.~\ref{fig:str_b} we demonstrate that this is indeed what we obtain following an interaction quench.}\\

 {Furthermore, in Fig.~\ref{fig:str_a}, it can be seen that though for very short strings the derivative dips broaden, making it difficult to extract the QCP, the signal becomes sharper with increasing string length for a fixed system size and time averaging window. This clearly demonstrate the advantage of using string observables to detect QCP in strongly chaotic systems over single-site observables. This is particularly interesting as measuring these string observables introduce no additional experimental challenges over single site observables in quantum simulators (as has been already demonstrated in Ref.~\cite{Zhang2017}). In Figs.~\ref{fig:str_c} and \ref{fig:str_d} we further verify the scaling of the time averaged observable $\mathcal{O}_n$ for local strings with string size $n$ and the finite-size scaling of the full LE, respectively, near the QCP in the limit of strongly non-integrable quenches.}\\

 {We further emphasize that the string observables also accurately detect any small shift in the critical point, which we demonstrate using weakly nonintegrable quenches in Fig.~\ref{fig:weak} and compare with the perturbatively obtained critical point in Eq.~\ref{eq:nonin_qcp}. }

\begin{figure*}[ht]
	\subfigure[]{
		\includegraphics[width=7.1cm,height=5.7cm]{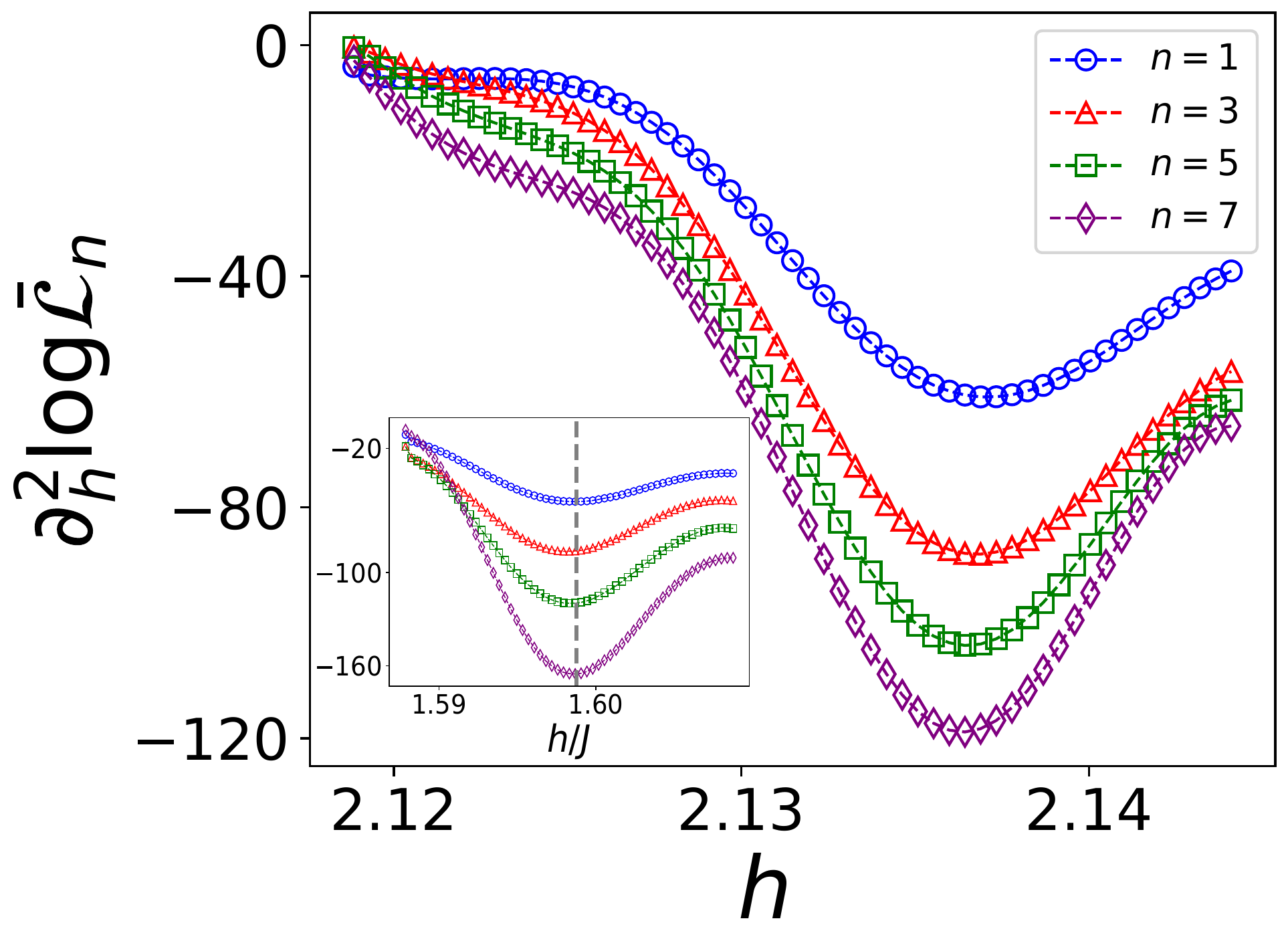}
		\label{fig:str_a}}
	\hspace{0.5cm}
	\subfigure[]{
		\includegraphics[width=6.8cm,height=5.6cm]{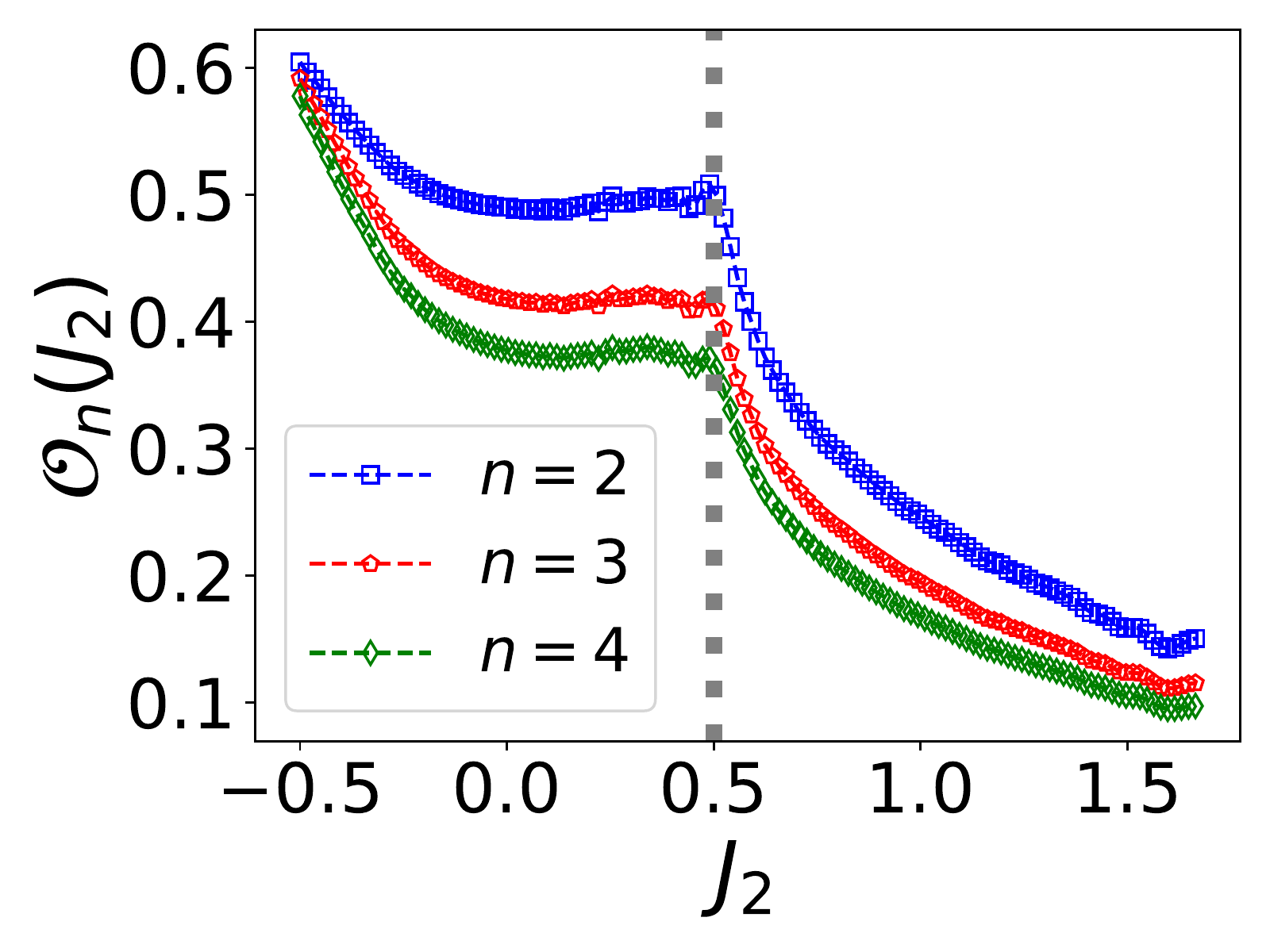}
		\label{fig:str_b}}

	\subfigure[]{
		\includegraphics[width=6.8cm,height=5.9cm]{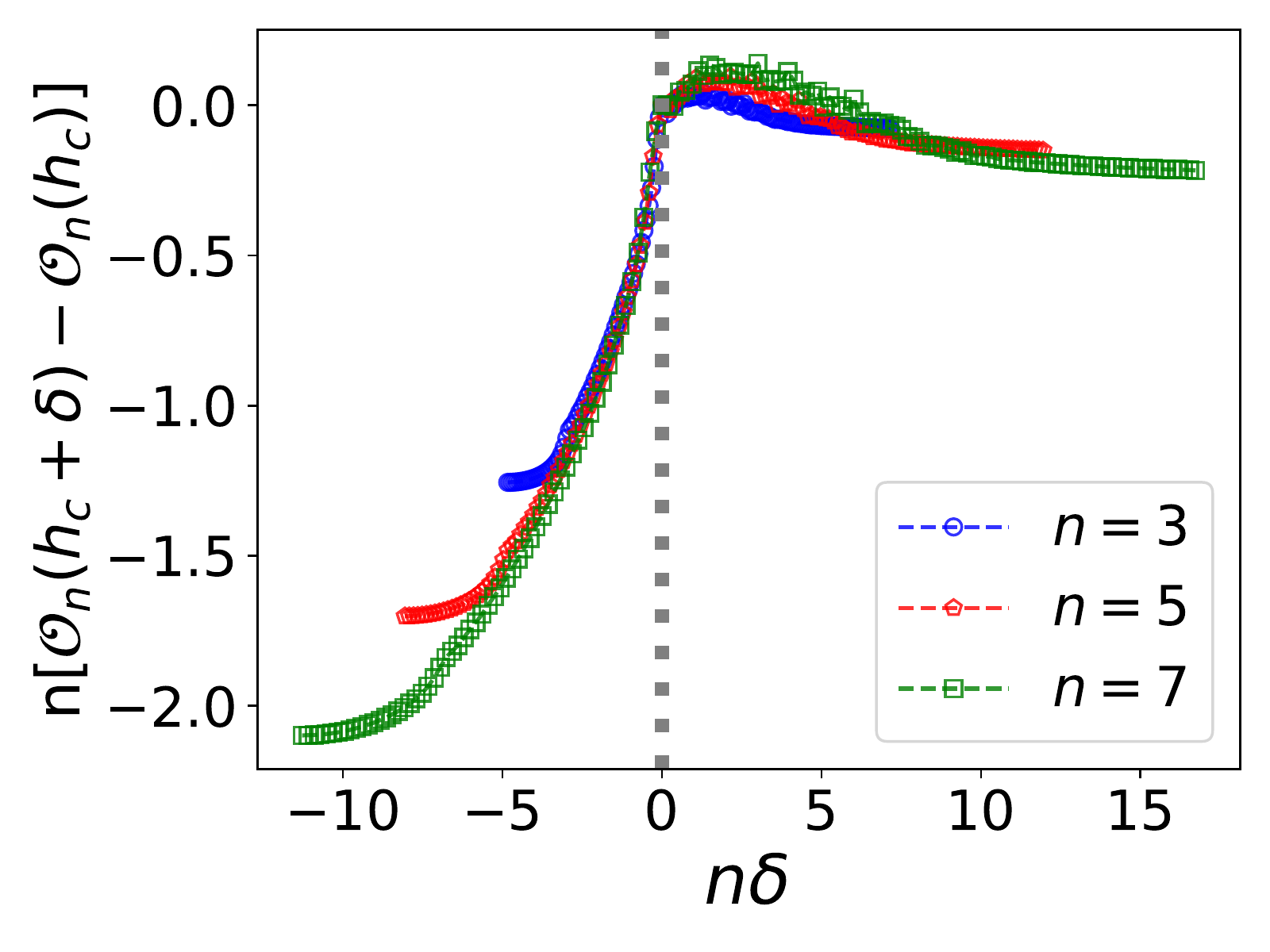}
		\label{fig:str_c}}
	\hspace{0.5cm}
	\subfigure[]{
		\includegraphics[width=7cm,height=6cm]{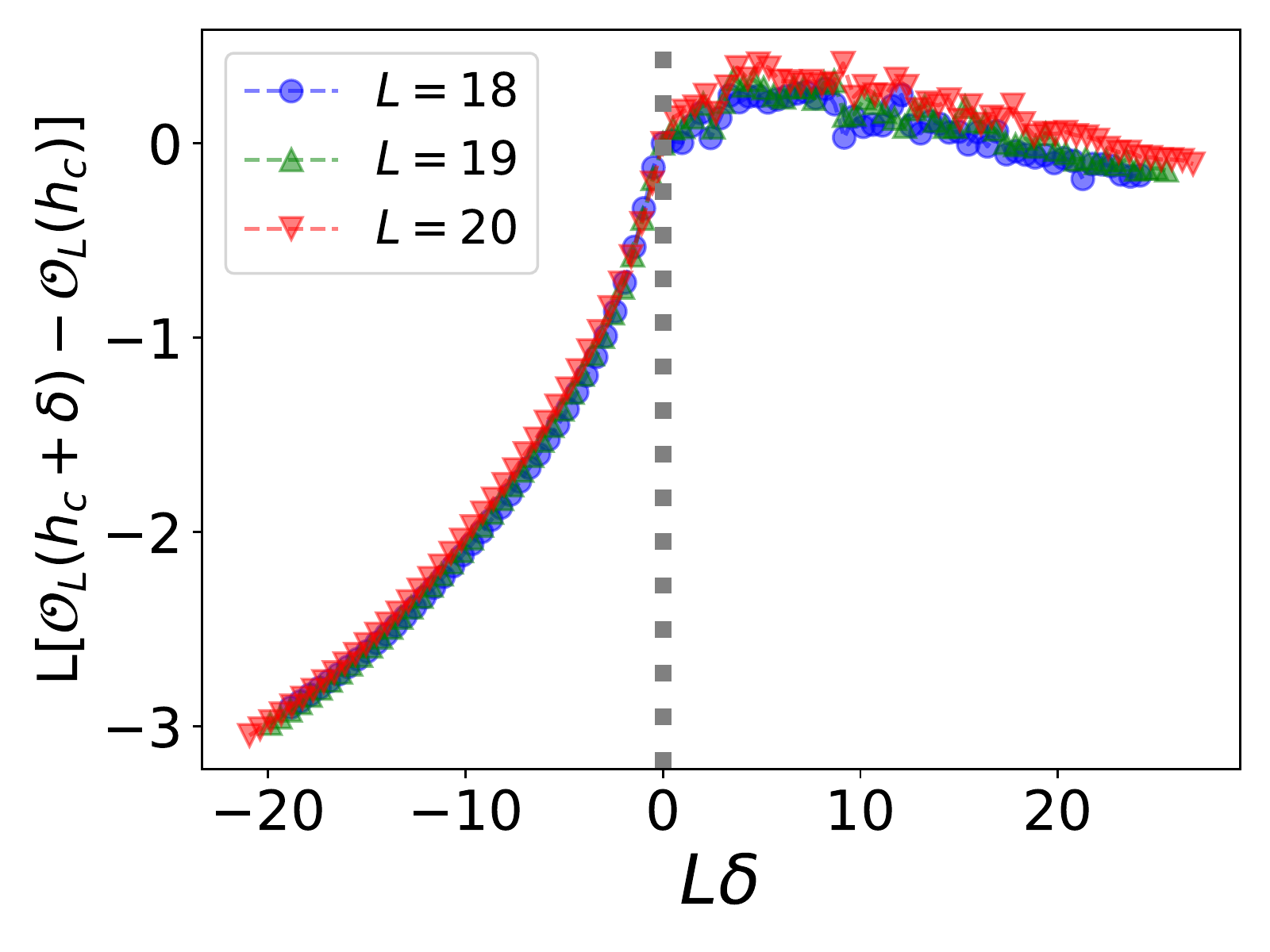}
		\label{fig:str_d}}
	
	\caption{ {(a) The second derivatives of the time averaged distributions $\log\bar{\mathcal{L}_n}$ develop a sharper dip near the QCP with increasing string length $n$ when the integrability breaking term $J_2=1.0$ and (Inset) $J_2=0.5$. The vertical dashed line corresponds to the extracted critical field $h_c^e=1.5987$ for $J_2=0.5$. (b) On quenching the next-nearest interaction $J_2$ while keeping the magnetic field fixed to $h=h_c^e$, the time averaged observables $\mathcal{O}_n$ as a function of the quenched interaction strength develop sharp signatures exactly at $J_2=0.5$. (c) Scaling collapse of $\mathcal{O}_n$ with string length (see Eq.~\eqref{eq:scaling_n}) and (d)  finite-size scaling collapse of the time averaged LE (see Eq.~\eqref{eq:scaling_1}) near the QCP against quenched field for $J_2=0.5$. In all the panels the simulation has been performed for $L=20$ spins with the fixed time averaging period $t\in[0,100]$. In all the panels the quench was started from a completely polarized initial state $\ket{\rightarrow\rightarrow...}$.}}
\end{figure*}

\begin{figure*}
	\subfigure[]{
		\includegraphics[width=6.5cm,height=5cm]{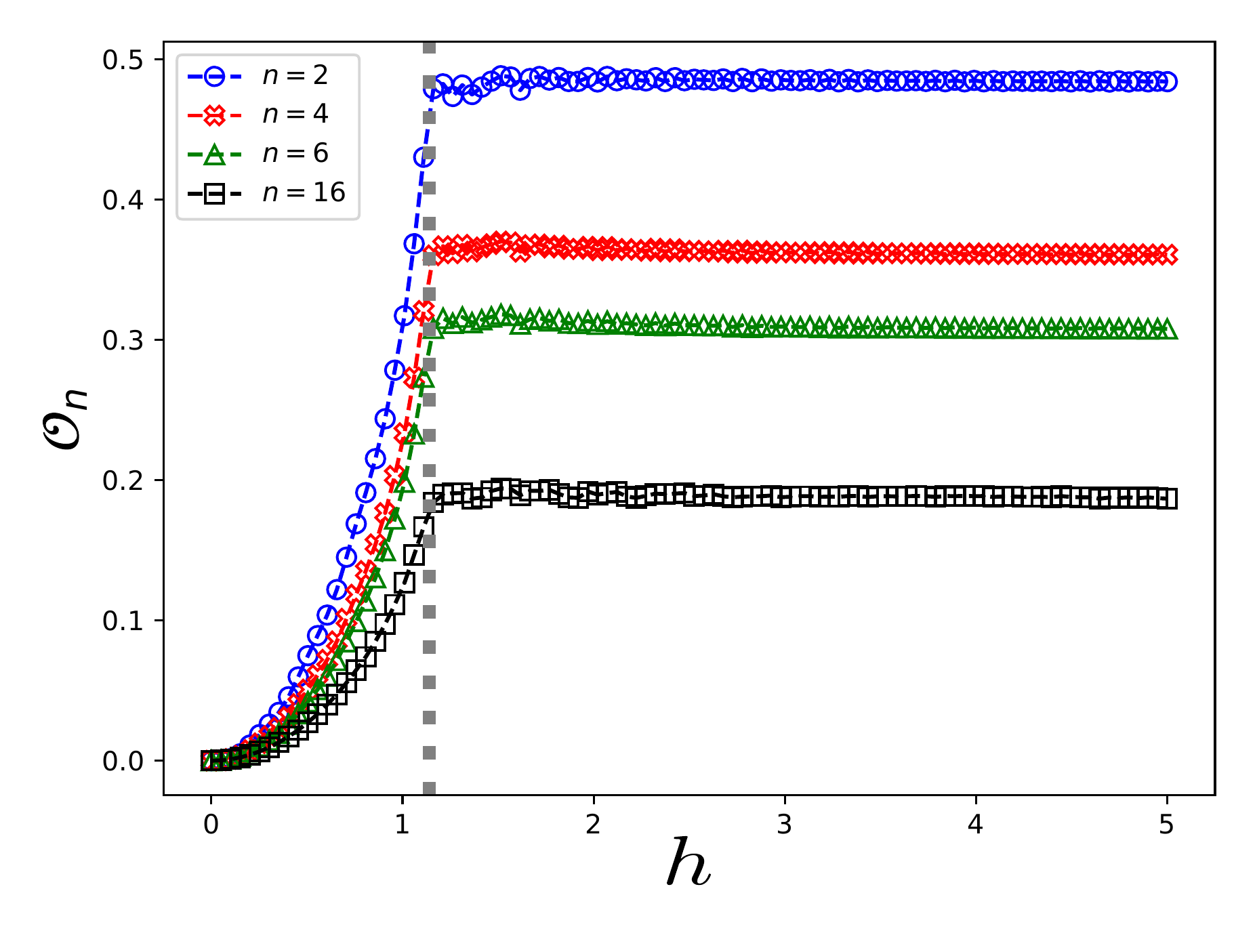}
		\label{fig:weak1}}
	\hspace{1cm}
	\subfigure[]{
		\includegraphics[width=6.5cm,height=5cm]{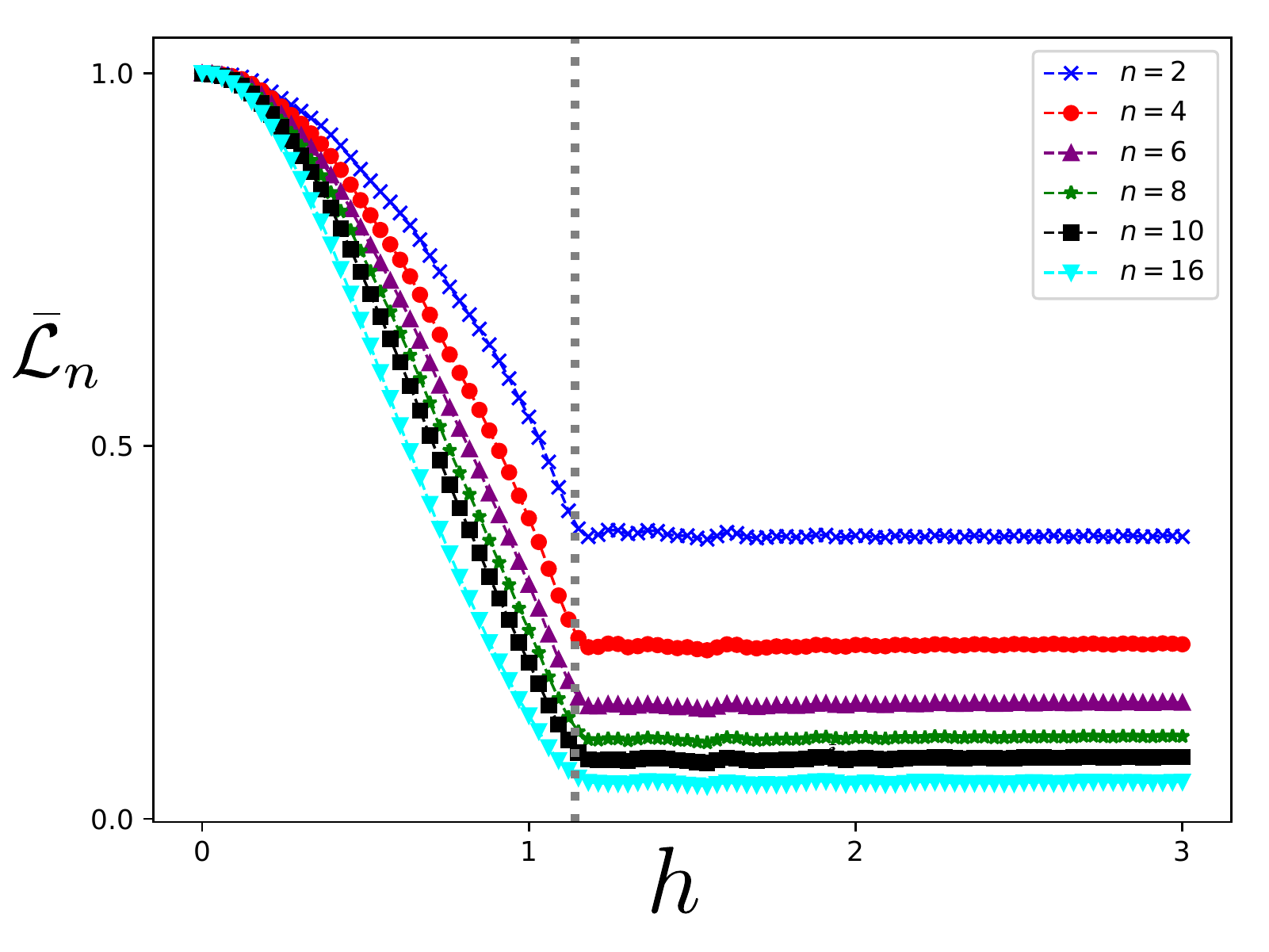}
		\label{fig:weak2}}
	
	\caption{ {(a) The observable $\mathcal{O}_n$ develop sharp signatures at the critical point even for a weakly chaotic quench with $J_2=0.1$. (b) The distributions $\bar{\mathcal{L}}_n$ detecting even a small shift in the critical point for $J_2=0.1$. In both the panels the quench was started from a completely polarized initial state $\ket{\rightarrow\rightarrow...}$ in a chain of $L=16$ spins. The vertical dashed line represents the critical point $h_c\approx 1.14$ obtained from the analytical perturbative expression in Eq.~\eqref{eq:nonin_qcp}.}}
	\label{fig:weak}
\end{figure*}

\section{Suppression of noise with increasing system size}
\label{sec:noise}

In Sec.~\ref{string_stat}, we noted that the time-averaged domain distributions in the paramagnetic phase changes very slowly with the quenched field $h$, apart from small fluctuations (see Eq.~\eqref{eq:Mprob} and Fig.~\ref{fig:9_cover}). We argue that this noise in the numerical data, particularly for quenches into the disordered phase, is due to a finite system size. In this regard, Fig.~\ref{fig:d_a} shows that for quenches into the paramagnetic phase, the long time averaged distribution $\mathcal{M}_1^+$ after $N=10$ uncorrelated measurement sequences, gradually approaches a slowly varying function of $h$ with increasing system sizes. Furthermore, the relative deviation of $\mathcal{M}_1^+$ from its mean over different quenched fields in the disordered phase,
\begin{equation}
	\sigma\left[\mathcal{M}_1^+\right]= \frac{\sqrt{{\rm Var}[\mathcal{M}_1^+]}}{\overline{\mathcal{M}_1^+}},
\end{equation}
decreases fast with increasing system size (see Fig.~\ref{fig:d_b}) regardless of the number of measurements $N$. Thus, in the thermodynamic limit, the functions $\mathcal{M}_n^+$ indeed approach a smooth distribution with the quenched field in the paramagnetic phase. This is also shown for an integrable quench in the thermodynamic limit in Fig.~\ref{fig:3a}.

\section{Overlap of a polarized initial state with the post-quench state}
\label{Sec: Appendix_2}

\begin{figure}
	\centering
	\includegraphics[width=7.5cm,height=5.5cm]{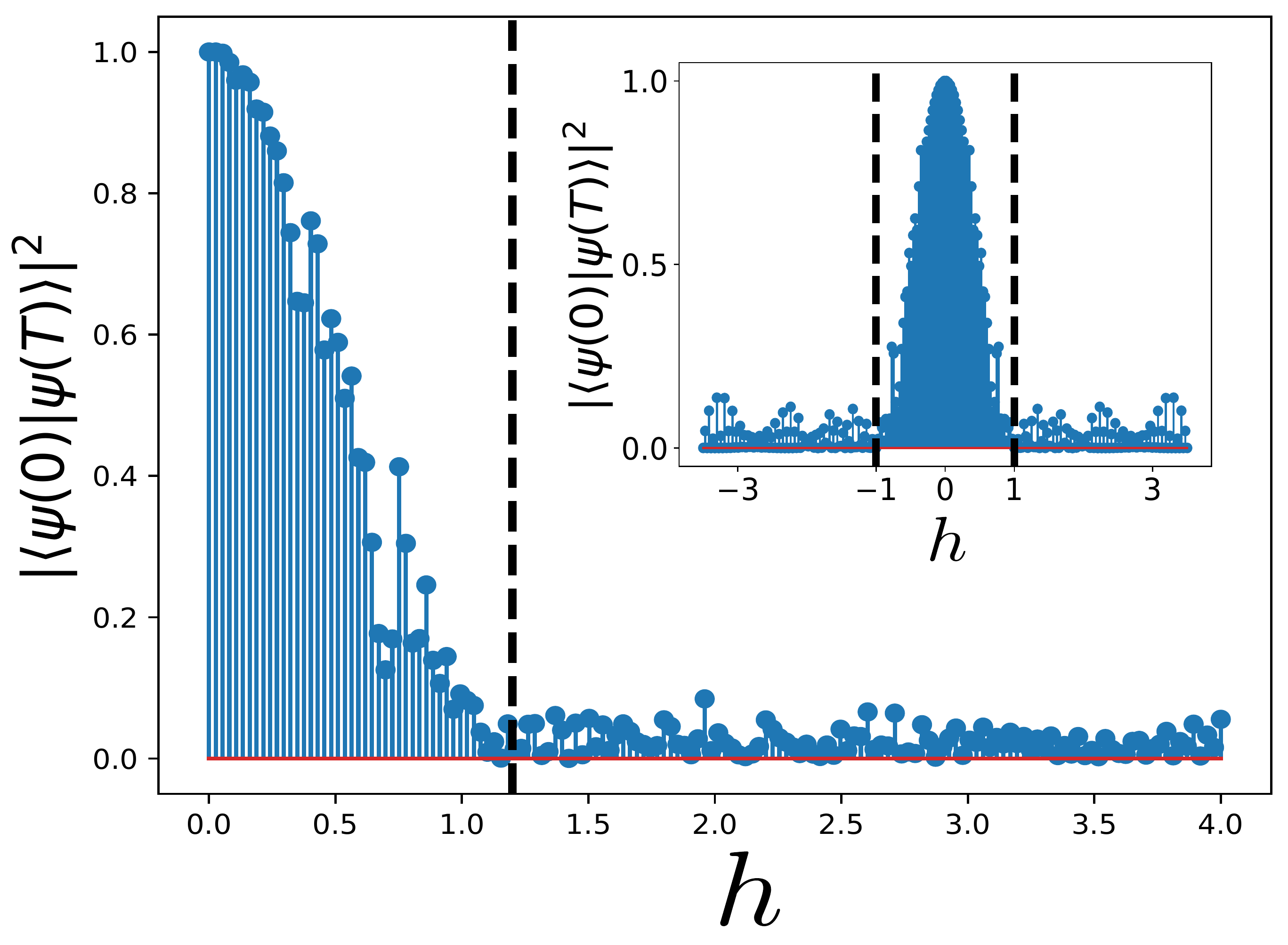}
	
	\caption{Overlap of the polarized initial state with the late time state following a quench in the nonintegrable ANNNI model ($J_2=0.1$) starting from a completely polarized state $\ket{\rightarrow\rightarrow...}$ to a transverse field $h$ in a system consisting of $L=20$ spins for $T=30$. (Inset) The corresponding distribution following an integrable quench. The vertical dashed lines indicate the analytical position of the critical point in both the plots.}
	\label{fig:11} 
\end{figure}
\begin{figure}
	\centering
	\includegraphics[width=7cm,height=6cm]{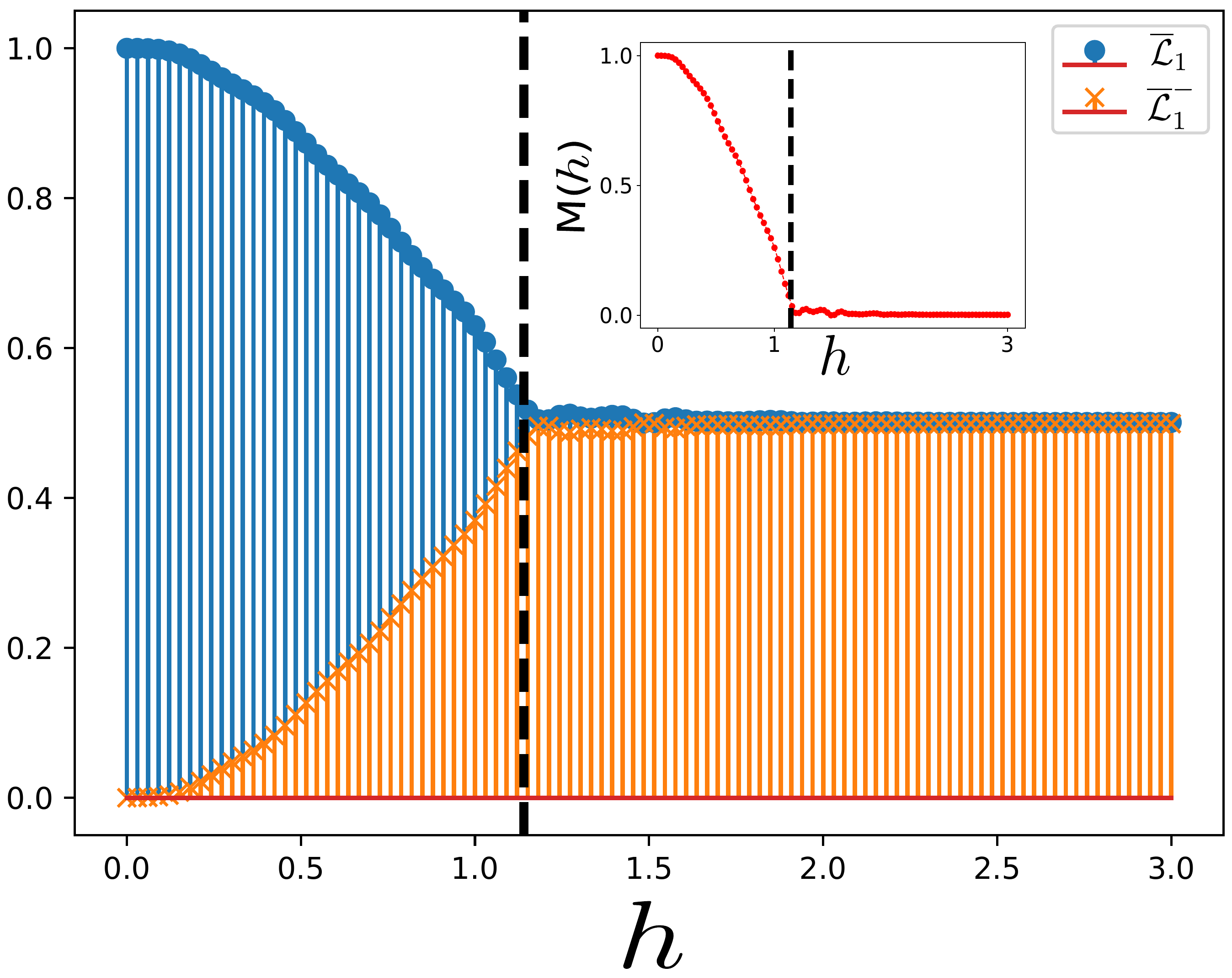}
	
	\caption{ Time averaged expectation of polarized single-spin projectors with respect to the quenched transverse field $h$, averaged up to $T=30$ following a quench in the chaotic ANNNI model ($J_2=0.1$). (Inset) The time averaged longitudinal magnetization density vanishes following quenches into the disordered phase. The time evolution was started from a polarized unentangled initial $\ket{\rightarrow\rightarrow...}$, in a system consisting of $L=18$ spins.}
	\label{fig:12} 
\end{figure}
In connection to our discussion in Sec.~\ref{string_stat}, in Fig.~\ref{fig:11} we calculate the overlap between the complete initial and time evolved states   $\lim\limits_{t\rightarrow\infty}|\braket{\psi(0)|\psi(t)}|^2$ with respect to the quenched transverse field following both nonintegrable and integrable quenches. As seen in Fig.~\ref{fig:11}, the completely polarized initial state has a significant overlap with the late time post-quench state for quenches in the ferromagnetic phase but the overlap becomes negligibly small following quenches into the paramagnetic phase. It is however important to note that at some finite time after quench, the single-shot distribution of larger domains can entail comparatively bigger temporal fluctuations as suggested from long relaxation times of longer string operators in Sec.~\ref{Sec:memory}.

\section{Time averaged local magnetization}
\label{Sec:Appendix_3}

As argued in Sec.~\ref{string_stat}, unlike quenches into the ordered phase, following quenches into the paramagnetic phase, domains of both polarizations contribute equally in the long time state. This can be explicitly seen by observing the behavior of finite but long time averaged local magnetization following a quench (see also Ref.~\cite{ettore20,halimeh21}). In Fig.~\ref{fig:12} we demonstrate this effect for a perturbative integrability breaking quench. We see that the time averaged expectation of single-spin projectors $\bar{\mathcal{L}}_1$ ($\bar{\mathcal{L}}^-_1$) into local $\rightarrow$($\leftarrow$)-spin polarizations, collapse into a single curve following quenches into the paramagnetic phase. This directly leads to the vanishing of local magnetization density,
\begin{equation}
	{\rm M}(h)=\bar{\mathcal{L}}_1-\bar{\mathcal{L}}_1^-,
\end{equation}
after finite but sufficiently long times following a weakly chaotic quench into the paramagnetic phase.

\section{Relaxation dynamics of string operators}
\label{Sec:Appendix_matele}

\begin{figure}[ht]
	\centering
	\includegraphics[width=8.25cm,height=6.25cm]{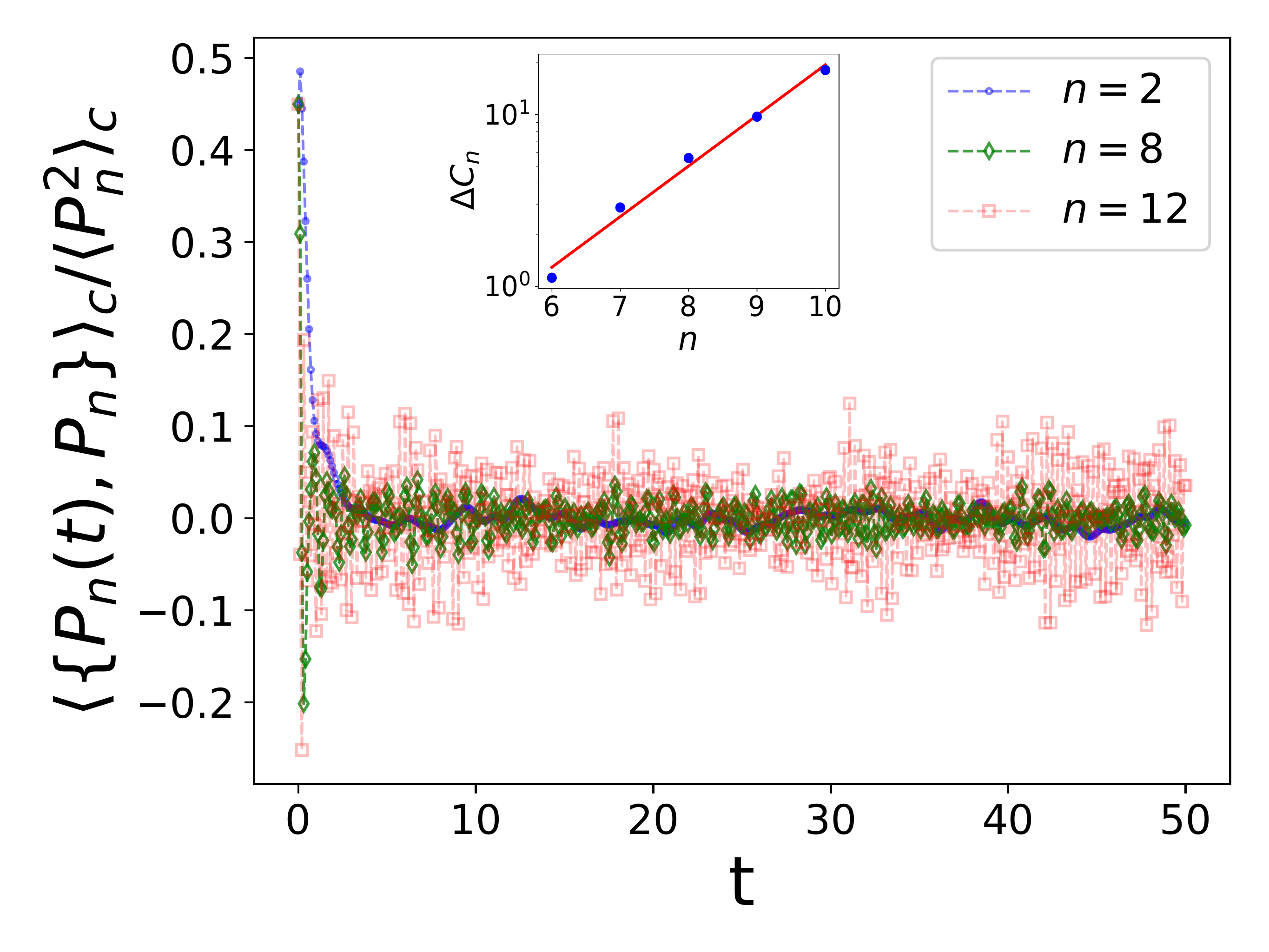}
	
	\caption{The connected auto-correlation function against time $t$ of string observables averaged over $20$ eigenstates about the center of the spectrum of the chaotic ANNNI Hamiltonian $H$ with $J_2=h=1$ for $L=14$ spins. After long times, longer string operators show faster oscillations with increasing amplitudes, indicating that they have comparatively longer lifetimes as compared to single-site observables. (Inset) The time averaged fluctuations for long string observables increases exponentially with string length.}
	\label{dyn_cor} 
\end{figure}

 In Sec.~\ref{Sec:memory} we observed that the low frequency spectral function of string operators approaches a sharply peaked distribution with increasing string length. This suggests late time oscillations in their real-time connected correlation functions about zero (see Eq.~\eqref{eq:realtime}), in an eigenstate $\ket{\phi_{\alpha}}$
\begin{equation}
{\rm C}_n^{\alpha}(t) = \frac{1}{2}\frac{\braket{\phi_{\alpha}|\{P_n(t),P_n(0)\}|\phi_{\alpha}}_c}{\braket{\phi_{\alpha}|P_n^2|\phi_{\alpha}}_c}.
\end{equation}
In Fig.~\ref{dyn_cor} we see that this is indeed what happens. Particularly, we plot the average connected correlation function ${\rm C}_n(t)$ over a few central eigenstates of the chaotic Hamiltonian. To further quantify the late time oscillations in the connected correlations, we calculate the variance,
\begin{equation}
\Delta {\rm C}_n = \frac{\sqrt{\overline{{\rm C}_n^2}-\overline{{\rm C}_n}^2}}{\overline{{\rm C}_n}},
\end{equation}
where the bar $\overline{\left(.\right)}$ represents time averaging. We find that for long string operators, the fluctuations increase almost exponentially with string length. 

\bibliographystyle{apsrev4-1}
\bibliography{bib_file}

\end{document}